\documentclass[longauth]{aa}  

\usepackage{graphicx}
\graphicspath{{./figures}}
\usepackage{txfonts}
\usepackage[hyphenbreaks]{breakurl}
\usepackage{hyperref}
\hypersetup{
    colorlinks = true,
    citecolor = {cyan},
}
\usepackage{xcolor}

\usepackage{subcaption}
\usepackage{threeparttable}

\def\photoz{photo-\textit{z}}

\def\Msol{{\rm M}_\odot}
\def\classic{\textsc{Classic}}
\def\lephare{\texttt{LePhare}}
\def\eazy{\texttt{EA$z$Y}}
\def\tractor{\texttt{The~Tractor}}
\def\farmer{\textsc{The~Farmer}}
\def\cfarmer{\texttt{The~Farmer}}

\def\Msol{\mathcal{M}_\odot}

\begin{document}

   \title{COSMOS2020: The galaxy stellar mass function\thanks{Data files containing sample IDs and key measurements are available for download:  \url{https://doi.org/10.5281/zenodo.7808832}.}}

   \subtitle{the assembly and star formation cessation of galaxies at $0.2<z\leq7.5$}

   \author{J.~R.~Weaver\inst{1,2,3}\fnmsep\thanks{Corresponding author, \email{john.weaver.astro@gmail.com}}
          \and
          I.~Davidzon\inst{1,2}
          \and
          S.~Toft\inst{1,2}
          \and
          O.~Ilbert\inst{4}
          \and
          H.~J.~McCracken\inst{5}
          \and 
          K.~M.~L.~Gould\inst{1,2}
          \and
          C.~K.~Jespersen\inst{6} 
          \and
          C.~Steinhardt\inst{1,2}
          \and
          C.~D.~P.~Lagos\inst{7,8,1}
          \and
          P.~L.~Capak\inst{1}
          \and
          C.~M.~Casey\inst{9}
          \and
          N.~Chartab\inst{10}
          \and
          A.~L.~Faisst\inst{11}
          \and
          C.~C.~Hayward\inst{12}
          \and
          J.~S.~Kartaltepe\inst{13}
          \and
          O.~B.~Kauffmann\inst{4}
          \and
          A.~M.~Koekemoer\inst{14}
          \and
          V.~Kokorev\inst{1,2}
          \and
          C.~Laigle\inst{5}
          \and
          D.~Liu\inst{15}
          \and
          A.~Long\inst{9}\fnmsep\thanks{NASA Hubble Fellow}
          \and
          G.~E.~Magdis\inst{1,16,2}
          \and
          C.~J.~R.~McPartland\inst{1,2}
          \and
          B.~Milvang-Jensen\inst{1,2}
          \and
          B.~Mobasher\inst{17}
          \and
          A.~Moneti\inst{5}
          \and
          Y.~Peng\inst{18,19}
          \and
          D.~B.~Sanders\inst{20}
          \and
          M.~Shuntov\inst{5}
          \and
          A.~Sneppen\inst{1,2}
          \and
          F.~Valentino\inst{1,2} 
          \and
          L.~Zalesky\inst{20}
          \and
          G.~Zamorani\inst{21}
          }

   \institute{
        Cosmic Dawn Center (DAWN), Denmark   
         \and 
        Niels Bohr Institute, University of Copenhagen, Jagtvej 128, 2200, Copenhagen N, Denmark
        \and 
        Department of Astronomy, University of Massachusetts, Amherst, MA 01003, USA
        \and
        Aix Marseille Univ, CNRS, CNES, LAM, Marseille, France  
        \and 
        Sorbonne Universit\'e, CNRS, UMR 7095, Institut d'Astrophysique de Paris, 98 bis bd Arago, 75014, Paris, France
        \and 
        Department of Astrophysical Sciences, Princeton University, Princeton, NJ 08544, USA
        \and 
        International Centre for Radio Astronomy Research (ICRAR), M468, University of Western Australia, 35 Stirling Hwy, Crawley, WA 6009, Australia
        \and 
        ARC Centre of Excellence for All Sky Astrophysics in 3 Dimensions (ASTRO 3D)
        \and 
        Department of Astronomy, The University of Texas at Austin, 2515 Speedway Blvd Stop C1400, Austin, TX 78712, USA
        \and
        The Observatories of the Carnegie Institution for Science, 813 Santa Barbara St., Pasadena, CA 91101, USA
        \and
        Caltech/IPAC, MS314-6, 1200 E. California Blvd. Pasadena, CA 91125, USA
        \and
        Center for Computational Astrophysics, Flatiron Institute, 162 Fifth Avenue, New York, NY 10010, USA
        \and
        Laboratory for Multiwavelength Astrophysics, School of Physics and Astronomy, Rochester Institute of Technology, 84 Lomb Memorial Drive, Rochester, NY 14623, USA
        \and 
        Space Telescope Science Institute, 3700 San Martin Dr., Baltimore, MD 21218, USA
        \and 
        Max-Planck-Institut f\"ur Extraterrestrische Physik (MPE), Giessenbachstr. 1, D-85748 Garching, Germany
        \and 
        DTU-Space, Technical University of Denmark, Elektrovej 327, 2800, Kgs. Lyngby, Denmark
        \and
        Physics and Astronomy Department, University of California, 900 University Avenue, Riverside, CA 92521, USA
        \and 
        Kavli Institute for Astronomy and Astrophysics, Peking University, 5 Yiheyuan Road, Beijing 100871, People’s Republic of China
        \and 
        Department of Astronomy, School of Physics, Peking University, 5 Yiheyuan Road, Beijing 100871, People’s Republic of China
        \and 
        Institute for Astronomy, University of Hawaii, 2680 Woodlawn Drive, Honolulu, HI 96822, USA
        \and 
        Istituto Nazionale di Astrofisica - Osservatorio di Astrofisica e Scienza dello Spazio, via Gobetti 93/3, I-40129, Bologna, Italy
             }

   \date{}

 
  \abstract
   {How galaxies form, assemble, and cease their star formation is a central question within the modern landscape of galaxy evolution studies. These processes are indelibly imprinted on the galaxy stellar mass function (SMF), and its measurement and understanding is key to uncovering a unified theory of galaxy evolution.}
   {We present constraints on the shape and evolution of the galaxy SMF, the quiescent galaxy fraction, and the cosmic stellar mass density across 90\% of the history of the Universe from $z=7.5\rightarrow0.2$ as a means to study the physical processes that underpin galaxy evolution.}
   {The COSMOS survey is an ideal laboratory for studying representative galaxy samples. Now equipped with deeper and more homogeneous near-infrared coverage exploited by the COSMOS2020 catalog, we leverage the large 1.27\,deg$^{2}$ effective area to improve sample statistics and understand spatial variations (cosmic variance) -- particularly for rare, massive galaxies -- and push to higher redshifts with greater confidence and mass completeness than previous studies. We divide the total stellar mass function into star-forming and quiescent subsamples through $NUVrJ$ color-color selection. The measurements are then fit with single- and double-component Schechter functions to infer the intrinsic galaxy stellar mass function, the evolution of its key parameters, and the cosmic stellar mass density out to $z=7.5$. Finally, we compare our measurements to predictions from state-of-the-art cosmological simulations and theoretical dark matter halo mass functions.}
   {We find a smooth, monotonic evolution in the galaxy stellar mass function since $z=7.5$, in general agreement with previous studies. The number density of star-forming systems have undergone remarkably consistent growth spanning four decades in stellar mass from $z=7.5\rightarrow2$ whereupon high-mass systems become predominantly quiescent (``downsizing''). Meanwhile, the assembly and growth of low-mass quiescent systems only occurred recently, and rapidly. An excess of massive systems at $z\approx2.5-5.5$ with strikingly red colors, with some being newly identified, increase the observed number densities to the point where the SMF cannot be reconciled with a Schechter function. 
   }
  {Systematics including cosmic variance and/or active galactic nuclei contamination are unlikely to fully explain this excess, and so we speculate that they may be dust-obscured populations similar to those found in far infrared surveys. Furthermore, we find a sustained agreement from $z\approx3-6$ between the stellar and dark matter halo mass functions for the most massive systems, suggesting that star formation in massive halos may be more efficient at early times.}


   \maketitle

%

\section{Introduction}

The galaxy stellar mass function (SMF) is defined as the number density of galaxies $\Phi(\mathcal{M}, z)$ in bins of stellar mass $\Delta\mathcal{M}$ at each redshift $z$, and is a fundamental cosmological observable in the study of the statistical properties of galaxies. Understanding its shape and evolution ultimately informs us about the growth of the baryonic content of the Universe, and can help infer the star formation activity and the availability of cold molecular gas across cosmic time. Its integral over $\mathcal{M}$ is the galaxy stellar mass density (SMD) $\rho_*(z)$, which describes the cumulative mass assembled by a given epoch.

Remarkable progress has been made since the first mass-selected measurements of the local $z\approx0$ SMF by \citet{Cole2001}. The intervening years have been marked by order-of-magnitude increases in sample size, photometric precision, and redshift accuracy which have enabled more precise determinations of the local SMF as well as groundbreaking extensions to increasingly higher redshifts \citep[e.g.,][]{ Fontana2006, Marchesini2009, Drory2009, peng10_quenching, Ilbert_2010, Pozzetti2010, Mortlock2011, Vulcani2013, muzzin13_mass, ilbert_mass_2013, Grazian2015, Mortlock2015, Duncan2014, Song2016, Bernardi2017, Davidzon17_mass, Wright2018, Santini2021, Stefanon2021, Adams2021,  Thorne2021, McLeod2021, Santini2022}. 

A coherent picture has emerged from these studies. For instance, the shape of the SMF for star-forming galaxies at $z<3$ is well described by the empirically constructed Schechter function \citep{schechter1976}, which features an exponential downturn at a characteristic stellar mass $\mathcal{M}^*$, with a low-mass end that declines with a slope $\alpha$; both seem to remain constant out to $z\approx2$ \citep{Marchesini2009, ilbert_mass_2013, Tomczak14, Davidzon17_mass, Adams2021}. This constancy in the shape of the SMF implies that star formation activity has acted consistently to transform baryons from cold gas into stars, and appears to have formed $>75\%$ of the total stellar mass of the Universe in $\approx10\,$Gyr. Only the normalization of the SMF, $\Phi^*(z)$, of star-forming galaxies has been confidently shown to evolve with cosmic time, and its rapid evolution at earlier times points directly to enhanced rates of mass growth at earlier epochs \citep[][and references therein]{Popesso2022}.

Which physical processes are responsible for the behavior of the SMF, and the extent of their influence at different cosmic epochs, are not well understood even in the low redshift Universe ($z<1$). Examples include feedback from supermassive black holes, supernovae, galaxy-galaxy mergers, stellar and gas dynamics, and gaseous inflows and outflows, all of which act on varying scales both physically and in time. As a result, significant effort has been undertaken to build detailed cosmological simulations to identify and understand the role of physical processes that underpin the observed shape and evolution of the SMF \citep[e.g.,][]{Somerville2015, Furlong2015_EAGLE, Lagos2018_SHARK, laigle19_horizonAGN, Pillepich2018_ILLUSTRIS, Dave2019_SIMBA, Lovell2021_FLARES}. Hence, precise measurements of the SMF and SMD are key observables utilized by most (but not all, see e.g., \citealt{Dubois2014}) large-scale cosmological simulations to tune input parameters such that the SMF of the local Universe ($z\approx0$) is recovered. Comparisons of predicted and observed SMFs at earlier cosmic ages can point to the physical processes at play and/or suggest recipes in need of refinement \citep{Torrey2014, Furlong2015_EAGLE}.

Owing to the challenges associated with large-scale simulations, purely analytical, data-driven models have enjoyed great popularity with the introduction of the first large galaxy samples with photometric redshifts (\photoz{}). For example, \citet{peng10_quenching} constructed a phenomenological model to explain the bimodality of galaxy types (star-forming and quiescent) as seen from their mass functions, megaparsec-scale environment, and star formation activity. That is, as a consequence of two mechanisms driving star formation cessation (often referred to as ``quenching''): massive galaxies cease forming stars irrespective of environment (``mass quenching''), and galaxies in dense environments cease forming stars irrespective of mass (``environmental quenching''). Several discrete mechanisms have been proposed to explain the latter, such as ram pressure stripping \citep[e.g.,][]{Gunn1972}, gas strangulation \citep[e.g.,][]{Larson1980, Balogh2000}, and dynamical heating of gas within haloes \citep[e.g.,][]{Gabor2015}. Proposed mechanisms for the former must, by definition, involve secular processes. This includes star formation cessation due to structural changes, or heating and/or ejection of gas by central super-massive black holes for the most massive galaxies -- radiative feedback from active galactic nuclei (AGN) -- or by supernovae for less massive systems as they are more weakly self-gravitating \citep[e.g.,][]{Shankar2006}. \citet{peng10_quenching} hypothesized that while environmental effects reproduce the single Schechter shape observed for star-forming, blue galaxies in the local Universe, it is through a combination of both environmental and mass effects that the two-component Schechter shape observed for quiescent, red galaxies is reproduced. A wide variety of extensions have been applied to this model to directly incorporate other measurements such as the gas fraction \citep{Bouche2010}, and wholly new models continue to be developed \citep[e.g.,][]{Peng2015, Belli2019, Suess2021, Varma2021}.

The observed distributions of stellar mass and redshift are not in themselves the intrinsic, underlying SMF. Instead, they are bridged by statistical inferences between carefully selected samples and a vast, practically unobservable parent population from which a sample is taken. Insufficient control of biases and systematics obscure such inferences by creating unrepresentative samples, and hence weaken comparisons with simulation and analytical models \citep{Marchesini2009, Fontanot2009}. Great effort has been expended over the past 20 years to strip away a number of these biases, and their obscuring effects \citep{Cole2001, Pozzetti2010, Marchesini2009, Bernardi2017, Davidzon17_mass, Leja2019, Adams2021}. Examples of these biases include sample incompleteness (including Malmquist biases and mass completeness), the so-called ``cosmic'' variance relating to sampling over- and under-dense regions of large scale structure, and Eddington bias which acts to overestimate the number density of the most massive galaxies \citep{malmquist1920, malmquist1922, Eddington1913}. While many studies incorporate only poisson noise \citep[][although this is steadily improving with time]{Fontanot2009}, other notable uncertainties exist including sample variance and effective volume size, as well as uncertainties on \photoz{} and stellar mass; the latter two items can introduce significant bias. Since the ultimate goal of any survey is to generalize the observed properties of a specific sample to that of the entire population, ignoring any of these important biases or uncertainties severely complicate this effort.

Studies of the SMF at increasingly higher redshift have been made possible due to advances in facilities and continued investment in deep, primarily photometric surveys of the distant Universe. Building on work by \citet{Songaila_1994} and \citet{Glazebrook_1995}, \citet{Cowie1996} secured the first mass-complete samples at $z\approx2-3$, selected in the near-infrared ($K$) to directly constrain the Balmer continuum at $\sim0.6\,\mu$m. Sampled nearer the stellar bulk at $\sim1-2\,\mu$m, their mass-selected samples enabled a higher degree of mass completeness, and in turn provided a more representative view of the high-$z$ Universe compared to optically selected forerunner surveys. More recently, precise mass estimates from similarly selected samples have been obtained by fitting observed spectral energy distributions \citep{Tomczak14, Martis2016, Straatman2015}, from the ground (with VISTA, UKIRT) and space (\textit{HST}/WFC3, \textit{Spitzer}/IRAC). Although limited in area, samples recovered by \textit{HST} have continued to pose new and significant challenges to existing paradigms by highlighting a series of stark changes in the shape of the SMF that indicates earlier galaxy populations were fundamentally different from those in the present day Universe. Additional studies at these early times ($z\gtrsim2$) by degree-scale surveys capable of finding the rarest sources have revealed the existence of surprisingly mature, massive quiescent galaxies whose mass has already been assembled by $z\approx4$ \citep[e.g.,][]{ilbert_mass_2013}, challenging the assumed timeline typical for galaxy assembly \citep{Steinhardt2016, Behroozi2018}. Limited numbers of them have been confirmed by detailed spectroscopic follow-up \citep[e.g.,][]{Glazebrook2017, Schreiber18_QGz4, Tanaka2019, Valentino20_QGz4, Forrest2020a, Forrest2020b}. Likewise, several studies have placed constraints on the massive end of the SMF at the highest-redshifts ($z>6$) from samples selected by their rest-frame UV luminosity \citep[e.g.,][]{Song2016, Harikane2016}, who utilized empirically calibrated $L_{\rm UV}-\mathcal{M}$ relations to convert directly measured UV luminosity into indirect estimates of stellar mass.

Although tantalizing, deriving mass estimates from UV luminosity involves significant uncertainties; only deep infrared photometry can directly measure the rest-frame stellar bulk ($\lambda>4000\,\AA$, although ideally $\sim 1-2\,\mu$m) required to more directly assess stellar mass at these redshifts. Only recently have deep ($K_s\approx25$) near-infrared photometric studies enabled the first mass-selected samples at $z>2$, although with mass uncertainties increasing with redshift \citep{Retzlaff2010, Fontana2014, laigle_cosmos2015_2016}. Importantly, photometric surveys of galaxies have -- and are expected to remain -- the primary means by which these measurements will be made; obtaining even elementary parameters for large ($N>100\,000$) samples with deep spectroscopy, while precise, is simply too expensive even when utilizing highly multiplexed instrumentation. For the time being, spectroscopy of the distant Universe will remain a follow-up exercise to strengthen larger photometrically selected samples. 

Of these deep photometric surveys, those with large areas have a key advantage: they probe a wider range of environments (i.e., density) compared to narrower surveys. As such, they provide more representative samples that are significantly less likely to be affected by field-to-field cosmic variances (and, for the same reason, are more suited to find rare, massive galaxies). Although one may resort to combining disparate data sets into a single analysis to combat this \citep[e.g.,][]{Moster2013, Henriques2015, Volonteri2016, McLeod2021}, the unknown systematics between survey selection functions and depths make the interpretation of these joined samples more uncertain.

Stitching together surveys of low- and high-$z$ samples carries similar concerns. A lack of uniform selection owing to different survey areas, detection bands, and \photoz{} determination strategies (i.e., dropout selection) can likewise complicate interpretations arising from such composite samples \citep[e.g.,][]{Leja2019, McLeod2021, Adams2021, Santini2022}. Worse, these photometric samples may have been processed with different spectral energy distribution (SED) fitting codes that assume different templates, emission line recipes, and dust attenuation laws. Differing choices of cosmology and initial mass function, while reversible, nonetheless add complexity. 

Accurate estimates of the galaxy stellar mass function at increasingly higher redshifts requires complementary deep observations from near-infrared selected samples to ensure both mass completeness and data reliability. Currently the widest field with near-infrared coverage of the requisite depth to probe large samples of $z\gg3$ galaxies is the Cosmic Evolution Survey \citep[COSMOS;][]{scoville_cosmic_2007}. This work takes advantage of the latest NIR-selected catalog of the COSMOS field, COSMOS2020 \citep{Weaver2022_catalog}. We adopt the ``\farmer{}'' catalog whose photometry was consistently measured across all bands by fitting galaxy profiles while taking into account PSF variations, depth, and source crowding. From the resulting \photoz{} and stellar masses we construct a consistently measured SMF from $z=0.2-7.5$, identify quiescent and star-forming systems, and study the build-up and assembly of stellar mass over 10\,Gyr of cosmic history. This includes a detailed study of key moments in the development of galaxy populations from the Era of Reionization ($z>6$) to the peak of star formation activity at Cosmic Noon ($z\sim2$), up to the rich bimodality of star-forming and quiescent galaxies observed at more recent times ($z\lesssim1$). 

This paper is organized as follows. Section~\ref{sec:data} introduces the dataset chosen for this analysis from which samples are drawn and possible uncertainties discussed in Section~\ref{sec:selection}. Section~\ref{sec:schechter} provides a brief overview of the Schechter function. Results are presented in Section~\ref{sec:results}, including the presentation and fitting of the total, star-forming, and quiescent mass functions. These are compared to literature measurements in Section~\ref{sec:discussion}, whereupon further discussion is had with regards to galaxy assembly, star formation cessation, and connections to dark matter halos. This work concludes in Section~\ref{sec:conclusions}.

These results are computed adopting a standard $\Lambda$CDM cosmology with $H_0=70$\,km\,s$^{-1}$\,Mpc$^{-1}$, $\Omega_{\rm m,0}=0.3$, and $\Omega_{\Lambda,0}=0.7$ throughout, such that the dimensionless Hubble parameter $h_{70} \equiv H_{0}/(70\,\mathrm{km}\,\mathrm{s}^{-1}\,\mathrm{Mpc}^{-1})=1$. Galaxy stellar masses, when derived from SED fitting, scale as the square of the luminosity distance (i.e., $D_L^2$); hence a factor of $h_{70}^{-1}$ is retained implicitly for all relevant measurements, unless explicitly noted otherwise (see \citealt{Croton2013} for an overview of $h$ and best practices). Estimates of stellar mass (henceforth $\mathcal{M}$) assume an \citet{ChabrierIMF} initial mass function (IMF). All magnitudes are expressed in the AB system \citep{oke_absolute_1974}, for which a flux $f_\nu$ in $\mu$Jy ($10^{-29}$~erg~cm$^{-2}$s$^{-1}$Hz$^{-1}$) corresponds to AB$_\nu=23.9-2.5\,\log_{10}(f_\nu/\mu{\rm Jy})$. 

\section{Data: COSMOS2020}
\label{sec:data}

We utilize the most recent release of the COSMOS catalog, COSMOS2020 \citep{Weaver2022_catalog}, comprised of $\sim1\,000\,000$ galaxies selected from a $izYJHK_s$ coadd with near-infrared depths approaching 26\,AB ensuring a mass-selected sample complete down to $10^{9}\,\Msol{}$ at $z\approx3$. This sample is complemented by extensive supporting photometry ranging from the UV to 8\,$\mu{\rm m}$ over a total area of 2\,deg$^2$, making it the widest near-infrared-selected multiwavelength catalog to these depths. This is made possible by the latest release of the UltraVISTA survey \citep[DR4;][]{mccracken_ultravista_2012, Moneti2019}, which is the longest running near-infrared survey to date, and complemented by even deeper optical $grizy$ data provided by Subaru's Hyper Suprime-Cam instrument \citep[HSC PDR2;][]{aihara_second_2019}. Flanking this core imaging are $u$ band measurements made by the CLAUDS survey from the Canada-France-Hawaii Telescope \citep{sawicki_cfht_2019} and deep Spitzer/IRAC imaging in all four channels, reprocessed to include nearly every exposure taken in COSMOS for use in this catalog \citep[][and references therein]{Moneti2022}. As in the previous COSMOS catalog by \citet{laigle_cosmos2015_2016}, several intermediate and narrow bands from both Subaru/Suprime-Cam and VISTA \citep{milvangjensen_2013} are used to provide precise determinations of photometric redshifts. 

In addition to the comprehensive set of data, photometry is performed by two methods to produce two separate catalogs. The catalog called \classic{} contains aperture photometry extracted with \texttt{Source Extractor} \citep{bertin_sextractor:_1996} for the optical/near-infrared and with IRACLEAN \citep{Hsieh_2012} for the infrared (as in \citeauthor{laigle_cosmos2015_2016}). The catalog called \farmer{} contains profile-fitting photometry extracted with \cfarmer{} \citep{Weaver2023_farmer}, which uses \tractor{} \citep{lang_tractor_2016} to construct and apply parametric models to estimate source fluxes and does so consistently across more than 30 bands, including all four IRAC channels. Each photometric catalog is paired with photometric redshift and physical parameter estimates from both \lephare{} \citep{arnouts_measuring_2002, ilbert_accurate_2006} and \eazy{} \citep{brammer08_eazy} resulting in four measurements of \photoz{} and $\mathcal{M}$ per source. The combination of comprehensive deep images, independent photometry techniques, and independent SED fitting enables an unparalleled control of systematics and uncertainties, and hence further insight when inferring the underlying true nature of galaxy populations. See \citet{Weaver2022_catalog} for details.

\section{Sample selection, completeness, and uncertainties}
\label{sec:selection}

\subsection{Selection function}
Galaxies in COSMOS2020 are selected from a near-infrared $izYJHK_s$ \texttt{CHI-MEAN} coadded detection image \citep{Szalay99_chidetection, Bertin2010}. The deepest band is $i$, with a 3$\sigma$ sensitivity limit at $\sim$27\,mag, and the shallowest is $K_s$ at $\sim$25\,mag\footnote{Computed at $3\sigma$ from 3\arcsec apertures randomly placed in regions without detected sources; see Table~1 of \citet{Weaver2022_catalog}.}. The reddest selection band, $K_s$ delivers mass-selected samples to $z\lesssim4.5$. Furthermore, the comparably deeper $K_s$ imaging (by $0.5-0.8$\,mag relative to COSMOS2015) translates directly into a higher degree of completeness where low-$\mathcal{M}$ samples are more mass complete down to lower masses. Taken together with IRAC, these images provide reliable detections of the reddest, oldest, and heavily dust obscured sources not seen in COSMOS2015.

Although the nominal area of the COSMOS survey is $2\deg^2$, the most secure region comprises $1.279\deg^2$ after removing contamination due to bright star halos and requiring an intersection of the deep near-infrared UltraVISTA imaging and the Subaru Suprime-Cam intermediate bands, where the \photoz{} performance is generally best. Compared to \citet{Davidzon17_mass} who use UltraVISTA DR2 \citep{mccracken_ultravista_2012} imaging to probe the near-IR, here we rely on UltraVISTA DR4 \citep{Moneti2019}, reaching greater and simultaneously uniform NIR depth across the entire field (especially $K_s$). These non uniform depths restricted the SMFs of \citeauthor{Davidzon17_mass} to only the four deepest stripes in UltraVISTA spanning $0.62\deg^2$. The gains leveraged in this work are hence five-fold. Firstly, the total number of recovered sources at all apparent magnitudes increases due to the additional area, improving statistical margins especially of rare sources. Secondly, the increased optical and near-infrared depths permit the recovery of fainter sources, improving flux completeness. Thirdly, the increased depth reduces photometric uncertainties leading to more precise, less biased \photoz{} and $\mathcal{M}$. Fourth, the consistent depth and improved imaging performance from Hyper Suprime-Cam provides significantly better photometry relative to Suprime-Cam. Lastly, the wider field makes it less likely to become biased due to specific structures along the line-of-sight (e.g., clusters, voids) and instead probes a greater variety of environments affecting the evolution of galaxies within them.

As presented in Fig.~13 of \citet{Weaver2022_catalog}, the best \photoz{} performance at $i>22.5$~AB is achieved by the combination of \cfarmer{} photometry and \photoz{} from \lephare{} with a $\delta z/(1+z)$ precision of $<1\%$ at $i\approx20$ and $<4\%$ at $i\approx26$~AB. For this reason, and to expedite comparisons with the most similar work in the literature \citep{ilbert_mass_2013, Davidzon17_mass}, this work adopts the photometry from \cfarmer{} paired with the \photoz{}, $\mathcal{M}$, and rest-frame magnitudes estimated by \lephare{}. This combination will be henceforth referred to as COSMOS2020, unless explicitly stated otherwise. Our \photoz{} estimates correspond to the column \texttt{LP\_zPDF}, defined as the median of the redshift likelihood distribution $\mathcal{L}(z)$ (often referred to as $p(z)$). We choose to adopt $\mathcal{M}$ estimates from \texttt{MASS\_MED} as the median of the mass likelihood distributions are generally less susceptible to template fitting systematics than \texttt{MASS\_BEST} which is taken at the minimum $\chi^2_{\rm SED}$. We find that the two agree at all redshifts within 0.01\,dex with a narrow scatter whose 68\% range is below 0.05\,dex.

Identification of stellar contaminants proceeds as described in Section~5.1 of \citet{Weaver2022_catalog}. Briefly, \lephare{} computes the best-fit stellar template to each SED from a range of templates, including white dwarf and brown dwarf stars. Sources that achieve a  $\chi^{2}$ from a stellar template fit lower than any galaxy template are removed. An additional criterion on morphology is used, identifying likely stars as point-like sources from the COSMOS \textit{HST}/ACS mosaics \citep{koekemoer_cosmos_2007} and Subaru/HSC \citep[PDR2,][]{aihara_second_2019} for the $i<23$ and $i<21.5$ AB, respectively. Bright, resolved sources with SED shapes similar to stars are not removed.

An initial sample of 636\,567 galaxies\footnote{Galaxies are selected by \texttt{lp\_type}\,=\,0, see catalog documentation.} in the range $0.2<z\leq6.5$ is selected from the contiguous $1.27\,{\rm deg}^2$ \texttt{COMBINED} region defined as the union of the UltraVISTA and Subaru/SC footprints but removing regions around bright stars\footnote{The \texttt{COMBINED} region corresponds to \texttt{FLAG\_COMBINED}\,=\,0.}. However, the source density at $6.5<z\leq7.5$ becomes noticeably greater in the ultra-deep stripes and so we restrict sources in this particular $z$-bin to only the $0.716\,{\rm deg}^2$ of the ultra-deep stripes also covered by Subaru/SC but away from bright stars, finding 1\,327 galaxies. The construction of a robust stellar mass function requires precise redshifts and accurate stellar masses. Since an infrared detection is required to accurately measure galaxy stellar masses at $z\gtrsim2-3$, where this study seeks to cover new ground,
227\,305 sources with $m_{\rm ch1}>26$\,AB (i.e., $S/N\lesssim5$) are removed; $\approx93\%$ of which fall below the mass completeness limit introduced in Section~\ref{subsec:completeness}. An additional 13\,644 sources with ambiguous redshifts are removed, quantified for simplicity by having $>32\%$ of their redshift probability lying outside $z_{\rm phot}\pm0.5$; $\approx\,70\%$ of which fall below the mass completeness limit. Finally, we remove a further 1\,850 sources with unreliable SED fits with reduced $\chi^{2}>10$. These three necessary cuts create a final sample of 395\,095 galaxies with precise redshifts and accurate stellar masses. We note that many excluded sources fail more than one of these three criteria (e.g., $90\%$ are undetected in the infrared); Fig.~\ref{fig:cuts} shows their cross-section. Although the rejected objects are potentially interesting, their deeply uncertain nature does not permit us to reliably incorporate them into our measurements without a substantially more complex approach (e.g., to model potential outliers, see \citealt{Hogg2010}). Overall the excluded sources are predominantly faint (median $K_s\sim$25.7\,AB) such that their best-fit masses are generally below our adopted mass completeness limit and their potential for bias minimal within the mass ranges examined in this work.

\subsection{Quiescent galaxy selection}
\label{subsec:qgsel}

Accurate identification of distant and/or unresolved quiescent galaxies has been a longstanding problem \citep[see e.g.,][]{leja_uvj_2019, Shahidi2020, Steinhardt_tnse}. While quiescent galaxies can be identified by their low star formation rates, SFR estimates from fitting broad-band photometry are subject to large uncertainties and are dependent on the particular SED template library \citep[][]{Pacifici2015, Leja2019, Carnall2019}. The other most successful approach is that of rest-frame colors, as the lack of blue O- and B-type stars in quiescent populations implies a dearth of UV emission and a red optical contintuum. Similar photometric UV to NIR SEDs are measured for heavily dust-enshrouded but otherwise star-forming systems, making the challenge of color-color selection that of distinguishing truly quiescent galaxies from dusty star-forming ones, in the absence of FIR measurements. Popular rest-frame color-color selections include ($U-V$) vs. ($V-J$), ($NUV-r$) vs. ($r-J$), ($NUV-r$) vs. ($r-K$), and ($FUV-r$, $r-J$) \citep[][respectively]{Williams2009, Ilbert_2010, Arnouts2013, leja_uvj_2019}, in addition to observed-frame methods such as ($B-z$) vs. ($z-K$) for sources at $1.4\lesssim \,z\lesssim\,2.5$ \citep{daddi04}. However, each implies a different definition of quiescence \citep[see][]{leja_uvj_2019, Shahidi2020}. These pairs of rest-frame colors may be measured directly from the observed photometry (where available) or derived from the best-fit templates that should largely agree with observed-band photometry, and can also be meaningfully extrapolated in cases where appropriate observed-frame data are not available. While the former may be adversely affected by uncertainties and systematics propagated from the observed photometry, the latter assumes that the SED template library or basis contains a suitable model. It is worth noting that efforts are being made to employ machine learning techniques to lessen the dependence and bias impact from model assumptions \citep[e.g.,][]{Steinhardt_tnse}, although no such method has gained comparable use as yet.

Recent work by \citet{leja_uvj_2019} has clarified the efficacy of these color-color selections, proposing to extend the typically adopted $U-V$ vs.\ $V-J$ baseline further into the UV on the basis of stronger correlation with intrinsic sSFR as measured in simulated photometry \citep[see also][]{Arnouts2013}. As such, catalogs containing reliable and deep $FUV$ and $NUV$ photometry have the advantage of being able to identify low-$z$ quiescent populations which though extrapolation can be consistently identified up to the highest redshifts even when the observed-frame $U$ band no longer corresponds to rest-frame $U$. At redder wavelengths the rest-frame $J$ band is crucial for measuring the rest-frame stellar bulk, however, it becomes easily redshifted out of most deep surveys with IRAC photometry by $z\approx2-3$ which renders quiescent galaxy selections highly dependent on template libraries even at these intermediate redshifts. 

In order to provide broad comparisons to other mass functions, in particular those measured in COSMOS \citep[e.g.,][]{ilbert_mass_2013, Davidzon17_mass}, this work selects quiescent galaxies following the criteria introduced by \citet{ilbert_mass_2013}:
\begin{equation}
    (NUV - r) > 3\,(r - J) + 1\,\mathrm{and}\,(NUV-r) > 3.1
\label{equ:nuvrj}
\end{equation}

\noindent whereby the slant line runs perpendicular to the direction of increasing sSFR and parallel to an increase in dust attenuation to separate truly quiescent systems from otherwise dusty star-forming contaminants. For clarity, we henceforth refer to this ($NUV-r$) vs. ($r-J$) selection as $NUVrJ$. Generally, this selection has been found to approximate a cut in sSFR~$\lesssim10^{-11}\,\mathrm{yr}^{-1}$ \citep{Davidzon2018}.

\begin{figure*}
   \centering
   \includegraphics[width=17cm]{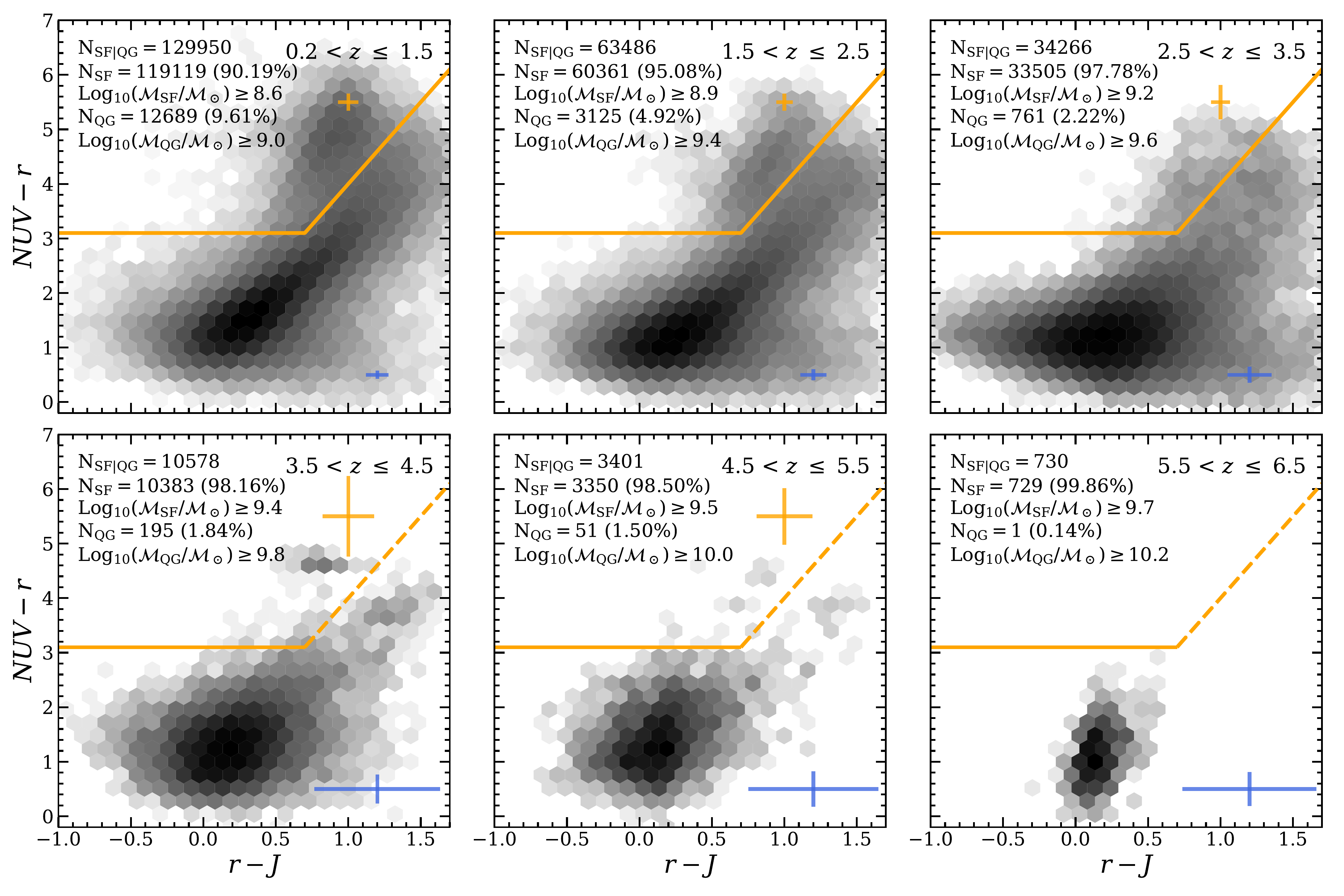}
     \caption{Galaxies are classified as either star-forming or quiescent in bins of redshift by selection in rest frame ${\rm NUV}-r$ and $r-J$ colors, following \citet{ilbert_mass_2013}. The rest-frame $J$ band is not directly measured at $z\gtrsim3.5$ and so the sloped part of the selection box is dashed  to indicate that the $r-J$ colors are strongly model dependent. Typical uncertainties for the rest-frame colors of the star-forming and quiescent subsamples are estimated as medians of the nearest observed-frame bands, consistent with the derivation of the rest-frame colors. See the text for details.}
     \label{fig:nuvrj}
\end{figure*}

It is important to note that \lephare{} computes rest-frame magnitudes by attempting to find an observed frame band which directly probes the rest-frame band, which for the reasons listed above is less model-dependent than simply adopting the rest-frame colors of the best-fit galaxy template (or sSFR) and hence conveys a view of the true variety of observed galaxies that is less biased by our assumptions. Although the rest-frame photometry are estimated using a $K$-correction and color-term, they are most strongly dependent on the best-fit template at redshifts where observed photometry are not well matched to the rest-frame band and are then effectively extrapolated from the best-fit template (see Appendix~1 of \citealt{ilbert2005} for details).

Unbiased sample selection becomes increasingly difficult with redshift. By $z\approx3$, rest-frame $J$-band fluxes are no longer directly measured even by IRAC channel~2\footnote{Channel~3 and 4 are too shallow to provide useful constraints at $z\gtrsim3$.} and so become increasingly model-dependent and uncertain. While rest-fame $NUV$ remains constrained even at $z>6$, rest-frame $r$ must be extrapolated by $z\approx5$ at which point differentiation between quiescent and dusty star-forming galaxies becomes statistically impossible. Because of this, and the expectation that quiescence is unlikely at such early times, selection of quiescent galaxies in this work is limited to $z\leq5.5$, noting that selection between $3<z\leq5.5$ is subject to a significantly higher degree of uncertainty. Advancement in this selection will only be made possible with deeper infrared imaging from facilities such as \textit{JWST} \citep{Rigby2022}.

Selection of quiescent galaxies is presented in Fig.~\ref{fig:nuvrj}, showing star-forming and quiescent galaxies whose masses are above their respective mass limits (see Section~\ref{subsec:completeness}). Error bars on the rest-frame colors are estimated from the quadrature addition of the observed-frame filter nearest to that of the rest-frame. Given that even rest-frame photometric uncertainties are generally inversely proportional to mass at given redshift, these error bars are most representative for median mass systems but are overestimated for bright, massive galaxies. In addition, uncertainties become increasingly underestimated at $z>3$ as the extrapolation based on the best-fit template is not propagated. However, it is already obvious that the dividing power of the slant line is significantly diminished at $z>3$ where the rest-frame $J$ band is no longer directly observed. Both the quiescent and dusty star-forming regions appear to be devoid of sources by $z\approx6$, such that no intrinsically quiescent (misclassified or not) are found at these early times. Whether this is an intrinsic feature of galaxy populations or merely a selection effect is discussed below in Section~\ref{subsec:completeness}. For these reasons this work restricts distinction between quiescent and star-forming galaxies to $z\leq5.5$. Galaxies at $z>5.5$ in this work should be considered star-forming.

\subsection{Completeness}
\label{subsec:completeness}
Accurate estimates of completeness are crucial when inferring general properties about a population from an otherwise incomplete sample. While advancements in near-infrared facilities have enabled breakthroughs in selecting representative sample of galaxies by measuring their stellar bulk, samples are still mass limited and these mass limits evolve with $z$, owing to the faintness of increasingly distant galaxies. As such, mass limits are highly dependent on accurate estimates of survey depths and their impact on the selection function as it relates to the detection of the lowest mass galaxies. 

There are three known populations which are expected to be missed by a near-infrared selection whose deepest images are on the blue-end of the selection function. Firstly, the faintest star-forming blue galaxies in the local Universe (e.g., dwarf irregular starbursts) may have detectable fluxes blueward of $i$, but they will not be included in existing $izYJHK_s$ selections as their near-infrared emission is too faint. However, their contribution is expected to be limited to only the low-mass end of the galaxy stellar mass function, and can henceforth be well characterized in large numbers by deeper small field surveys \citep[e.g., CANDELS:][]{Grogin_2011, Koekemoer_2011}. 

Secondly, quiescent galaxies are more difficult to detect than star-forming galaxies owing to a lack of rest-frame UV and blue optical emission, meaning their detection relies upon deep near-infrared imaging at both low and high-redshifts. For this reason, an $izYJHK_s$ detection will be insufficient to detect the quiescent systems compared to star-forming ones of the same mass, meaning that the quiescent sample will be as complete as the star-forming sample at higher masses. We compute their respective mass completeness limits separately and apply them consistently throughout.

Lastly, and similar to quiescent galaxies, the most heavily dust obscured star-forming galaxies ($A_V\gg5$) at high redshift ($z\gtrsim2$) will not present appreciable optical or near-infrared fluxes to be detected in COSMOS2020, but unlike quiescent galaxies are ubiquitously and efficiently detected in far-infrared, submillimeter, and radio surveys \citep[e.g.,][]{Schreiber2018, Wang2019, Sun2021, Jin2019, Fudamoto2020, Fudamoto2021,  Casey2021, Shu2022}. The nature and extent of this recently discovered population remains difficult to quantify, owing to a combination of complex selection functions and serendipitous detections. The majority are likely to be missed by an $izYJHK_s$ selection function even at the depths of COSMOS2020 (see discussions in Section~\ref{subsec:massivegal}).

Although mass completeness estimates are presented in \citet{Weaver2022_catalog}, they are derived for a comparably less secure sample than used in this work. For this reason the mass completeness limits are rederived in identical fashion following the method of \citet{Pozzetti2010}. A critical advantage of this method is that it does not rely on the theoretical mass distribution of galaxies fainter than the magnitude limit, but assumes that those just above the threshold share similar properties with the undetected ones, modulo a rescaling factor as detailed below.  This contrasts with  studies that estimate mass completeness either through injection-recovery of simulated sources, or extrapolation of the observed distribution itself below the magnitude limit. The latter approach, in particular, may underestimate the stellar mass limit if the galaxy sample is sparse due purely to astrophysical reasons and not genuine incompleteness. In our case, the sample in each $z$-bin is first cleaned by discounting the 1\% worst-fit sources via $\chi^2$. The stellar masses $\mathcal{M}$ of the 30\% faintest galaxies in channel~1 are then rescaled such that their observed channel~1 apparent magnitude $m_{\rm ch1}$ matches the IRAC sensitivity limit $m_{\rm lim}=26$:
\begin{equation}
    {\rm Log}_{10}(\mathcal{M}_{\rm resc}) = {\rm Log}_{10}(\mathcal{M}) + 0.4\,(m_{\rm ch1}-26.0) \,.
\end{equation}

\noindent The limiting mass $\mathcal{M}_{\rm lim}$ is then taken to be the 95$^{th}$ percentile of the $\mathcal{M}_{\rm resc}$ distribution. Finally 2$^{\rm nd}$ order expansions in $(1+z)$ are fitted to each $M_{\rm lim}$ per $z$-bin up to $z=7$ for the total sample, and $z=5$ for star-forming and quiescent samples, to produce a smoothly evolving limiting mass. Limits above which samples are $\sim70\%$ mass complete (see justification below) are derived consistently for the total sample from $0.2<z\leq7.5$ and for the star-forming, and quiescent samples from $0.2 < z \leq 5.5$:

\begin{equation}
    {\rm Total:} -3.23\times10^{7}\,(1+z) + 7.83\times10^{7}\,(1+z)^2
    \label{equ:mc_total}
\end{equation}
\begin{equation}
    {\rm Star-forming:} -5.77\times10^{7}\,(1+z) + 8.66\times10^{7}\,(1+z)^2
    \label{equ:mc_sfg}
\end{equation}
\begin{equation}
    {\rm Quiescent:} -3.79\times10^{8}\,(1+z) + 2.98\times10^{8}\,(1+z)^2
    \label{equ:mc_qg}
\end{equation}

\begin{figure*}
   \centering
   \includegraphics[width=21cm]{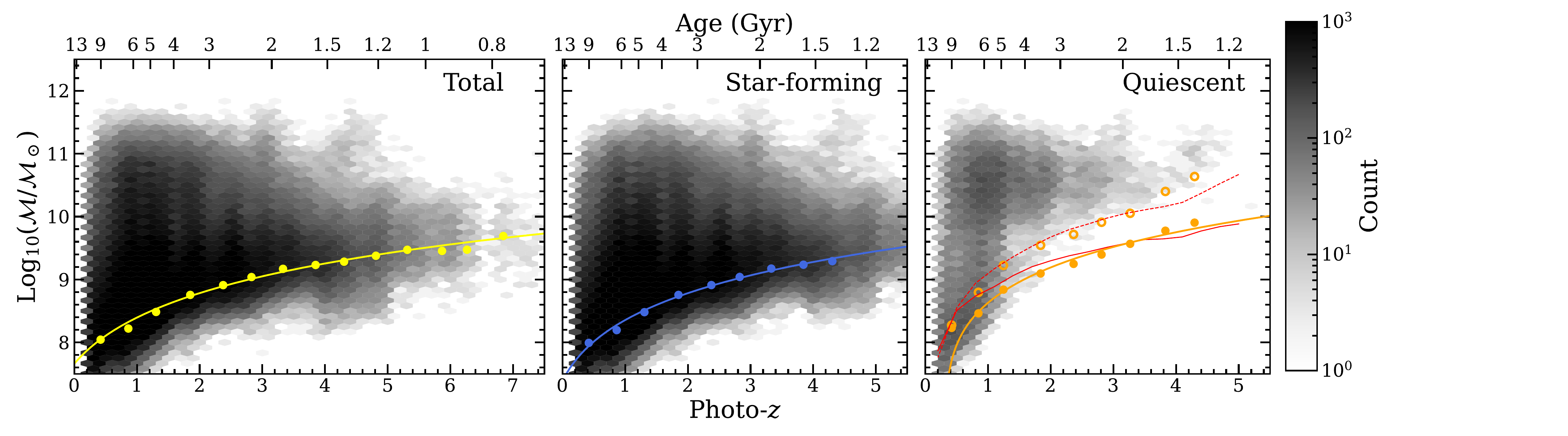}
     \caption{Stellar mass completeness limits. The three panels show, from left to right, the density of the total, star-forming, and quiescent samples with their respective mass completeness limits. The limits are derived in discrete $z$-bins following \citet{Pozzetti2010}  with magnitude limits adopted from IRAC channel~1 (filled circles), then interpolated with Eqs.~\ref{equ:mc_total}, \ref{equ:mc_sfg}, and \ref{equ:mc_qg}, respectively, resulting in the solid lines. For quiescent galaxies, consistent estimates using $K_s$ are also shown (empty circles). To verify these estimates we compute conservative mass limits from simulated spectra following a delta-like burst at $z=15$ which are normalized separately to the $K_s$ and channel~1 magnitude limits; they are shown by the dotted and solid red lines, respectively. See the text for details.}
     \label{fig:masscomp}
\end{figure*}

\noindent and are shown in Fig.~\ref{fig:masscomp}. Despite the more conservative selection adopted in this work, the derived mass completeness limits are essentially identical to those derived in \citet{Weaver2022_catalog}, which indicates the robustness of these limits against sample selections. For reference, at $z\approx1$ the quiescent sample is complete at $\sim0.25$\,dex in mass above the star-forming sample, growing to $\sim0.45$\,dex above by $z\approx3-5$.

There remains an additional incompleteness arising from the fact that mass completeness is derived from IRAC channel~1 photometry, and yet our $izYJHK_s$ selection function does not include sources which are only identified in IRAC. As discussed in \citet{Davidzon17_mass}, despite this drawback it is also disadvantageous to estimate the mass completeness from any one of the six $izYJHK_s$ detection bands either. For instance, the reddest, $K_s$, samples the rest-frame stellar bulk only out to $z\lesssim2$, and will tend to underestimate stellar masses at $z\gtrsim2$ until $z\approx4.5$ when it becomes a UV tracer. Thankfully, it is possible to estimate the impact of this additional incompleteness by examining a sample of galaxies common to this work and those of the comparably deeper $\sim$200\,arcmin$^2$ CANDELS-COSMOS catalog of \citet{Nayyeri2017}. This analysis is performed and discussed at length in \citet{Weaver2022_catalog}. In summary, 75\% of CANDELS sources at $M_{\rm lim}$ are recovered by both \farmer{} and \classic{} COSMOS2020 catalogs, which combined with the choice of $M_{\rm lim}$ being the 95$^{th}$ percentile of $M_{\rm resc}$ implies that $M_{\rm lim}$ of the total sample corresponds to a mass completeness of $\sim70$\%.

As shown in Fig.~\ref{fig:masscomp}, the mass limit of the total sample is almost identical with that of the star-forming sample at all redshifts. This is unsurprising as star-forming galaxies generally dominate galaxy demographics. As in our $NUVrJ$ selection from Fig.~\ref{fig:nuvrj}, we see the lack of quiescent systems at $z>5.5$ and so we consider the star-forming subsample to be statistically equivalent to the total sample at $z>5.5$.

Deriving a consistent mass completeness limit for quiescent galaxies is more challenging. It is well known that the predominantly older, redder stellar populations of quiescent systems imply higher mass-to-light ratios than found in star-forming galaxies which in turn imply a lower degree of mass completeness at the same flux limit. Fig.~\ref{fig:masscomp} shows that our channel~1 derived mass completeness falls $\gtrsim0.2$\,dex lower than the bulk of the quiescent sources. Taken at face value, this would appear to indicate a lack of quiescent systems at low-intermediate masses at $z>2$. Again, we have cause for concern as channel~1 is not included in our $izyJHK_s$ detection strategy, and hence our selection function is liable to miss IRAC-bright sources that are not present in bluer bands. Furthermore, due to the predominantly red optical spectral slopes of quiescent galaxies, continuum emission in $K_s$ should be lower than in channel~1, implying that one would need a deeper $K_s$ magnitude limit to detect the same sources from shallower channel~1 imaging. In other words, it is expected that there are red sources visible only in channel~1 that are not included in our $izYJHK_s$ detected catalog. Furthermore, by $z\approx4.5$ $K_s$ no longer traces mass but rather UV continuum (channel~1 does so by $z\approx8$). Worse, the UV continuum in these systems is expected to be faint (even given `frostings' of star formation in $NUVrJ$-selected post-starburst systems). Therefore, while deriving a mass completeness from channel~1 magnitude limits may be suboptimal due to possible selection effects with regards to a $izYJHK_s$-selected sample, we cannot turn to $K_s$ as it is no longer a mass indicator at $z\gtrsim4.5$.  

One solution is to modify our selection function by incorporating $izYJHK_s$ undetected IRAC-only sources into our sample. While a systematic search for IRAC only sources is ongoing, their identification is made difficult not only because of the significantly lower resolution and consequently higher source crowding in necessarily deep IRAC images, but these sources also lack optical/NIR data which is, by definition, insufficiently deep to identify low-$z$ interlopers. Deep mid-infrared (MIR) data redward of IRAC over sufficient degree-scale areas do not currently exist, making a determination of redshift, mass, and quiescent nature of these IRAC-detected sources problematically uncertain.

For the time being, whether or not this absence of intermediate-low mass quiescent systems is astrophysical cannot be determined. Literature measurements do not yield more mass-complete quiescent samples as the UltraVISTA DR4 NIR depths are now similar to those from even the deepest small-field NIR imaging, $H_{\rm 160}\approx25.9$ \citep[e.g., CANDELS,][]{Tomczak14}, and even so, comparisons are hampered by field-to-field and photometric systematics. Comparisons with other stellar mass functions measured more consistently in COSMOS \citep[e.g.,][]{Davidzon17_mass, ilbert_mass_2013} are still affected by systematics from comparably less certain measurements from previous, shallower data, despite the lessened impact of cosmic variance. Additionally, comparisons with \citet{Davidzon17_mass} are complicated by the fact that they refit the photometry of \citet{laigle_cosmos2015_2016} to produce new $z$ and $\mathcal{M}$ estimates. Modifications include expanding the redshift baseline from $z\leq6$ to $z\leq8$ and adding in additional SED templates to probe star-bursting galaxies with rising star formation histories as well as dust-obscured systems. As demonstrated by a recent comparison by \citet{Lustig2022}, these modifications produce lower number densities of massive galaxies compared to the original measurements. 

We turn to theoretical frameworks to investigate this further. Namely, we use Flexible Stellar Population Synthesis \citep[\texttt{FSPS};][]{Conroy2009, Conroy2010} to estimate the stellar masses produced by extraordinarily old stellar population produced by a delta-burst evaluated at $z=15$.  We assume a \citet{ChabrierIMF} IMF with ${\rm log}_{10}(Z/Z_{\odot})\,=\,-0.3$. At each epoch (i.e., $z$) we normalize the evolved model spectrum to match our channel~1 observed frame magnitude limits (26.0\,mag) and estimate the corresponding stellar mass shown by the solid red curve in Fig.~\ref{fig:masscomp}. This exercise is repeated consistently with the $K_s$ magnitude limit (25.5\,mag) shown by the dashed red curve.

We caution however, that the assumption that these models accurately describe real high-$z$ quiescent galaxies is becoming increasingly dubious. Such a system formed in a monolithic delta-burst just following the big bang cannot reach quiescence (as defined by $NUVrJ$) above $z\approx5.3$, and yet remarkably mature systems at $z\approx4-5$ have already been reported in the literature \citep[e.g.,][]{Schreiber18_QGz4,Tanaka2019, Valentino20_QGz4}. Worse, the spread of mass-to-light ratios found in quiescent systems means that a single mass completeness limit for all quiescent galaxies at a given redshift is ill-defined even for consistently selected (e.g., $UVJ$, $NUVrJ$, sSFR) samples, resulting in a non negligible selection effects. 

Nonetheless, we stress that the difference in completeness between the effectively flux-complete $K_s$-derived mass limit and the mass limit derived from channel~1 magnitudes is only $\sim0.3$\,dex, which is typically less than a single bin in our analysis. In light of this considerable uncertainty, we adopt the optimistic quiescent galaxy mass completeness limits derived via \citet{Pozzetti2010} from channel~1 magnitudes with the caveat that the lowest mass bins in each measurement at $z\gtrsim2$ are potentially incomplete. We indicate the $K_s$ mass limit for quiescent samples throughout. Also, we note that our quiescent mass limit, although consistently determined in \citet{Davidzon17_mass}, is more conservative despite the deeper NIR data. As will be discussed in Section~\ref{subsec:sfqg_smf}, we attribute this to an overestimate of the quiescent galaxy mass completeness by \citeauthor{Davidzon17_mass}. 

The difference in mass completeness between the star-forming and quiescent samples presents an additional complication. Because the star-forming galaxies can be reliably detected to lower masses than quiescent galaxies from our $izYJHK_s$ selection function, the low-mass regime of the total SMF at a given epoch, as defined in this work, does not actually include contributions from the lowest-mass quiescent systems. Therefore the total SMF is simply the star-forming SMF at masses below the quiescent mass limit. The effect from the undetected low-mass quiescent systems is almost certainly negligible: a simple extrapolation from the evolution in the number density of $z\approx0.2-2.5$ low-mass quiescent galaxies predict a number density at least 10$\times$ lower than that of the star-forming galaxies at $z>3$. Furthermore, given that general uncertainties are still above the 10\% level at low-masses even at $z\approx2$ (Fig.~\ref{fig:errors}), any bias arising in the total number density from undetected low-mass quiescent systems is lower than other sources of error.

\subsection{Derivation of the $1/V_{\mathrm{max}}$ correction}
\label{subsec:vmax}
Intrinsically faint galaxies at any given redshift are more likely to be missed by survey selection functions compared to brighter sources. For a NIR-selected sample, this equates to a mass bias by which low-luminosity, low-mass galaxies can only be detected in smaller volumes (i.e. out to lower $z$) relative to brighter, more massive ones which could be detected if they were at higher $z$. This is the well known Malmquist bias \citep{malmquist1920, malmquist1922}.

The most straight-forward approach to correct for such a bias is the $1/V_{\rm max}$ method of \citet{schmidt1968}, which has enjoyed significant popularity owing to its simplicity. Briefly, the $1/V_{\rm max}$ method statistically corrects for selection incompleteness  by weighting each detected object by the maximum comoving volume in which it can be observed, given the characteristics of the telescope survey. The $V_{\rm max}$ estimate per individual object is computed after finding the maximum redshift $z_{\rm max}$ by which the best-fit SED would no longer be observable\footnote{$z_{\rm max}$ estimates are computed with \texttt{ALF} \citep{ilbert2005}.} because of the survey's  flux limit. On the other hand, the minimum redshift ($z_\mathrm{min}$) should be the one at which the source would become too bright and saturate the camera, although in practice is the lower boundary of the $z$ bin in which the considered galaxy lies ($z_\mathrm{low}$). Therefore, for the $i$-th galaxy inside the bin  $z_\mathrm{low}<z<z_\mathrm{high}$, the maximum observable volume is 
\begin{equation}
    V_{\mathrm{max}, i} = \frac{4\pi}{3} \frac{\Omega}{\Omega^{\rm sky}} (D_{\rm cov}(\min(z_{\mathrm{max}, i},z_\mathrm{high}))^3 - D_{\rm cov}(\max(z_{\mathrm{min}, i},z_\mathrm{low}))^3) \, ,
\end{equation}

\noindent where $\Omega$ is the solid angle subtended by the sample, $\Omega^{\rm sky} \equiv 41\,253 \deg^2$ is the solid angle of a sphere, and $D_{\rm cov}(z)$ is the comoving distance at $z$ \citep{Hogg1999}. If $z_\mathrm{max}$ exceeds the upper boundary of the redshift bin, the latter is used instead, meaning that the brightest sources are often assigned a weight that corresponds to the full volume of that redshift slice. As such, this correction is expected to be significant for only the faintest, lowest-mass sources in a given $z$-bin. While it is nonparametric and does not assume a functional form of the SMF, the $1/V_\mathrm{max}$ technique does assume that samples are drawn from a uniform spatial distribution, which is not accurate in the case of over- or under-dense environments \citep{efstathiou1999}. However, the assumption of a uniform spatial distribution is expected to be problematic only at $z<1$, where large-scale structures have fully assembled, or in narrower surveys that can be biased by structures at smaller scales.  Other methods exist which do not make this assumption such as STY \citep{sandage1979} and SWML \citep{Efstathiou1988}, a parametric and nonparametric maximum likelihood method, respectively. Already, \citet{Davidzon17_mass} found that the constraints provided by COSMOS2015 were sufficiently strong for the $1/V_{\rm max}$ method as well as more complex methods (e.g., STY, SWML) to essentially converge. With even stronger constraints and larger effective area provided by COSMOS2020, we can expect even better agreement, with minimal advantages to the more complex methods. More extensive discussions on  strengths and weaknesses of the various approaches can be found in the literature \citep[e.g.,][]{ilbert2004, binggeli1998, johnston2011, takeuchi2000, weigel2016}.

\subsection{Further considerations of uncertainty}
\label{subsec:uncertainty}
\begin{figure}
   \centering
   \includegraphics[width=\columnwidth]{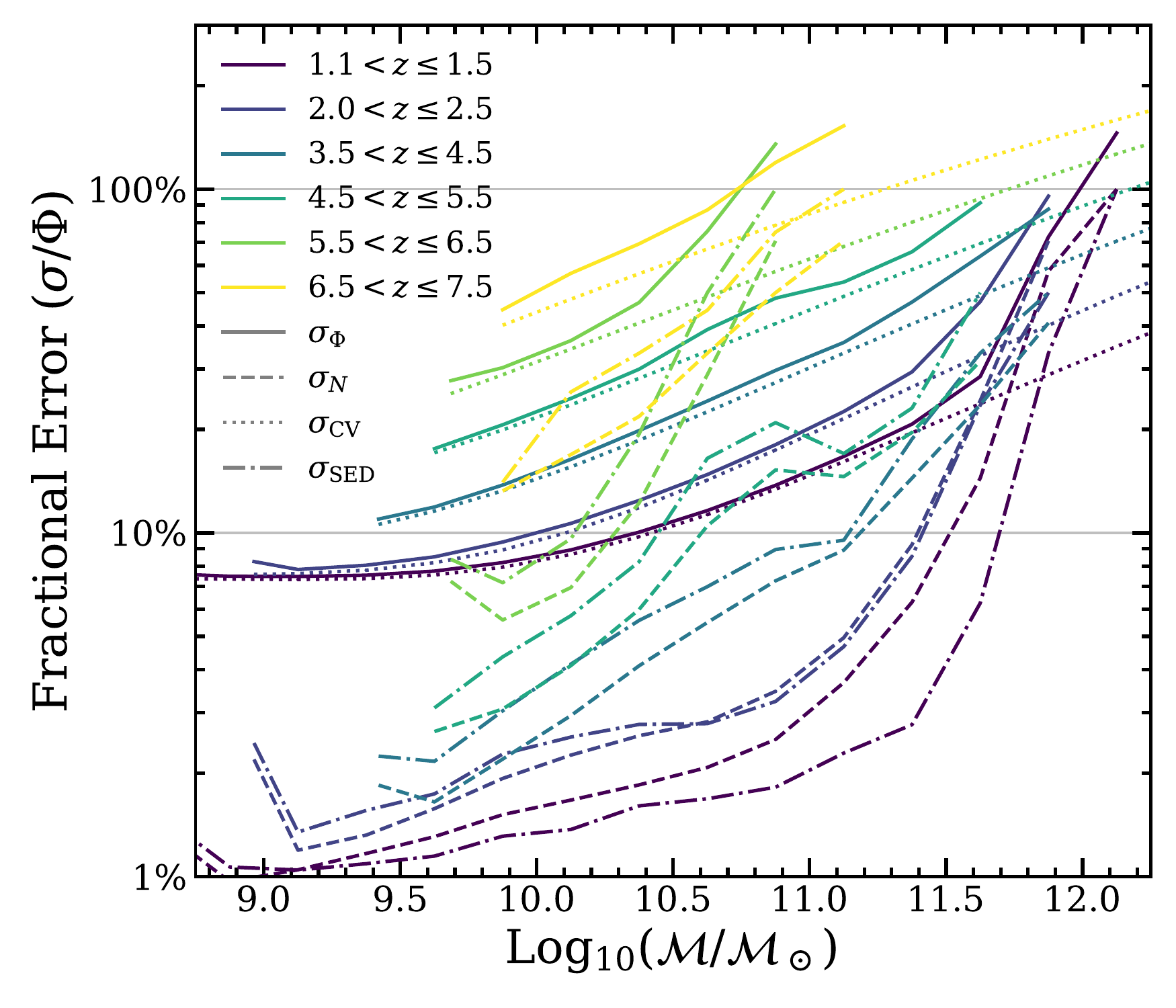}
     \caption{Adopted estimates for the total uncertainty $\sigma_{\Phi}$ (solid) as a function of stellar mass at several redshifts (by color), for mass-complete samples only. Contributions include uncertainties from Poisson noise $\sigma_{N}$ (dashed), Cosmic Variance $\sigma_{\rm CV}$ (dotted), and SED fitting $\sigma_{\rm SED}$ (dash-dot). 10\% and 100\% levels are indicated for reference.}
     \label{fig:errors}
\end{figure}

We adopt a statistical error budget on the SMF number density $\Phi$ consisting of the quadrature addition of Poisson noise ($\sigma_N$), cosmic variance fluctuations ($\sigma_{\rm cv}$), and uncertainties on masses induced by SED fitting ($\sigma_{\rm SED}$) such that $\sigma_{\Phi} = (\sigma_N^2 + \sigma_{\rm cv}^2 + \sigma_{\rm SED}^2)^{1/2}$. Fig.~\ref{fig:errors} shows the composition of the total error budget from $z=1.1$ to 6.5 as a function of stellar mass for mass-complete bins.

\subsubsection{Poisson noise}
Poisson noise arises from processes wherein the abundance of a discrete quantity (or counts) is measured. Although in the limit of many events a Poisson process becomes indistinguishable with that of a Gaussian, in small number counts it can be the dominant source of uncertainty. Here we compute the Poisson error $\sigma_{N}$ for each mass bin as $\sqrt{N}$ where $N$ is the number of objects in that bin. These values are recomputed for the star-forming and quiescent subsamples separately. As shown in Fig.~\ref{fig:errors}, the fractional contribution from $\sigma_{N}$ increases with mass and redshift with the largest contribution at $\mathcal{M}>10^{11.5}\Msol{}$.

The discrete nature of Poisson processes allows us to also provide upper limits on bins containing zero detected galaxies. Following Table~1 of \citet{Gehrels1986}, the statistical upper limit on $\Phi\,(N=0)$ for a given observed volume $V$ is $\sigma_{N, {\rm limit}} = 1.841 / V$. See \citet{Ebeling2003} and \citet{weigel2016} for details and further discussions.

\subsubsection{Cosmic variance}
It is well established that galaxy properties are correlated with environmental density (i.e., clustering). Galaxy clusters, while being an important laboratory for galaxy evolution, are not typical of galaxy environments. Because of their density, they impart a higher overall normalization to the stellar mass function. More noticeably, they tend to inflate the massive end of the mass function as they preferentially contain the most massive systems. This environmental field-to-field bias (so-called ``cosmic variance'') is a topic of intense study, and is a key component to accurately assessing sample uncertainties when trying to infer universal or intrinsic properties of galaxies. 

There are many published methods to estimate cosmic variance, based on numerical simulations \citep{BluetidesBhowmick2020, Ucci2021}, analytical models calibrated to observations solved either using linear theory \citep{moster2011cosmic, Trapp2020} or on forward simulation corrections to linear theory \citep{Steinhardt2021}, and observationally \citep[e.g.,][]{Driver2010CV_observational}. \citet{T+S_theory} combines results from cosmological simulations with analytical predictions.

Cosmic variance $\sigma_{\rm cv}$ is estimated following \citet{Steinhardt2021}, who adapt the methods of \citet{moster2011cosmic} which, importantly, scale with stellar mass (up to $10^{11.25}\,\Msol{}$) and are commonly adopted for use in $0.1\lesssim z \lesssim3.5$ measurements as that is the redshift range in which the \citeauthor{moster2011cosmic} calculator was devised. 
However, above $z\approx3.5$ these estimates become increasingly underestimated, so \citeauthor{Steinhardt2021} use linear perturbation theory to extend this work more reasonably to the early Universe, while maintaining agreement at $z<3$.
Although environmental density has known covariance with star formation \citep[e.g.,][]{Davidzon2016, Bolzonella2010}, we assume cosmic variance is equivalent between star-forming and quiescent subsamples. As shown in Fig.~\ref{fig:errors}, the fractional contribution from $\sigma_{\rm CV}$ is dominates the error budget for low-$\mathcal{M}$ systems at all redshifts and becomes progressively more important at high-$\mathcal{M}$ with increasing $z$.

\subsubsection{SED fitting uncertainties and bias}
\begin{figure}
   \centering
   \includegraphics[width=\columnwidth]{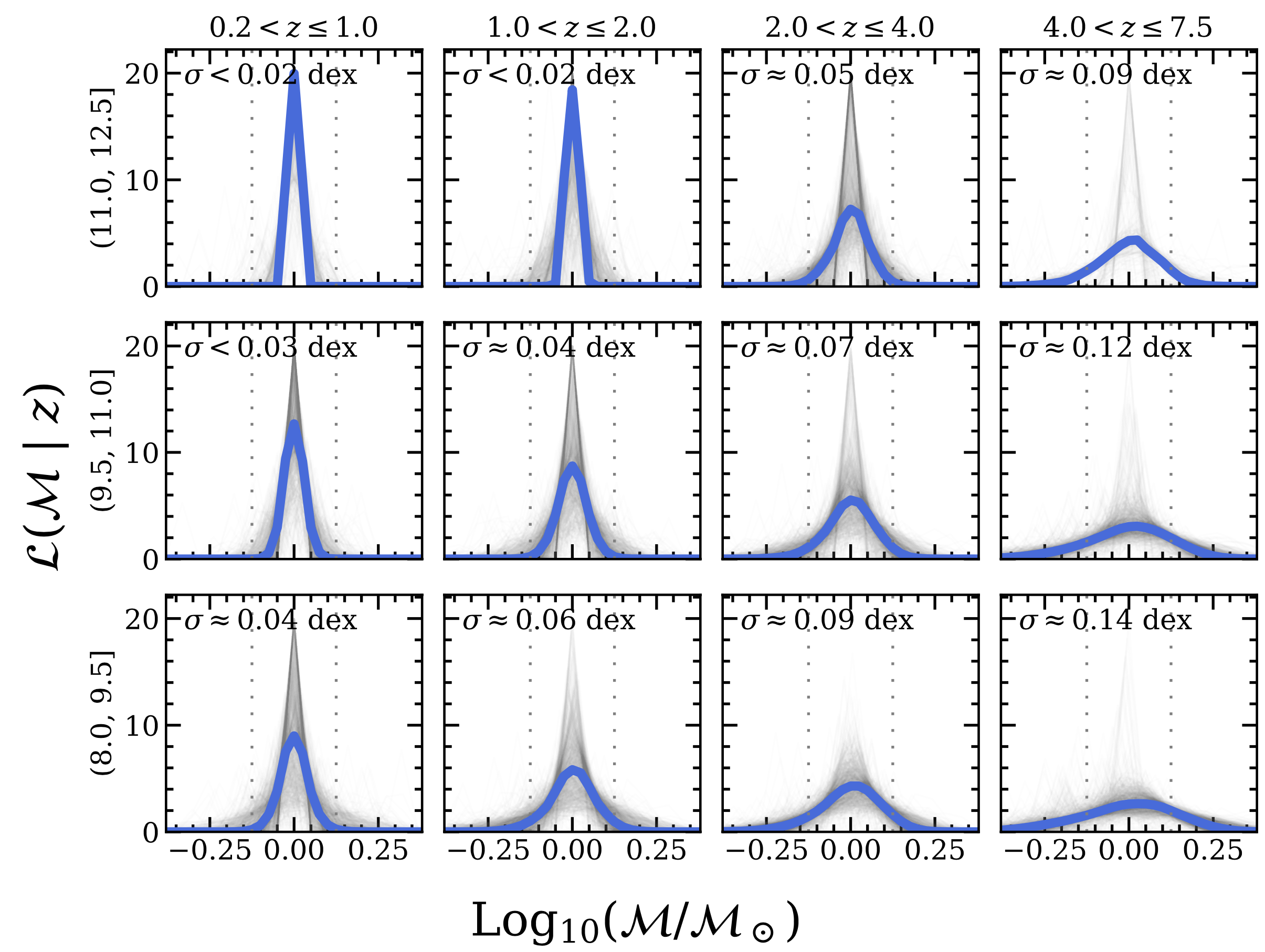}
     \caption{Likelihood distributions of galaxy stellar mass are derived from SED fitting with \lephare{} and assume a fixed redshift. Individual distributions (gray) are summarized by a median stack (blue) grouped by redshift and stellar mass (indicated in ranges of $\mathrm{log}_{10}(\mathcal{M}/\Msol{})$). Estimates of standard deviation $\sigma$ are shown. The size of a typical mass bin used in this work is 0.25\,dex, indicated by the pair of dotted gray lines in each panel.}
     \label{fig:pdfm}
\end{figure}

Another consideration is the uncertainty on the stellar mass estimate provided by the SED fit. Fig.~\ref{fig:pdfm} shows the likelihood distributions on stellar mass at fixed redshifts and masses produced by \lephare{} sampling at $\Delta \log_{10}(\mathcal{M}/\Msol{}) = 0.025$\,dex. Trends with width of the likelihood distributions indicate that mass is best constrained for low-redshift, massive (i.e., bright) sources. Although there is nonzero skewness and kurtosis in individual cases, the overall median distribution is symmetric. This is expected, as the uncertainty on stellar mass is essentially a measurement of the range of allowable templates and their normalization in the fitting procedure, and thus $\sigma_\mathrm{SED}$ scales with photometric uncertainties. However, these likelihood distributions on stellar mass should be treated as lower limits as they do not take into account any covariance with redshift.

While these typical per-bin distributions can be valuable, especially for injecting noise into measurements from simulations, attempting to compute its contribution to the SMF, $\sigma_{\rm SED}$, using the typical width in a given bin is suboptimal as the wings of neighboring mass-bins contribute asymmetrically. To address this, we use the individual mass likelihood distributions to draw 1\,000 independent realizations of the galaxy stellar mass function and thereby directly estimate the variance produced by the mass uncertainties, which we take as the 68\% range about the median number density per bin of mass. As shown in Fig.~\ref{fig:errors}, the contribution from $\sigma_{\rm SED}$ become dominant only at $\mathcal{M}>10^{11.5}\Msol{}$, in some cases becoming larger than unity. They are comparable to contributions from $\sigma_{N}$ across the entire mass range.

It is important to note that this does not account for systematic biases arising from SED fitting, such as assumptions of the stellar initial mass function\footnote{For example,  $\mathcal{M}$ computed assuming \citet{SalpeterIMF} is on average 0.24\,dex larger than those computed assuming \citet{ChabrierIMF}.}, potential \photoz{} offsets, and assumptions as to the star formation histories that all propagate into the stellar mass determination. However, concerning the latter case, we show in \citet{Weaver2022_catalog} that the combination of \cfarmer{} and \lephare{} achieves a subpercent \photoz{} bias even for faint, high-$z$ sources ($-0.004\,\delta z/1+z$ at $25<i<27$) improving over other works including COSMOS2015. Systematic errors cannot be combined with random errors, and so additionally complicate measurements of the SMF. Given this indication of relatively low bias arising from the \photoz{} and significantly better constrained SEDs relative to previous measurements, we omit these considerations in the present work. See \citet{Marchesini2009} and \citet{Davidzon17_mass} for detailed discussions of various sources of bias and their effect on the SMF.

\subsubsection{Eddington bias}
The number of low-mass galaxies is orders of magnitude larger that of the highest-mass systems, and so a randomly chosen galaxy is overwhelmingly likely to be lower-mass. If even a small fraction of such truly low-mass systems scatter toward high-mass (owing to a $\mathcal{M}$ overestimate) it can significantly change the poisson-dominated high-mass number density estimate. The converse situation, while depleting the high-mass end, would have virtually no effect on the low-mass estimates. This is the well known Eddington bias \citep{Eddington1913}. While generally understood to mean that there is a net bias leading to the overestimation of the density of massive galaxies, a small, but highly asymmetric uncertainty on the mass of low-mass systems can similarly generate such a bias that affects the shape of the low-mass regime. See \citet{Grazian2015} for further discussions.

Effectively correcting for Eddington biases has been a leading point of discussion in recent literature, generally favoring the convolution of the fitting function with a kernel that describes the uncertainty in stellar mass \citep[e.g.,][]{Davidzon17_mass, ilbert_mass_2013}. Recently, \citet{Adams2021} compared the effect of using three different forms for the convolution kernel finding a significant difference in the inferred intrinsic SMF. Alternative approaches have been proposed, e.g., \citet{Leja2019} developed a nonparametric formalism for incorporating $\sigma_\Phi$ into an unbinned Likelihood fitting, whereby multiple realizations of the parent catalog are made, each time sampling stellar mass from the mass likelihood distributions of each galaxy. In this work we primarily adopt the traditional convolution kernel method to estimate Eddington bias. At the same time, we also fit the mass function using the method of \citeauthor{Leja2019}, and so follow their approach where relevant.

\section{The Schechter function}
\label{sec:schechter}

Galaxy luminosity and stellar mass functions can be described empirically by the  parametric formulation first introduced by \citet{schechter1976} in the context of the local Universe. Since then, the Schechter function  has been adopted ubiquitously  in  statistical studies of galaxy mass assembly.  
It is more convenient to express the number density of galaxies per logarithmic mass bin $d\,{\rm log}\,\mathcal{M}$ as given by \citet{weigel2016}:

\begin{equation}
\begin{split}
    n_{gal} & = \Phi\,d\,{\rm log}\,\mathcal{M} \\
 & = {\rm ln\,(10)}\,\Phi^*\,e^{-10^{{\rm log}\mathcal{M} - {\rm log}\mathcal{M}^*}} \times (10^{{\rm log}\mathcal{M} - {\rm log}\mathcal{M}^*})^{\alpha+1}\,d\,{\rm log}\,\mathcal{M}\;,
\end{split}
\label{equ:ss}
\end{equation}

\noindent which describes a power law of slope $\alpha$ at  masses  smaller than  a characteristic stellar mass $\mathcal{M}^*$, whereupon the function is cut off by a high-mass exponential tail. The overall normalization is set by $\Phi^*$, which corresponds to the number density at $\mathcal{M}^*$. Although $\Phi\,(\mathcal{M})$ is expected to evolve smoothly with redshift such that $\Phi\,(\mathcal{M}, z)$ can be mapped from secondary functions $\mathcal{M}^*(z), \Phi^*(z), \alpha(z)$, most literature applications fit for each parameter independently at each redshift. A notable exception is \citet{Leja2019}. 

Many studies have since found evidence that the total galaxy population at low-$z$ is better described as the coaddition of two Schechter functions \citep[e.g.,][]{peng10_quenching} whereby the low-mass and high-mass end acquire individual normalizations ($\Phi^*_1$ and $\Phi^*_2$) and low-mass slopes ($\alpha_1$ and $\alpha_2$) but retaining a single characteristic stellar mass $\mathcal{M}^*$. This so-called `Double' Schechter form is as follows:
\begin{equation}
\begin{split}
    \Phi\,d\,{\rm log}\,\mathcal{M} &= {\rm ln\,(10)}\,e^{-10^{{\rm log}\mathcal{M} - {\rm log}\mathcal{M}^*}} \\
    & \times {\Big [} \,\Phi_1^*\,(10^{{\rm log}\mathcal{M} - {\rm log}\mathcal{M}^*})^{\alpha_1+1} \\
    & + \Phi_2^*\,(10^{{\rm log}\mathcal{M} - {\rm log}\mathcal{M}^*})^{\alpha_2+1}{\Big ]}\,d\,{\rm log}\,\mathcal{M}\;.
\end{split}
\label{equ:ds}
\end{equation}

The Double Schechter function has been found to be as suitable description in most of the studies in the local Universe and up to $z\approx2$. At higher redshifts, it is  less clear whether this is a better description of $\Phi$ than a single Schechter function. Moreover, deviations have been recently been reported at $z>7$ \citep[e.g.,][]{Stefanon2021} in which the observed stellar mass function is better described by a power law than by any Schechter profile. 

\begin{figure*}[ht]
	\centering
	\includegraphics[width=\textwidth]{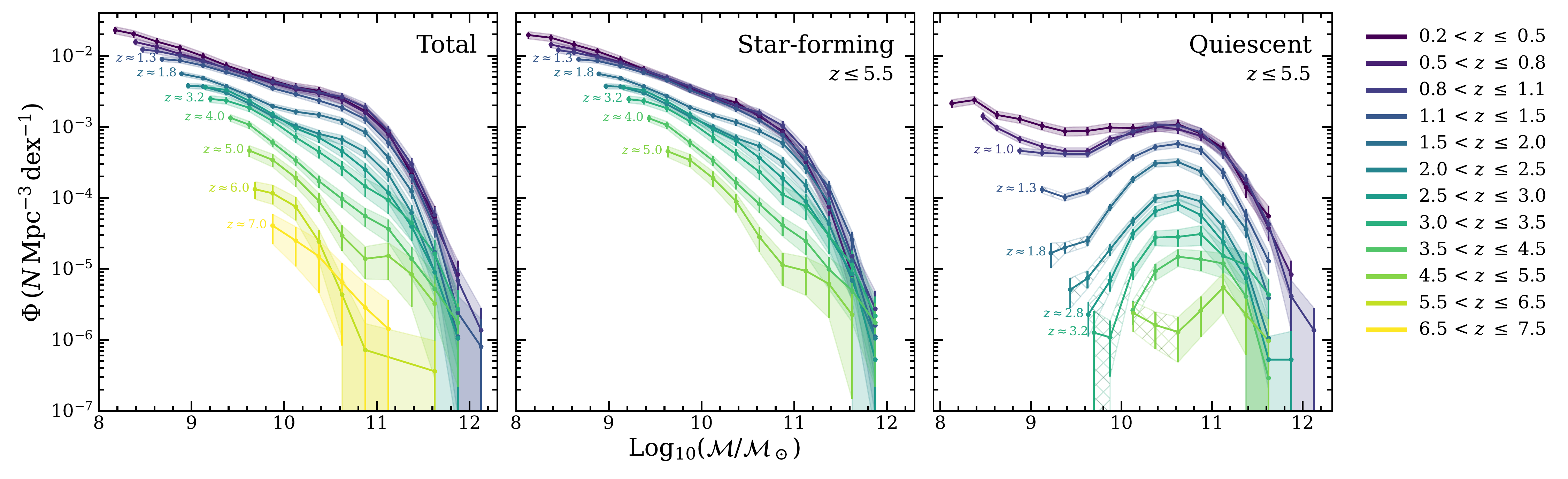}
	\caption{Evolution of the galaxy stellar mass function over 12 redshift bins ($0.2<z\leq7.5$) for the total, star-forming, and quiescent samples. Mass incomplete bins based on the channel~1 limiting magnitude are not shown. For the quiescent sample, bins that are fully incomplete based on the $K_s$ limiting magnitude are indicated by cross hatching.
	}
	\label{fig:mainfig}
\end{figure*}

\section{Results}
\label{sec:results}
Now having established the selections and methods adopted in this work, we present the resulting measurement of the SMF for the total, star-forming, and quiescent samples. We investigate also the evolution of number densities and quiescent fractions at fixed mass, and then fit the SMF for each sample with several methods to derive the evolution of the best-fit Schechter parameters.

\subsection{The total galaxy stellar mass function}
\label{subsec:total_smf}
We measure the SMF for our total (star-forming and quiescent) sample divided in 12 redshift bins from $z=0.2$ to $7.5$ in fixed bins of stellar mass $\Delta\mathcal{M}=0.25$ above the mass limit for that $z$-range\footnote{The range of the lowest-$\mathcal{M}$ bin at each $z$ is adjusted to include its respective mass limit at its lowest extent. Bins of width $<0.05$\,dex are discarded.}. Shown in the left panel of Fig.~\ref{fig:mainfig}, the shape and normalization of the SMF changes considerably over the $\sim\,10$\,billion years corresponding to this redshift range.  At $z\lesssim3$, the SMF features a smooth, monotonically decreasing number density at low-$\mathcal{M}$ that flattens before falling off steeply at $\mathcal{M}\approx10^{11}\,\Msol{}$, around the so-called `knee' of the function. Its overall shape and normalization remains roughly constant until $z\sim1.5$ indicating a lack of mass growth at recent times, consistent with the decline of the cosmic star formation rate \citep{Madau_2014}. However, by $z\sim1.5$ the normalization decreases dramatically. The number density at the knee is only $\sim1\%$ of its $z\approx0.5$ value with the fastest growth occuring on the low mass end, consistent with mass ``downsizing in time'' \citep[][]{Cowie1996, Neistein2006, Fontanot2009}. The slope of the low mass end steeps with time and is expected to be driven by physical processes including outflows, star formation efficiency, slope of the main sequence, and mergers (see discussions in \citealt{Peng2014}). At the same time, the knee itself becomes difficult to distinguish as the SMF takes the form of a smooth power law, and we become increasingly unable to constrain the low-mass end of the SMF due to worsening mass completeness (mass incomplete points are omitted in the  figure). As expected from Fig.~\ref{fig:errors}, the overall uncertainty grows significantly with increasing redshift and mass. We note that at a few epochs (e.g., $z\approx5$), the SMF is not strictly monotonic, likely driven by systematics rather than a real physical phenomenon (see discussions in Section~\ref{sec:discussion}).

\begin{figure*}[t]
	\centering
	\includegraphics[width=\textwidth]{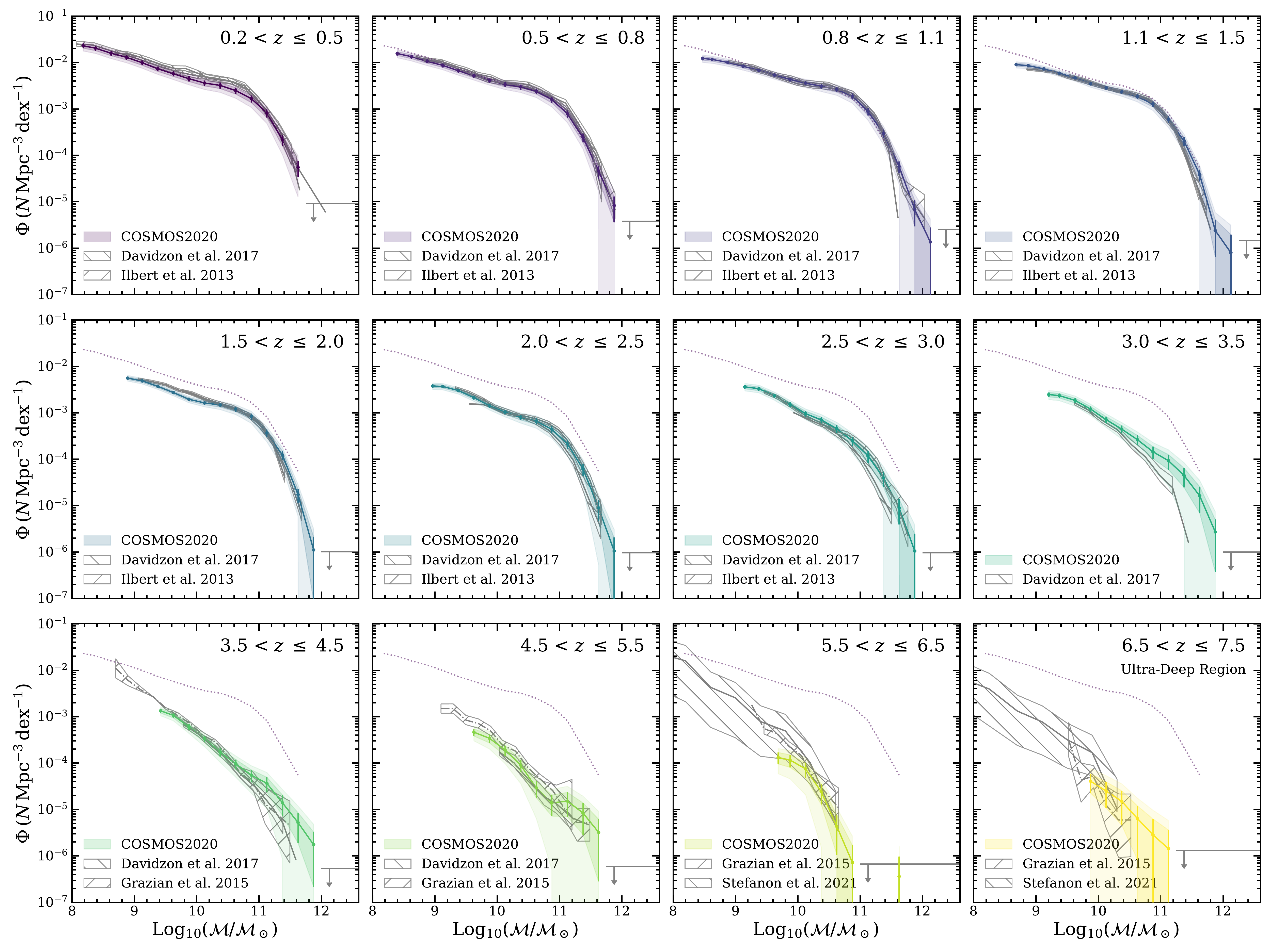}
	\caption{Evolution of the galaxy stellar mass function across 12 redshift bins ($0.2<z\leq7.5$). The $0.2<z\leq0.5$ SMF from the first redshift bin is repeated in each panel for reference shown by the purple dotted curve. Two other COSMOS stellar mass functions from \citet{ilbert_mass_2013} and \citet{Davidzon17_mass} are shown for comparison, along with \citet{Grazian2015} from UDS/GOODS-S/HUDF and the recent work of \citet{Stefanon2021} from GREATS at $z>6$. Mass incomplete measurements are not shown. Upper limits for empty bins are shown by the horizontal gray line with an arrow.
 	}
	\label{fig:comp_lit_total}
\end{figure*}

The evolution of the total SMF measured in this work is compared with literature results in Fig.~\ref{fig:comp_lit_total}. We begin at $z\approx0.2$ by comparing with two SMFs  previously measured in the same field: \citet{ilbert_mass_2013} and \citet{Davidzon17_mass}; both use \photoz{} and $\mathcal{M}$ computed with \lephare{} over COSMOS, out to $z\approx4.0$ and $z\approx5.5$, respectively, with the nearly same redshift binning scheme as we use. We note one exception that \citeauthor{ilbert_mass_2013} bins sources at $3.0<z<4.0$ whereas \citeauthor{Davidzon17_mass} uses $3.0<z\leq3.5$ and $3.5<z\leq4.5$, and so we opt to follow the scheme of \citeauthor{Davidzon17_mass} as it allows a comparison up to higher redshift and omit the comparison with the highest-$z$ measurement of \citeauthor{ilbert_mass_2013}. 
Additionally, \citeauthor{Davidzon17_mass} only considered sources in the ultra-deep regions of COSMOS2015 (corresponding to UltraVISTA (DR2) as the spatial inhomogeniety of the NIR bands implies a significant variation in selection function and mass completeness between the deep and ultra-deep regions. Thankfully, COSMOS2020 (corresponding to UltraVISTA DR4) contains nearly uniform NIR coverage ($\Delta \approx 0.4\,\mathrm{mag}$, \citealt{Moneti2019}) and so can leverage an area almost $2\times$ larger out to at least $z\approx6.5$, beyond which the source density becomes clearly different between the deep and ultra-deep stripes. Thus, for the $6.5<z\leq7.5$ bin we use the 0.72\,deg$^2$ subset of our primary area covered by the UltraVISTA deep stripes.

At $z<2.5$, in the range that all three can be directly compared, we find excellent agreement with \citeauthor{ilbert_mass_2013} and \citeauthor{Davidzon17_mass}. This is unsurprising, as measurements from \citeauthor{Davidzon17_mass} at $z<2.5$ are adopted directly from COSMOS2015 and computed nearly identically as \citeauthor{ilbert_mass_2013} but with deeper imaging. Our measurements are similar, but with visibly less structure around the knee especially between $0.8<z\leq1.1$ where the volume density of sources is slightly higher, with the greatest increase at $\mathcal{M}\sim10^{10}\,\Msol{}$. However, at $z\approx2.5$ \citeauthor{Davidzon17_mass} predicts a lower volume density than either \citeauthor{ilbert_mass_2013} or our measurements, which lie in agreement. This is not surprising, as \citeauthor{Davidzon17_mass} expanded the template library of \lephare{} to include both starbursting and dust-obscured systems (see Sect.~3 of \citeauthor{Davidzon17_mass}) and so our work is naturally more similar to that of \citeauthor{ilbert_mass_2013} -- a trend made clearer when looking at quiescent systems in the Section~\ref{subsec:sfqg_smf}. At $z>3.0$, we observe a significantly higher volume densities of massive galaxies compared to \citeauthor{Davidzon17_mass}, although the general shape of the low-mass end of the SMF remains similar. However, if we similarly limit our SMF to only the Ultra-Deep region we find it decreases to about the existing lower $1\sigma$ limit at $\mathcal{M}>10^{11}\mathcal{M}_\odot$. We speculate that this may be due to the presence of (proto-)clusters preferentially in the Deep region at $3<z\leq4$ \citep[see][]{Brinch2023}, one of which has been recently spectroscopically confirmed by \citet{McConachie2021}. Constraints from \citet{Grazian2015} can be introduced at $3.5<z\leq4.5$, although at a significantly higher degree of caution as \citeauthor{Grazian2015} uses a combination of smaller CANDELS fields (GOODS-South, UDS, HUDF) with an overall area of $\sim\,12\times$ smaller than the present study which implies a higher degree of uncertainty from cosmic variances and Poisson noise, as well as a reduced constraint on rare, high-mass systems found only in larger volumes. They are, however, completely independent as \citeauthor{Grazian2015} do not include the CANDELS-COSMOS field. Where a comparison is possible ($3.5<z\leq7.5$), our results are consistent with those of \citeauthor{Grazian2015} within the stated uncertainties. Similarly at $z>5.5$, we are able to compare with the recent measurements of the low-to-intermediate mass regime from \citet{Stefanon2021}, with which the present work also is consistent within the stated uncertainties. 

We do not include comparisons to stellar mass functions inferred from $L_{\rm UV}$-selected samples with stellar masses estimated from their UV luminosities \citep[e.g.,][]{Song2016, Harikane2016} as UV-bright sources are only a portion of the general galaxy population and conversations between $L_{\rm UV}$ and $\mathcal{M}$ are hazardous. Moreover, such studies often rely on color-color selections which are less certain than \photoz{}, in addition to being susceptible to a number of systematic effects when translating a UV luminosity function to a SMF by means of a $L_{\rm UV}-\mathcal{M}$ relation. We refer the reader to \citet{Davidzon17_mass} for basic comparisons and discussion.

While the smaller area surveys used by \citet{Grazian2015} and \citet{Stefanon2021} are more mass complete at lower masses, COSMOS2020 contains the widest contiguous NIR imaging at these depths and consequently provides a larger volume than any previous study (including \citeauthor{Davidzon17_mass}) and so introduces new constraints on the rarest systems at all redshifts $z\leq7.5$ such as massive galaxies. Yet it appears that we are quickly approaching the statistical limit beyond which a larger survey is needed to find more massive systems, if they exist (e.g., \textit{Euclid}, see discussions in McPartland et al. in preparation). The nature of these sources, their authenticity, and their potential Eddington bias is further assessed in Section~\ref{subsec:massivegal}.

\subsection{The star-forming and quiescent galaxy stellar mass functions}
\label{subsec:sfqg_smf}

\begin{figure*}[ht]
	\centering
	\includegraphics[width=\textwidth]{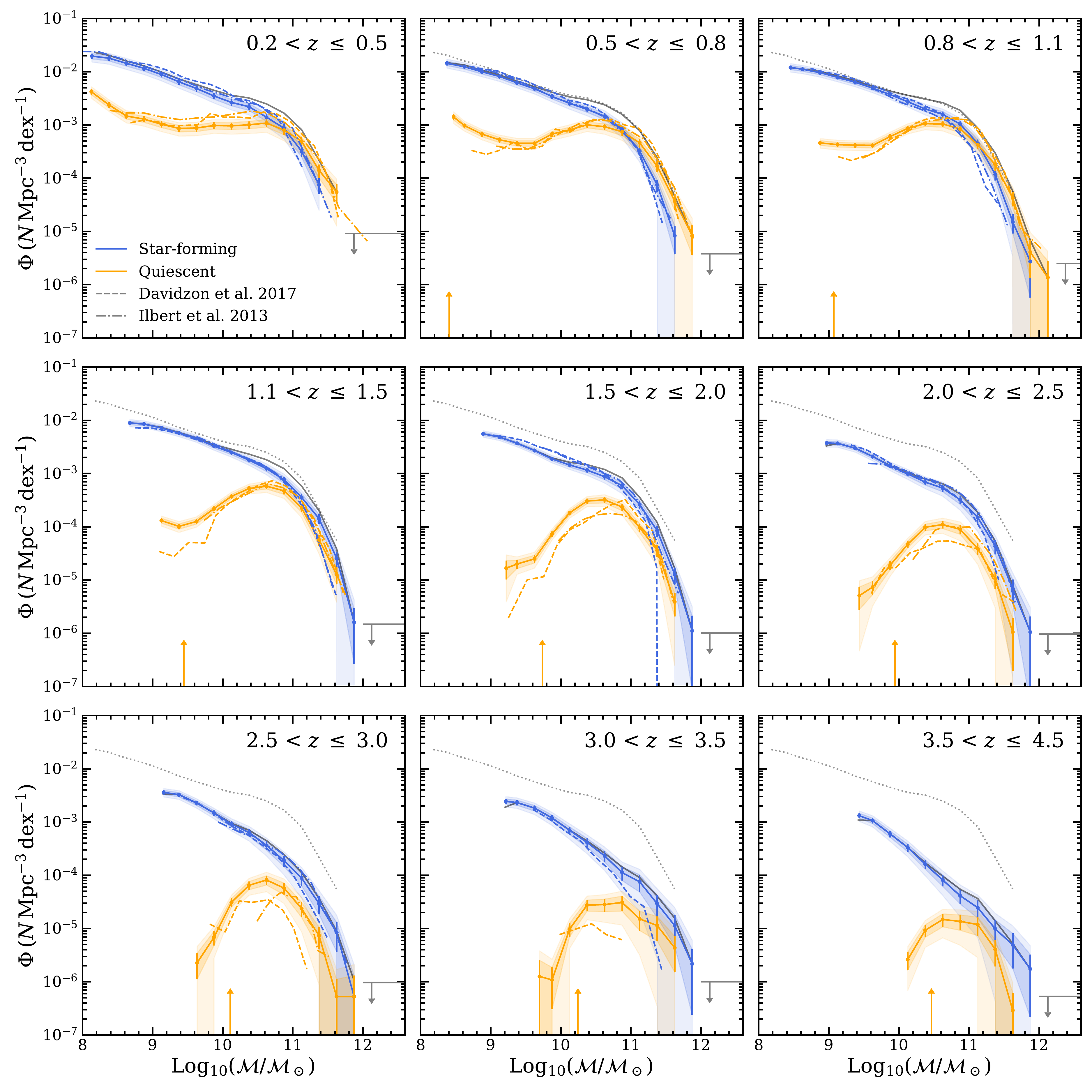}
	\caption{Evolution of the star-forming (blue) and quiescent (orange) galaxy components of the galaxy stellar mass function in nine bins of redshift ($0.2<z<4.5$). Uncertainty envelopes correspond to 1 and 2$\sigma$ limits. For comparison, we show other literature studies in COSMOS from \citet{ilbert_mass_2013} and \citet{Davidzon17_mass} that adopt similar selections and methodologies. For reference, the total SMF is shown in each bin (solid gray) and the $0.2<z<0.5$ total SMF is repeated in each panel (purple dotted). Mass incomplete measurements as defined by the channel~1 limiting magnitude are not shown, with the mass limits corresponding to the $K_s$ limiting magnitude shown by orange arrows. Upper limits for empty bins are shown by the horizontal gray line and arrow.
	}
	\label{fig:comp_lit_sfqg}
\end{figure*}

In Section~\ref{subsec:qgsel}, we used an $NUVrJ$ color-color selection to distinguish quiescent systems from star-forming ones and then estimated their respective mass completeness limits, which will differ since their $\mathcal{M}/L$ ratios are not the same. The corresponding SMFs are shown in Fig.~\ref{fig:mainfig} and compared to literature in Fig.~\ref{fig:comp_lit_sfqg}. Fitting results are discussed in Section~\ref{subsec:intrinsic} and details can be found in Appendix~\ref{app:fitting}; their corresponding Schechter parameters are reported in Tables~\ref{table:fit_sfg} and \ref{table:fit_qg} for the star-forming and quiescent samples, respectively. Fractional uncertainties on mass and cosmic variance are adopted from the total sample, leaving only the impact of poisson noise differentiated between the star-forming and quiescent samples (see Section~\ref{subsec:uncertainty}).

We can follow the development of quiescent systems out to $z\approx5$, although with significant uncertainty at $z\gtrsim4$. As evidenced by Fig.~\ref{fig:nuvrj}, only $\sim200$ of quiescent systems are found at $z\gtrsim4$, dropping precipitously to only one candidate by $z\approx6$. This is partly driven by mass completeness. At $z<3$ the difference between the IRAC channel~1 mass limit and the effective mass completeness dictated by our $izYJHK_s$ selection function is $\lesssim0.2\,$dex in mass. This difference grows to $\sim0.4\,$dex at $z\sim4.5$ when $K_s$ falls blueward of the Balmer break causing the mass completeness to shift upwards to higher masses, indicated by the hatched region of the SMF of $4.5<z\leq5.5$ quiescent systems in Fig.~\ref{fig:mainfig}. At this point the identification of quiescent systems is reliant on only a few bands, and is dependent on the particular SED templates (as suggested by the two distinct clusters in the upper right corner of the $4.5<z\leq5.5$ $NUVrJ$ diagram), making such a distinction hazardous and susceptible to interlopers.

\begin{figure*}[t]
	\centering
	\includegraphics[width=18cm]{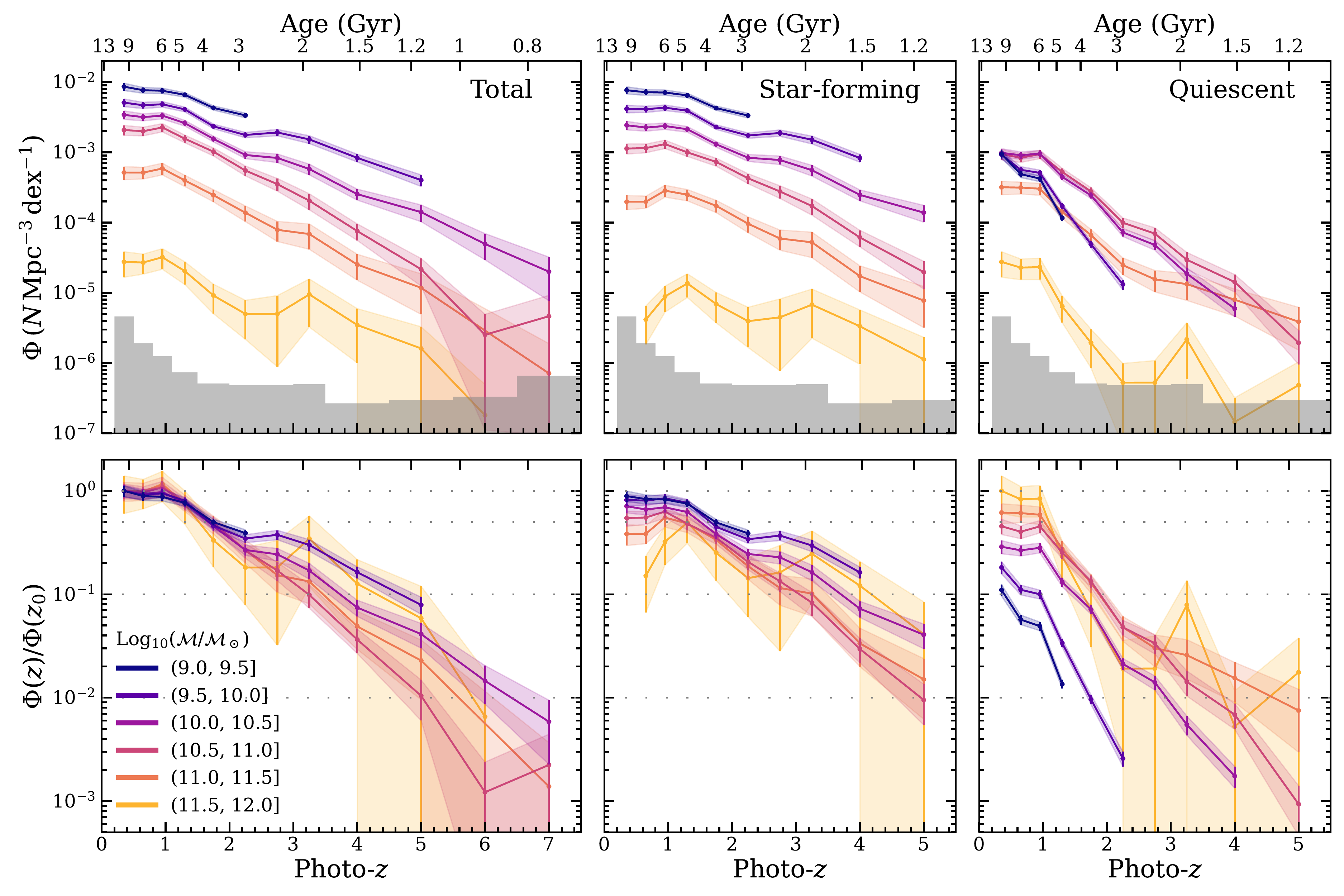}
	\caption{The evolution of the volume number density $\Phi$ of the total, star-forming, and quiescent samples at fixed mass, showing only mass complete bins. 1$\sigma$ uncertainties are indicated by the colored envelopes. Gray shaded regions indicate upper limits corresponding to $\Phi(N=0)$ at each redshift bin. The lower panel shows the evolution of the fractional change in the volume number density at fixed mass, relative to $z_0\equiv0.2<z\leq0.5$ of the total sample.}
	\label{fig:phivz}
\end{figure*}

Our measurements of the star-forming and quiescent SMFs are generally in good agreement with other literature measurements in COSMOS, namely \citet{Davidzon17_mass} and \citet{ilbert_mass_2013} as they are the only NIR-selected, mass-complete samples from which star-forming and quiescent subsamples are identified by $NUVrJ$. Other selections may induce additional systematics, and other separation methods (e.g., $UVJ$, $BzK$, sSFR) implicitly adopt a different criteria for quiescence \citep[see][]{Davidzon2018, leja_uvj_2019}. As shown in Fig.~\ref{fig:comp_lit_sfqg}, our measurements indicate similar high-$\mathcal{M}$ quiescent number densities compared to \citeauthor{ilbert_mass_2013} and \citeauthor{Davidzon17_mass}, but find significantly more low-mass lowest-$\mathcal{M}$ quiescent galaxies compared to \citeauthor{Davidzon17_mass} (which are not measurable from the data used by \citealt{ilbert_mass_2013}). Since our sample is derived from comparably deeper NIR data, it is expected to be complete down to lower masses relative to \citeauthor{Davidzon17_mass}. Given the increased quiescent galaxy number densities near the mass completeness limit in our work, we conclude that the 70-80\% completeness threshold of \citeauthor{Davidzon17_mass} is underestimated by a factor of $\sim2-10$ across this $z$-range, and is more likely only $15-35\%$ complete. We note, however, that this in agreement with worst-case scenario discussed by the authors (see Section~4.2 of \citealt{Davidzon17_mass}).

The SMFs of star-forming and quiescent galaxies have remarkably different shapes and evolutionary histories. As shown by Fig.~\ref{fig:mainfig}, star-forming galaxies at $0.2<z\leq0.5$ follow a double Schechter form (Equ.~\ref{equ:ss}) with a characteristic stellar mass $\mathcal{M^*}\approx10^{10-11}\,\Msol{}$ and yet by $z\approx3$ it appears to flatten into a smooth powerlaw-like form with a lower overall normalization. Meanwhile the SMF of $0.2<z\leq0.5$ quiescent galaxies follows the form of a double Schechter (Equ.~\ref{equ:ds}) with a low-mass upturn and a similarly positioned $\mathcal{M}^*$ \citep[see][]{Moutard2016} but the form appears to loose its lower-mass Schechter component around $z\sim2$. Although this may be physical, the fact that the quiescent sample is less mass complete at a given $z$ means that the low-mass end of the total sample reflects only the contribution from star-forming systems. However, the contribution from low-mass quiescent systems, if they exist, can be expected to be $<\,1\%$. 

This evolutionary picture is perhaps more easily understood by Fig.~\ref{fig:phivz}. Here galaxies have been binned by mass instead of redshift, allowing for a more direct comprehension of the growth, or lack thereof, in the number density of galaxies at fixed mass. We first notice a growth rate in the number densities that is strikingly consistent across all masses from $z=7.5\rightarrow1$ (i.e., similar slopes in each mass bin), constituting $\sim\,5$\,Gyr or $\sim\,36\%$ of the history of the Universe, consistent with recent findings by \citet{Wright2018}. This consistent growth is made especially clear in the lower left panel where the growth is relative to that of the total sample at $z_0\equiv0.2<z\leq0.5$. Although the uncertainties are considerable, there may be a hint that the growth is fastest (i.e., the slope is higher) for systems at $\mathcal{M}\sim10^{10.5-11.0}\Msol{}$ or at the `knee' of the mass function where star formation is hypothesized to be the most efficient \citep{Moster2013, Behroozi2013, Wechsler2018_ARAA}. However, we note that Eddington bias remains uncorrected and so a growth in the bias strength with redshift may produce this apparently slow evolution (as suggested by Fig.~\ref{fig:pdfm}). While the least massive systems are always more common than more massive ones, this is simply a consequence of the monotonic shape of the SMF.

Another interesting feature is that the number density of $\mathcal{M}=10^{9.5-10.0}$ and $\mathcal{M}=10^{10.0-10.5}\Msol{}$ systems appear to decrease mildly at $z\sim2.5$. There are fewer such systems in this work relative to both \citeauthor{Davidzon17_mass} and \citeauthor{ilbert_mass_2013}, and so this may be indicative of an incompleteness in our selection, or alternatively a bias in $z$ and/or $\mathcal{M}$. Determining the origin of this difference is nontrivial, and so we simply caution against over-interpretation of this feature.

The highest-$z$ constraints are difficult to interpret due to the incompleteness of low-mass systems (which are omitted) and the rarity of the most massive sources. The fact that the constraints at $z\gtrsim5$ overlap with the $\Phi(N=0)$ upper limit region in gray indicates that even the comparably large, deep volume of COSMOS is insufficient to provide robust measurements of the number density of massive galaxies at these early times. Future constraints will be derived from even larger volumes, which are expected to be delivered soon by near-infrared wide-area deep surveys such as \textit{Roman} and \textit{Euclid}.

This nearly uniform growth in the number density of sources slows down at $z\lesssim1$ in all but the least massive bin, with almost no growth for $\mathcal{M}>10^{10.5}\,\Msol{}$ systems. To understand this further, we examine the growth of the number densities of the star-forming and quiescent subsamples. As seen in the middle column of Fig.~\ref{fig:phivz}, star-forming systems maintain their monotonic mass-ranked order. In addition, their number density grows rapidly until $z\approx1$, when it starts to decrease. As shown in the rightmost column, quiescent galaxies follow a different pattern. Instead of a monotonic, mass ranked growth in number density, quiescent systems around the knee at $10^{10.5}\,\Msol{}$ are always the most numerous, with a slower rate of growth compared to the least massive and most massive bins. While the evidence for massive quiescent systems at $z>2$ is hampered by the limited volume of COSMOS, we can confidently see that they appear to grow quickly in number at $z<2$. By $z\approx1$ the number density of the most massive ones ($\mathcal{M}=10^{11.5-12.0}\mathcal{M}_\odot$) stabilizes at $\approx3\times10^{-5}\,{\rm Mpc}^{-3}\,{\rm dex}^{-1}$. Reasons for this are discussed in Section~\ref{subsec:massivegal}.

\begin{figure}
  \centering
  \resizebox{\hsize}{!}{\includegraphics{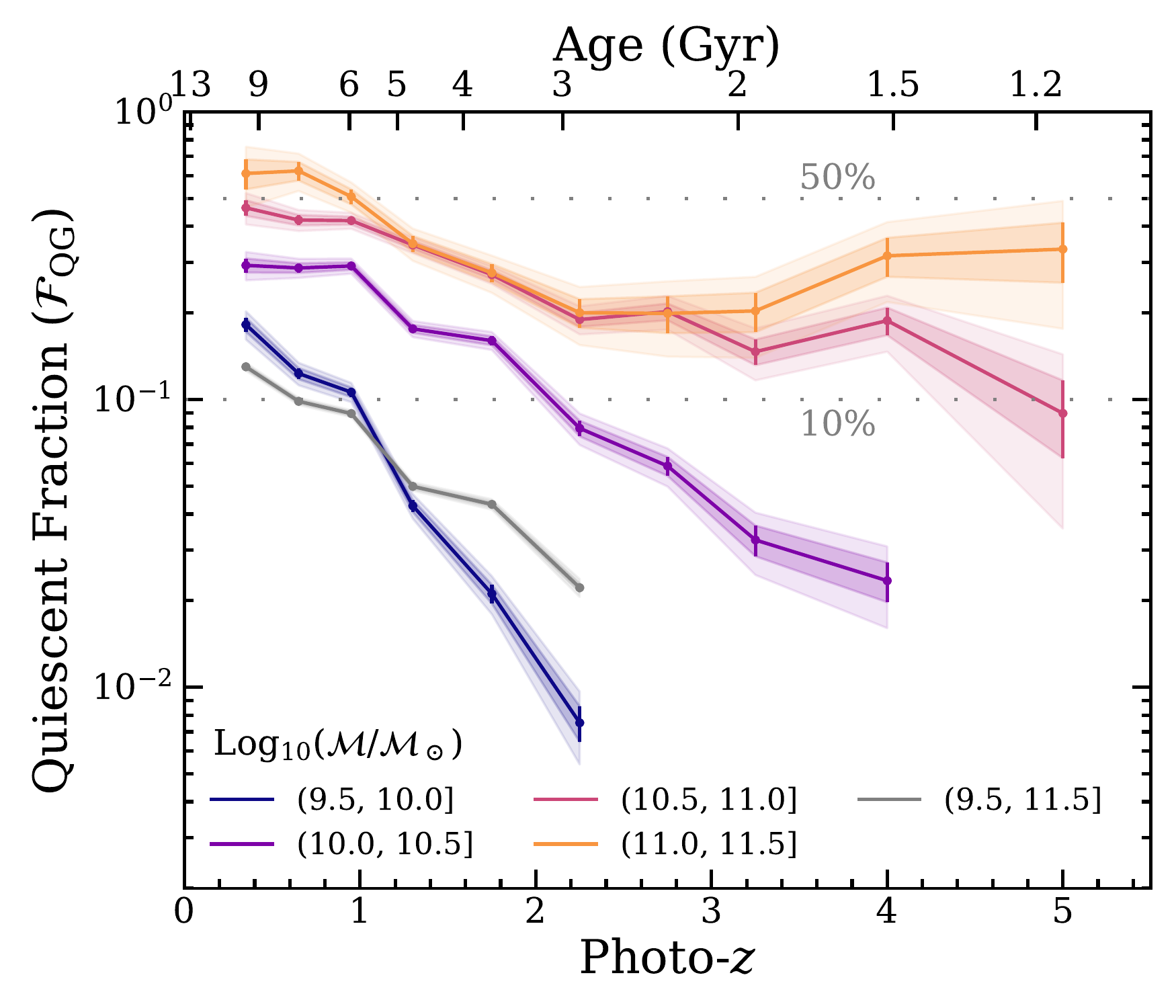}}
  \caption{Evolution of the fraction of quiescent galaxies as a function of redshift in four bins spanning $10^{9.5}<\mathcal{M}\leq10^{11.5}\,\mathcal{M}_\odot$. Uncertainty envelopes correspond to 1 and 2$\sigma$ limits. Points which are mass incomplete are not shown.}
  \label{fig:qg_frac_evol}
\end{figure}

Fig.~\ref{fig:qg_frac_evol} shows the evolution of the fractional contribution of quiescent systems in four 0.5\,dex bins at fixed mass and one wider 2\,dex bin. The quiescent fraction of low mass systems increases with time monotonically, and at a higher rate of growth than those of the highest mass, at least for $z<2$ where comparisons can be made. The quiescent fraction of the most massive systems at any redshift is more affected by both Eddington bias and poisson noise compared to low-mass systems and and so should be taken as an upper limit.  Nevertheless, it seems that between 20$-$30\% of $\mathcal{M}>10^{11}\,\Msol{}$ galaxies are quiescent from $z\approx5\rightarrow1$, growing above 50\% and plateauing at $z<1$.

\subsection{Inferring the intrinsic mass function}
\label{subsec:intrinsic}

\begin{figure*}[t]
	\centering
	\includegraphics[width=18cm]{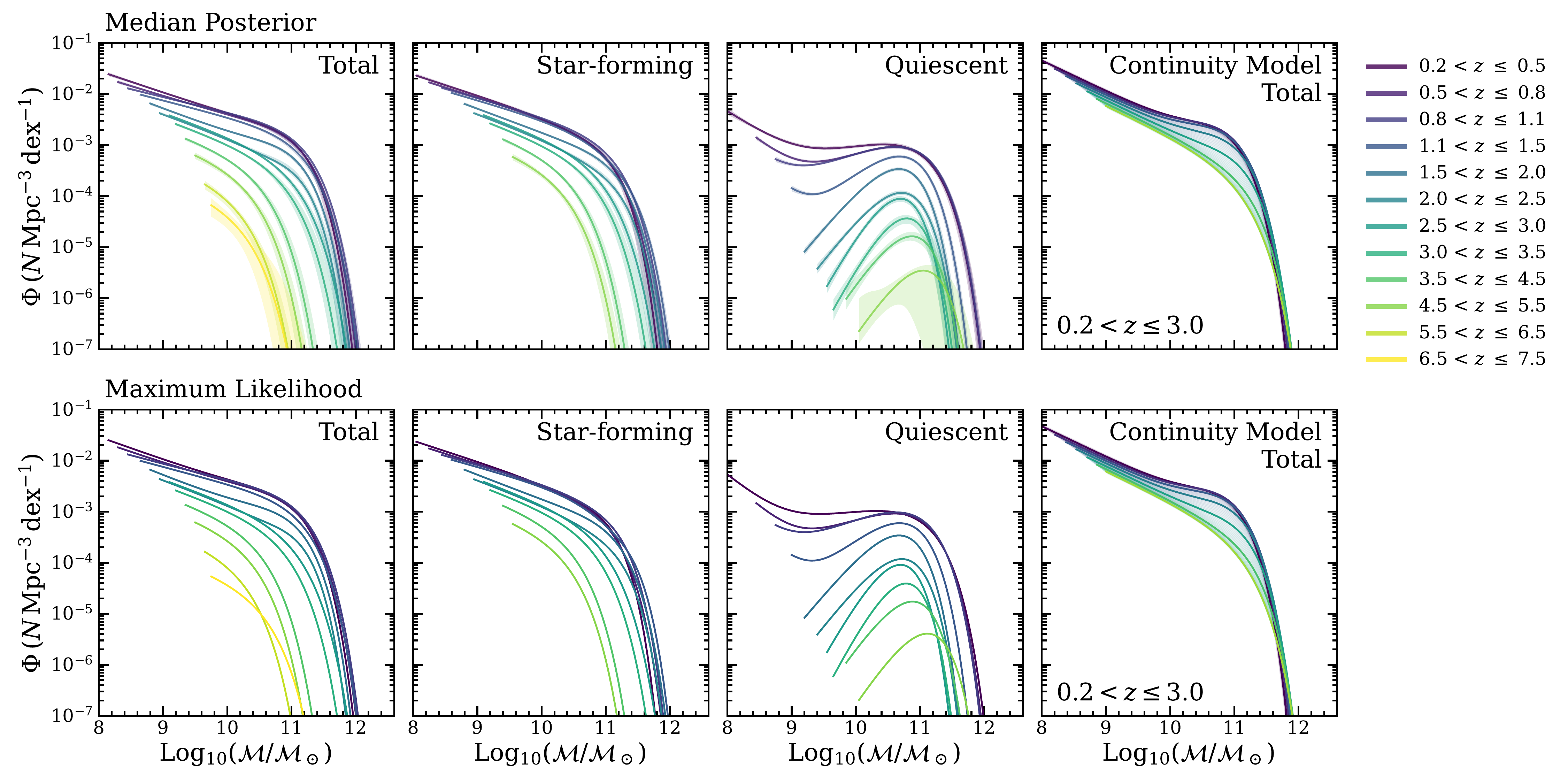}
	\caption{Summary of inferred galaxy stellar mass function evolution. Best-fit parameters are estimated by a Markov Chain Monte Carlo fitting with a Schechter function at fixed redshift convolved with a redshift dependent kernel from which the inferred intrinsic mass function is recovered with the unconvolved Schechter fit (colored solid curves) for the total (leftmost), star-forming (center left), and quiescent (center right) samples. Estimates are shown corresponding to both the median posterior parameter values including 1$\sigma$ envelopes (upper row) and the single set of maximum likelihood parameters (bottom row). The rightmost panel shows the results of fitting the Double Schechter continuity model of \citet{Leja2019} to the total SMF between $0.2<z\leq3.0$ where the fits are reliable; see later in Section~\ref{subsec:intrinsic} for details.}
	\label{fig:fit_summary}
\end{figure*}

So far we have accounted for three key sources of statistical uncertainty in Section~\ref{subsec:uncertainty}. However, Eddington bias remains a source of systematic uncertainty that has not yet been removed. 

To do so, we fit the observed total, star-forming, and quiescent SMFs with Schechter functions convolved with a kernel which attempts to describe the mass uncertainty for a given mass and redshift interval. This approach is common in the literature \citep{ilbert_mass_2013, Davidzon17_mass, Adams2021}. We adopt a two component parametric kernel of the form introduced by \citet{ilbert_mass_2013}:

\begin{equation}
\mathcal{D}(\mathcal{M}) = \frac{1}{2\pi}exp \left(\frac{-\mathcal{M}}{2\sigma_{\rm Edd}^2}^2 \right) \times  \frac{\tau_{\rm Edd}}{2\pi}\frac{1}{(\tau_{\rm Edd}/2)^2+\mathcal{M}^2} 
\end{equation}

\noindent which describes a core Gaussian component with a constant width $\sigma_{\rm Edd}$ in product with a Lorenztian component which provides wide wings that are crucial to capturing catastrophic errors in SED measurements. The other free parameter $\tau_{\rm Edd}$ is redshift dependent such that $\tau_{\rm Edd}$ = $\tau_c\,(1+z)$, widening the wings with increasing $z$.  Instead of fitting the kernel directly to the $\mathcal{L}(\mathcal{M}\,|\,z)$ distributions (as in \citealt{Davidzon17_mass} and \citealt{Adams2021}), we determine the kernel parameters from directly fitting the total SMF leaving the kernel free to vary independently at each redshift. However, both $\sigma_{\rm Edd}$ and $\tau_{\rm c}$ are nuisance parameters and leaving them free to vary is may produce over-fitting of any one SMF at a given $z$. As expected, the resulting parameter distributions with respect to $z$ contain outliers on either side of the median particularly where the kernel shape is most degenerate with that of the intrinsic SMF; we adopt the median values $\sigma_{\rm Edd}=0.2$ and $\tau_{\rm c}=0.1$. The resulting kernel is significantly wider than the $\mathcal{L}(\mathcal{M}\,|\,z)$ distributions shown earlier in Fig.~\ref{fig:pdfm}. This is because the latter assume perfect redshifts and therefore underestimate the full $\mathcal{M}$ uncertainty as there is likely to be a covariance with $z$. In addition, there is a mild evolution of $\mathcal{L}(\mathcal{M}\,|\,z)$ with $\mathcal{M}$, which we have omitted from our kernel form for the sake of simplicity and to avoid overfitting. Interestingly, while $\sigma_{\rm Edd}\approx0.2$ is smaller than that found by \citeauthor{Davidzon17_mass} (0.35), $\tau_{\rm c}\approx0.1$ is conspicuously larger (0.04, same as in \citealt{ilbert_mass_2013}). If we instead fit to the $\mathcal{L}(\mathcal{M}\,|\,z)$ distributions directly, we find $\sigma_{\rm Edd}\approx0.05$ and $\tau_{\rm c}\approx0.04$. These are much smaller than \citeauthor{Davidzon17_mass}, which may be explained in part by the slightly different approach they used to estimate $\mathcal{L}(\mathcal{M})$ that more directly incorporates uncertainties on \photoz{} (see Section~4.1 of \citeauthor{Davidzon17_mass}). Nevertheless, we opt for the pessimistic case and fix the kernel shape and evolution with $z$ to $\sigma_{\rm Edd}\approx0.2$ and $\tau_{\rm c}\approx0.1$ for the subsequent analysis of the total, as well as the star-forming and quiescent SMFs; their shapes are shown in the inset panels in Figs.~\ref{fig:fit_total}, \ref{fig:fit_sfg}, and \ref{fig:fit_qg}, respectively.

At the present, precisely determining the correct, intrinsic shape of the Eddington bias, and its evolution with $z$ as well as $\mathcal{M}$ is problematically difficult. In addition, the results stemming from these kernel-convolved fits suffer a degree of degeneracy with the kernel parameters (fixed or unfixed). We are not alone in issuing this caution; although the SMF measurements of \citet{Davidzon17_mass} is similar to those of \citet{Grazian2015}, their different choice of convolution kernel caused their inferred Schechter parameters to differ. Recently, \citet{Adams2021} explored the impact of the assumed shape of the kernel, as well as various other systematics at $z<2$, highlighting the full extent of the problem. Pushing measurements of the SMF to higher redshifts, where fewer constraints are available, naturally increases the influence of the kernel shape, and so the results derived here for $\alpha$, $\mathcal{M}^*$, and $\Phi^*$ (Tables~\ref{table:fit_total}, \ref{table:fit_sfg}, \ref{table:fit_qg}) should be taken in conjunction with our assumed kernel and its evolution with $z$. 

When fitting each sample (total, star-forming, and quiescent) we assume a double Schechter parametric form (Equ.~\ref{equ:ds}) and move to a single Schechter form (Equ.~\ref{equ:ss}) when it achieves a lower $\chi^{2}$ per degree of freedom. The point at which this change occurs, and when various parameters are fixed, are different for each of the three samples as it depends on not only their shape but also $\mathcal{M}_{\rm lim}$. The fitted points follow from before: we account for minor incompleteness on the low-mass end using the $1/V_{\rm max}$ correction and include only mass complete $\mathcal{M}$-bins adopting the uncertainty budget from Section~\ref{subsec:uncertainty}. We proceed in two stages, fitting first using a simple and efficient $\chi^2$ minimization routine \citep[\textsc{minuit},][]{James1975} whose resulting best-fit parameters are used to set the initial positions of walkers in a secondary and longer Markov Chain Monte Carlo (MCMC) optimization \citep[\textsc{emcee},][]{emcee}. It is generally appropriate to initialize walkers at well defined initial positions, assuming each chain is linearly independent (in this case scattering them by 1\% of the initial parameter values) and allowed to converge well past its respective autocorrelation length such that it is not sensitive to those initial conditions \citep{Hogg2010}. Flat priors are used for each parameter, with generous limits that are typically not encountered during a given fit. 500 MCMC walkers are initialized to seek the state of maximum likelihood derived similarly to the minimum $\chi^2$, and do so following the "stretch move" \citep{Goodman2010} until converged, defined as having every chain iterate for at least 10$\times$ their autocorrelation length, and every autocorrelation length having changed by $<1$\% of their previous value. We do not see any significant differences by using a different move style (e.g., Red-Blue, Metropolis-Hastings), suggesting that the data provide sufficient constraining power. Although we include the original $\chi^2$ results throughout, we focus on the MCMC results which provide full posterior sampling, which are taken from the last 90\% of each chain to avoid only the initial burn-in when the parameters are only beginning to converge. The MCMC and $\chi^2$ methods generally find consistent results. Fitting solutions corresponding to both the median posterior and maximum likelihood for the total, star-forming, and quiescent samples is shown in Fig.~\ref{fig:fit_summary}. An alternative version binned in redshift (as opposed to total, star-forming, or quiescent sample) is included for reference in Fig.~\ref{fig:fit_summary_alt} and discussed in Appendix~\ref{app:fitting}.

\begin{figure*}[t]
	\centering
	\includegraphics[width=18cm]{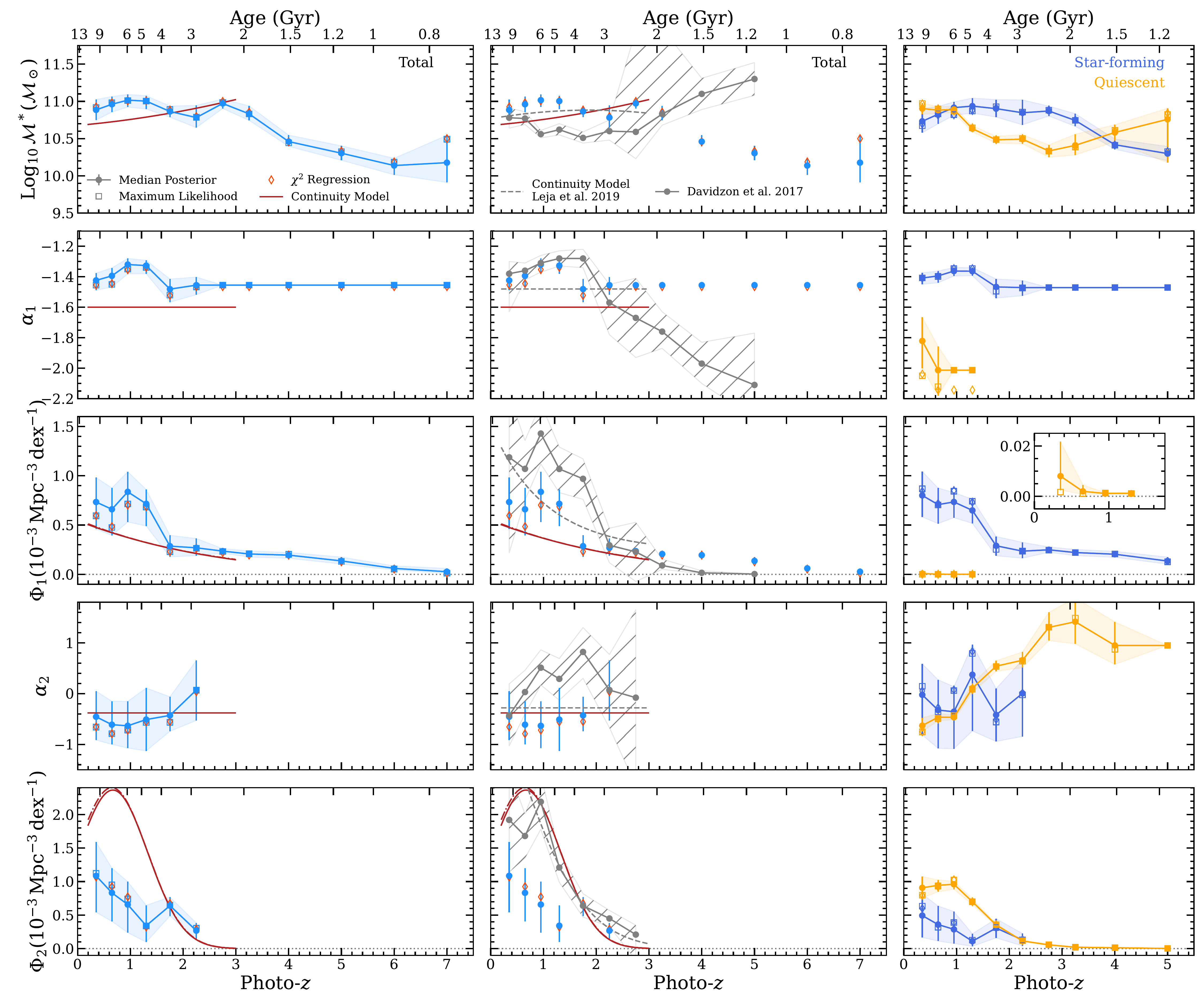}
	\caption{Schechter parameter evolution with redshift. \textit{Left:} Evolution of best-fit Schechter parameters computed from fits to the total SMF from $\chi^2$ regression (red diamonds), as well as likelihood methods using fixed redshifts bins (blue; median posteriors as filled points with 68\% error bars, maximum likelihood parameter set as unfilled squares), and our continuity model fit (maroon curve; median posterior). Error bars are not shown when parameters are fixed. \textit{Center:} Same points as on the left panels compared to literature values from \citet{Davidzon17_mass} (gray points with hatching) and \citet{Leja2019} (gray dashed curves). \textit{Right:} Evolution of the best-fit Schechter parameters computed from fits to the star-forming (blue) and quiescent (orange) SMFs from the $\chi^2$ regression (unfilled colored diamonds) and likelihood methods using fixed redshifts bins (median posterior as filled points with $\pm34$\% envelope, maximum likelihood as unfilled squares).
	}
	\label{fig:param_evol}
\end{figure*}

The evolution of the Schechter parameters inferred from the total SMF are shown in the leftmost panels Fig.~\ref{fig:param_evol}, and compared with \citeauthor{Davidzon17_mass} in the center panels. Table~\ref{table:fit_total} records the best-fit parameters, with detailed fits shown in Fig.~\ref{fig:fit_total}. 

As evidenced by the low $\chi^2$ values, the double Schechter well describes the observed SMF at $z\leq3$. No parameters are fixed other than those of the kernel. While a single Schechter finds a better fit at $z>3$, the increasing mass completeness limit weakens the constraints on $\alpha_1$. To avoid overfitting, we fix $\alpha_1$ to the value from the best-fit value at $2.5<z\leq3.0$ and do so for the $\chi^{2}$ and MCMC methods independently but to excellent agreement. All of the unfixed search parameters are generously bounded such that the chains are not likely to encounter them. However, for the sake of physicality especially at high-$z$, we choose to bound the search space for the normalization $\Phi\,(z>0.5)$ (both $\Phi_1$ and $\Phi_2$ where applicable) to below the upper $68\%$ on the posterior distribution of $\rm max\,(\Phi_1, \Phi_2)$ of the previous mass bin.

It is important to note that from $\chi^2$ minimization we obtain a single set of parameters that have minimized the $\chi^2$ as well as their symmetric, Gaussian uncertainties. On the other hand, MCMC provides only posterior distributions that  can be used to estimate parameter uncertainties but do not imply a unique definition of ``best-fit parameters''.  Commonly, the median is the best-fit statistic of choice, bounded by a 68\% one-parameter confidence interval (which ignores covariance). However, for highly skewed posteriors the median may lie out in the wings and is therefore not a typical value for that parameter. In this case, the most obvious choice may be the parameters corresponding to the model which has found the maximum likelihood. However, MCMC cannot provide an uncertainty on these maximum likelihood parameters, limiting its use. Worse, the model uncertainty cannot be derived from the posterior widths, as they also are covariant and so the resulting error envelopes will be meaningless. The most obvious choice then is to compute the medians and widths of posterior distributions of the chains after burn-in. However, the curve traced by the median is not guaranteed to reflect the actual model function, and the 68\% confidence envelopes may not contain the maximum likelihood nor the median posterior. We therefore strongly advise that models are reconstructed using the maximum likelihood parameters, and that caution should be exercised when interpreting best-fit parameters from median posteriors. Thankfully, the situation is less severe for symmetric posteriors, which for the total SMF are generally symmetric at $z\lesssim6$. At 
$6.5<z\leq7.5$, the posteriors are highly skewed as the relatively weak constraints produce widespread parameter covariance. The maximum likelihood results in all cases appear reliable. The evolution of the fitted Schechter models for both the median posterior and maximum likelihood is shown in Fig.~\ref{fig:fit_summary}. 

In addition to the fits at fixed redshift, we also use the method of \citet{Leja2019} to fit both the shape \textit{and} evolution of the SMF simultaneously. This aptly named ``continuity model'' is fitted on the unbinned distribution of mass-complete sources in $z$ and $\mathcal{M}$\footnote{This is not formally an STY method, as the normalization is constrained as a fitted parameter.}. Double Schechter parameters $\mathcal{M}^*$, $\Phi_1$, and $\Phi_2$ are treated as continuous functions over time described by second order polynomial expansions in $z$. The slopes $\alpha_1$ and $\alpha_2$ are assumed to be constant. These 11 parameters are joined by a minor parameter $\sigma_{\rm eff}$ which attempts to capture the cosmic variance. The effects of Eddington bias are incorporated by resampling the entire catalog by 10 random draws from their $\mathcal{L}(\mathcal{M}\,|\,z)$. Only the mass-complete points are fitted, which allows sources near the mass limit to scatter in and out of the SMF realizations. Importantly, this method accounts for the significant covariances in the Schechter function between adjacent redshift bins which is neglected when binning in redshift, and exploits it to provide generally tighter parameter constraints; see \citeauthor{Leja2019} for details. 

We employ \textsc{emcee} to determine the posterior distributions and maximum likelihood fit of the continuity model, noting the aforementioned caveats. Initially we tried to fit the entire SMF out of $z=7.5$, but the fits were inaccurate due to the limited range of evolution which can be described by a second order expansion. Already known from \citet{Davidzon17_mass}, $\Phi$ rises quickly before slowing down toward $z\sim0$. The second order expansion in $z$ can either fit the quick rise or the slow down, but not both. We therefore only consider galaxies at $z\leq3$ in our continuity model fit, as \citeauthor{Leja2019} do, and leave modifications to the fitting functions to future work. The expansion shapes are determined not by their coefficients, but rather by the amplitudes of three anchor points at fixed redshifts. The location of these anchor points is largely arbitrary, and so we follow \citeauthor{Leja2019} in choosing $z=0.2$, 1.6, and 3.0. Given the larger parameter space, we randomly initialize 100 walkers which explore the space until converged as before. We cannot apply the $1/V_{\rm max}$ correction directly to the continuity model, and so to remain conservative we exclude sources in the lowest 0.25\,dex from the $\mathcal{M}_{\rm lim}$ given in Equ.~\ref{equ:mc_total} for the total sample.

The evolution of the Schechter parameters inferred from the star-forming and quiescent SMFs are shown in the rightmost panels in Fig.~\ref{fig:param_evol}. Tables~\ref{table:fit_sfg} and \ref{table:fit_qg} record the best-fit parameters, with fits shown in Figs.~\ref{fig:fit_sfg} and \ref{fig:fit_qg}, respectively. Adapting the continuity model to the star-forming and quiescent SMFs is left to future work.

The treatment for the star-forming SMF fit is similar to that of the total form. We begin with a double Schechter form at $0.2<z\leq0.5$, and transition to a single Schechter form at $2.5<z\leq5.5$ with $\alpha_1$ fixed. The best-fit model describes the observed SMF reasonably well until $z\approx3.5-5.5$ where an excess of high-mass star-forming systems is observed that cannot be described by a single Schechter form. Possible reasons for this are discussed in Section~\ref{subsec:massivegal}. 

The quiescent SMF behaves noticeably differently from that of the star-forming sample and therefore requires a different strategy. We begin with a double Schechter form at $0.2<z\leq1.5$ but fix $\alpha_2$ at $1.1<z\leq1.5$ as the constraints on the low-mass regime deteriorate. With no significant low-mass constraints, we transition to a single Schechter at $1.5<z\leq5.5$. We note that this disregards a possible low-mass contribution and so the extrapolation to low-masses is likely an underestimate at $z>1.5$, depending on the growth of low-mass quiescent galaxies beyond our mass limit (see \citealt{Santini2022}). The normalization of the quiescent SMF continues to steadily decline with $z$. Given the uncertainty about sample completeness at $4.5<z\leq5.5$, we necessarily fix $\alpha_1$ to ensure a reliable fit of the secure measurements at high-$\mathcal{M}$. Consequently, we find a mild rise in the expected $\mathcal{M}^*$ from $z\approx3\rightarrow5$.

The evolution of the inferred total, star-forming, and quiescent SMFs resulting from the fixed redshift and continuity model fits are shown in Fig.~\ref{fig:fit_summary}. Furthermore, the evolution of the Schechter function parameters inferred from each sample are shown in Fig.~\ref{fig:param_evol}. Although the evolutionary physics inferred from the fitting will be discussed in Section~\ref{subsec:evolution}, we briefly describe the immediate result here. In general, parameters derived at fixed redshift from the $\chi^2$, maximum likelihood, and median posterior agree well for the total, star-forming, and quiescent SMFs, with few exceptions.

The characteristic mass is found to be $\mathcal{M}^*\approx10^{10.7-11.0}\,\Msol{}$ with very little significant evolution until $z\approx3$ when it begins to decrease with increasing $z$, with the continuity model suggesting a potentially increasing value with $z$. This contrasts with results from \citet{Davidzon17_mass}, who find an initial decrease in $\mathcal{M}^*$ which increases at $z>3$. The results of the continuity model of \citet{Leja2019} agree well with the lack of evolution suggested by our measurements, although in some tension with our continuity model results. $\mathcal{M}^*$ of the star-forming sample similar in shape but generally larger than that of the quiescent sample. The likeness to the total $\mathcal{M}^*$ at fixed redshift is proportional to the sub sample that dominates around the knee in that $z$ interval. 

The low-mass slope of the low-mass Schechter component $\alpha_1$ is roughly constant with time, although may experience a maximum at $z\approx1$, in agreement with \citeauthor{Davidzon17_mass}. However, whereas \citeauthor{Davidzon17_mass} finds a steeply decreasing $\alpha_1$ at $2<z\leq5$, we find a constant evolution which we then fix at $z>2.5$ to avoid mounting degeneracies as the low-$\mathcal{M}$ constraints are lost; this is also where a single Schechter becomes the preferred form. The value of $\alpha_1$ in good agreement with that found from our continuity model fit, as well as that of \citeauthor{Leja2019} who formed a composite SMF with significantly lower-$\mathcal{M}$ and hence $\alpha_1$ constraints. While $\alpha_1$ slope of the star-forming sample is similar to that of the total sample, the $\alpha_1$ slope of the quiescent sample is poorly constrained and appears to decrease with redshift (i.e. steepen with time), consistent with the rapid appearance of low-$\mathcal{M}$ quiescent systems at late times.

The normalization of the low-mass Schechter component $\Phi_1$ has little evolution at $0.2<z\leq1.0$, afterwards rapidly decreasing until it appears to approach zero asymptotically. Although \citeauthor{Davidzon17_mass} finds slightly larger values of $\Phi_1$ at low-$z$, both measurements generally agree on the rapid decline of the low-mass normalization. $\Phi_1$ derived from the continuity model finds still lower values and hence a more gradual evolution\footnote{We note that a second order expansion cannot describe an exponential tail as seen here, which justifies $z\sim3$ as the rightmost domain limit reachable by our 12 parameter continuity model.}. Yet, $\Phi_1$ derived by \citeauthor{Leja2019} agrees well with our fixed redshift measurements. As with $\alpha_1$, the $\Phi_1$ normalization of the star-forming sample is similar to that of the total sample and $\Phi_1$ of the quiescent sample remains significantly smaller and decreases with redshift.

The low-mass slope of the high-mass Schechter component $\alpha_2$ is statistically consistent with being constant with $z$, but may rise slightly toward $z\approx2$. This is broadly comparable with \citeauthor{Davidzon17_mass} within the stated uncertainties. $\alpha_2$ derived from our continuity model agrees well with the fixed redshift measurements, and with that of \citeauthor{Leja2019}. Interestingly, at $z\lesssim3$ we find that $\alpha_2$ - $\alpha_1 \approx 1$, which is predicted by \citet{peng10_quenching} as an indicator of mass quenching. The $\alpha_2$ slope of the star-forming sample is similar to that of the total sample, with $\alpha_2$ of the quiescent sample changing from negative at $z\lesssim1$ to positive at $1\lesssim z \lesssim 2$ where it appears to stabilize. 

Lastly, the normalization of the high-mass Schechter component $\Phi_2$ generally declines over $z=0.2\rightarrow3.0$. Interestingly, the continuity model finds an initial increase in $\Phi_2$ toward $z\approx1$, declining afterwards in agreement with the fixed redshift estimates. We find lower values compared to those of \citeauthor{Davidzon17_mass} and \citeauthor{Leja2019}, which are in general agreement in part because they both use COSMOS2015. This suggests that the different evolution in $\Phi_2$ between this work and the literature may be driven by subtle differences between the catalogs, sample selection, assumed Eddington kernels, or a combination of the three. The $\Phi_2$ normalization of the star-forming sample is subdominant to that of the quiescent sample with both decreasing steadily with time, consistent with the large fraction of quiescence found in massive systems out to $z\approx2.5$.

Overall we find good agreement with the most directly comparable literature measurements from \citet{Davidzon17_mass} and \citet{Leja2019}. The implications for these evolutionary trends, and their relation to the growth of massive quiescent galaxies, are explored in detail in Sections~\ref{subsec:massivegal} and \ref{subsec:rise_qg}.

\section{Discussion}
\label{sec:discussion}

In this section we focus on some of the open issues in galaxy evolution that can be addressed from a statistical perspective, using the results of Section~\ref{sec:results} as a starting point. Besides the assembly of $z\approx3-5$ massive galaxies and their quiescent fractions, we also discuss derived measurements such as the cosmic stellar mass density $\rho_*$, as well as the relation between stellar and dark matter halo mass functions. We include comparisons with simulations to provide physical insight and also to highlight areas for improvement from both sides.

\subsection{Cosmic stellar mass density evolution}
\label{subsec:evolution}

\begin{figure}[]
  \centering
  \includegraphics[width=0.926\columnwidth]{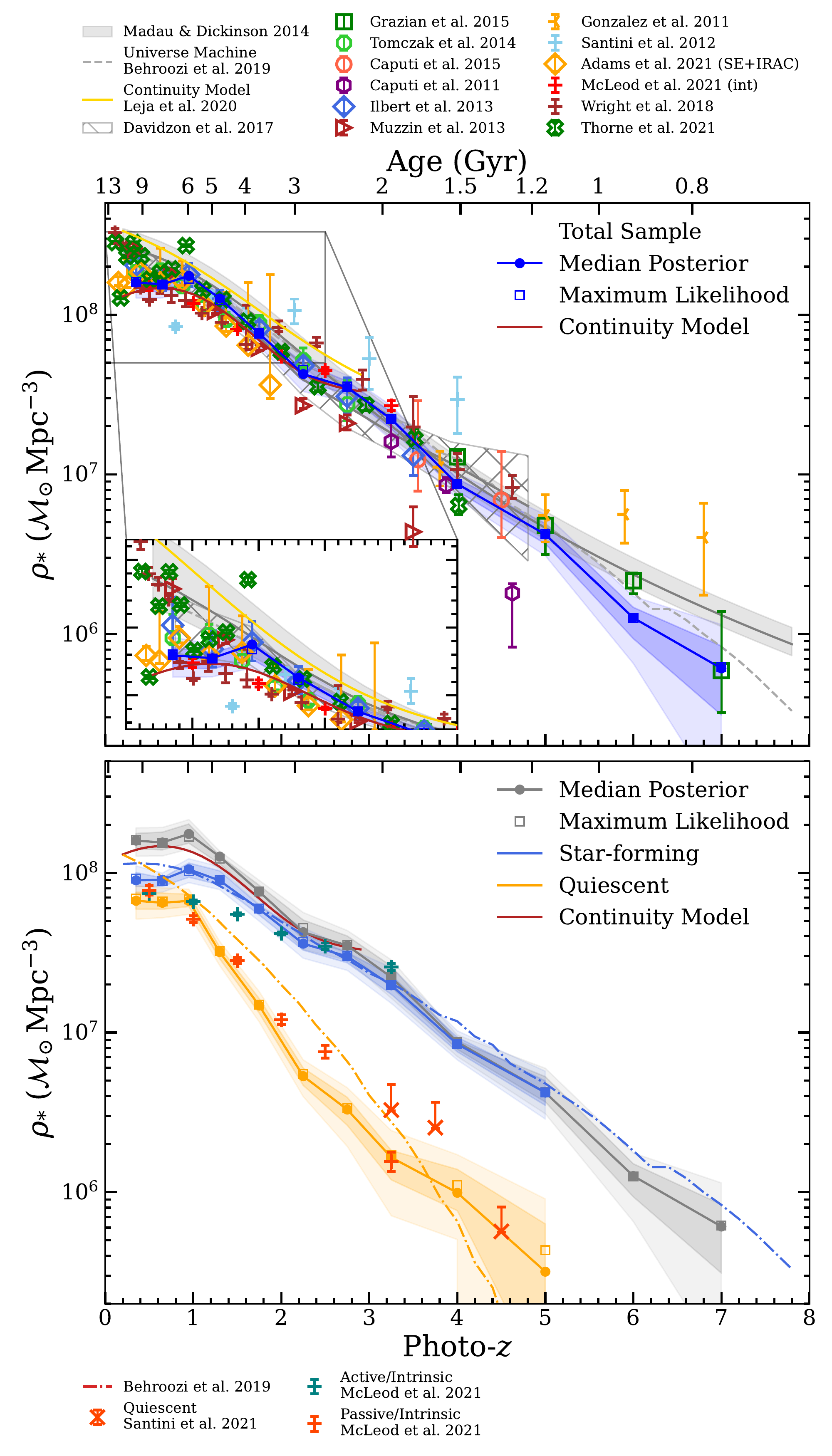}
  \caption{The $0.2<z\leq7.5$ cosmic stellar mass density. \textit{Upper:} Evolution of the cosmic stellar mass density of the total sample computed from the best-fit likelihood models (blue) integrated above $10^{8}\,\Msol{}$. Literature results from observational studies of mass-selected samples \citep{Grazian2015, Tomczak14, caputi15_highz_mass, Caputi2011, ilbert_mass_2013, muzzin13_mass, Santini2012, Adams2021, McLeod2021, Wright2018, Thorne2021} and mass inferred from rest UV measurements \citep{Gonzalez2011}. By integrating their SFRD functions, we can plot $\rho_*$ from \citet{Behroozi2013} and \citet{Madau_2014}. In both cases we assume a return fraction of 41\% (based on Chabrier’s IMF, see Section~6.1 of \citealt{Ilbert_2010}). For \citet{Madau_2014}, we include a shaded area based on return fractions between 25-50\% (the latter value is similar to the one given by Salpeter’s IMF). \textit{Lower:} Evolution of the cosmic stellar mass density of the total (gray, repeated from above), star-forming (light blue), and quiescent (orange) samples compared to literature measurements \citep{Behroozi2019, Santini2021, McLeod2021}.
  }
  \label{fig:massdensity}
\end{figure}

Galaxy mass assembly and growth is inextricably related to star formation. As such, reconciling direct measurements of galaxy mass growth with the behavior of the star formation main sequence \citep[e.g.,][]{Brinchmann2004, Noeske2007, Daddi2007, Salim2007, Whitaker2012, Whitaker2014} is of great interest. Meaningful assessment of empirical models of galaxy growth that link the two \citep[e.g.,][]{peng10_quenching, Behroozi2019} depends on unbiased, accurate measurements of the integrated mass density $\rho_*$ and its evolution with $z$.

We integrate the SMF measurements presented in Section~\ref{sec:results} to derive an estimate of the stellar mass density ($\rho_*$) for each bin of $z$. Although definitions vary, $\rho_*$ is commonly integrated down to $10^8\,\Msol{}$. Since our $\mathcal{M}_{\rm lim}$ at all redshifts is larger than $10^8\,\Msol{}$, we integrate into the extrapolated low-$\mathcal{M}$ regime of our (unconvolved) Schechter models. The resulting $\rho_*$ for the total sample in Fig.~\ref{fig:fit_summary} is compared with other literature measurements in Fig.~\ref{fig:massdensity}, converting to a Chabrier IMF where relevant. All SMF-based literature measurements of $\rho_*$ have been reintegrated consistently to the same mass limit. We also show $\rho_*$ derived from integrating the star formation rate density (SFRD) function of \citet{Madau_2014}, assuming a 41\% return fraction corresponding to Chabrier's IMF (see Section~6.1 of \citealt{Ilbert_2010}).

We find remarkably good agreement with previous observational studies. At $z\lesssim3$, the agreement seems to be limited by systematics as these are generally well measured, secure samples. However, $\rho_*$ at $z\gtrsim3$ is dominated by significantly less certain measurements, due both to the size of the samples and their typically noisy photometry. Our measurements place $\rho_*$ near the midpoint of the scatter, in closest agreement with \citet[][$z\leq5$]{Davidzon17_mass} and \citet[][$z\leq7$]{Grazian2015}. The relatively large uncertainties on $\rho_*$ beyond $z\approx4$ stem from the increasing degeneracy of $\mathcal{M}^*$ and $\phi$ with decreasing low-$\mathcal{M}$ constraints ($\alpha$ being fixed at $z>1.5$; see Fig.~\ref{fig:fit_total}).

At $z\gtrsim5$, our estimated median $\rho_*$ as well as that of \citeauthor{Grazian2015} is lower than the SFRD-derived predictions of \citet{Madau_2014}. This discrepancy could suggest that while star formation is high, the mass growth is lagging behind. As intriguing as this is, we stress that the measurements are consistent within $1\sigma$, that the SFRD constructed in \citeauthor{Madau_2014} are constrained by only two surveys at $z>5$ available at the time \citep{Bouwens2012a, Bouwens2012b, Bowler2012}, and that our measurements assume a fixed low-mass slope $\alpha$. Further work utilizing larger samples complete to lower masses will be required to confidently evaluate this divergence at $z>7$.

In addition to $\rho_*$ of the total sample, the lower panel of Fig.~\ref{fig:massdensity} includes $\rho_*$ for the star-forming and quiescent samples. While stellar mass is overwhelmingly concentrated in star-forming systems from $z\approx7\rightarrow3$, the mass density of quiescent systems grows rapidly until flattening at $z\approx1$, consistent with Fig.~\ref{fig:phivz}. From $z=1\rightarrow0$, there is no growth in $\rho_*$ for either star-forming and quiescent systems where the former remains larger than the latter.

While $\rho_*$ of the star-forming sample is well constrained out to lower masses, the same cannot be said about $\rho_*$ of the quiescent sample as its SMF is less mass complete in comparison. Since we cannot directly determine its low-$\mathcal{M}$ shape, we assume that the low-$\mathcal{M}$ Schechter component effectively vanishes at $z\gtrsim1.5$. If this is not the case, then we underestimate $\rho_*$ for the quiescent sample. This may help explain the differences observed with respect to \textsc{Universe Machine} from \citet{Behroozi2019} who report generally larger values of $\rho_*$ for their quiescent sample, which features a significant quiescent population at low $\mathcal{M}$ (see Fig.~\ref{fig:sfqg_sims}). Although in agreement at $0.2<z\leq0.5$, \citeauthor{Behroozi2019} report higher quiescent fractions at $<10^{10.7}\Msol{}$ with redshift, finding more than $10\times$ as many $10^{9}\,\Msol{}$ by $z\approx1$. \textsc{Universe Machine} primarily defines quiescence as sSFR$ < 10^{-11}\,{\rm yr}^{-1}$, which according to \citet{Davidzon2018} is comparable our $NUVrJ$-selection, and so this overproduction of quiescent galaxies may indeed be at odds with our observations. Although not readily available, a consistently $NUVrJ$-selected sample from \textsc{Universe Machine} would clarify this sensitive comparison. Comparisons with the results of \citet{Santini2021} and \citet{McLeod2021}, although generally in agreement, are also complicated by differences in selection. Whereas \citet{Santini2021} adopts a novel SFR-driven selection, \citet{McLeod2021} adopts a $UVJ$ selection following \citet{Carnall_2018}. Therefore, we stress that these estimates of $\rho_*$ for our quiescent sample are particular to $NUVrJ$-selected quiescent systems, and our assumption of a single Schechter description at $z>1.5$ may drive our relatively low estimates. Finally, despite differences in selection and constraining power, these observational studies collectively indicate a general agreement as to the evolution of $\rho_*$ for quiescent systems.

\subsection{Comparison to simulations}
\label{subsec:compsims}

\begin{figure*}[t]
	\centering
	\includegraphics[width=18cm]{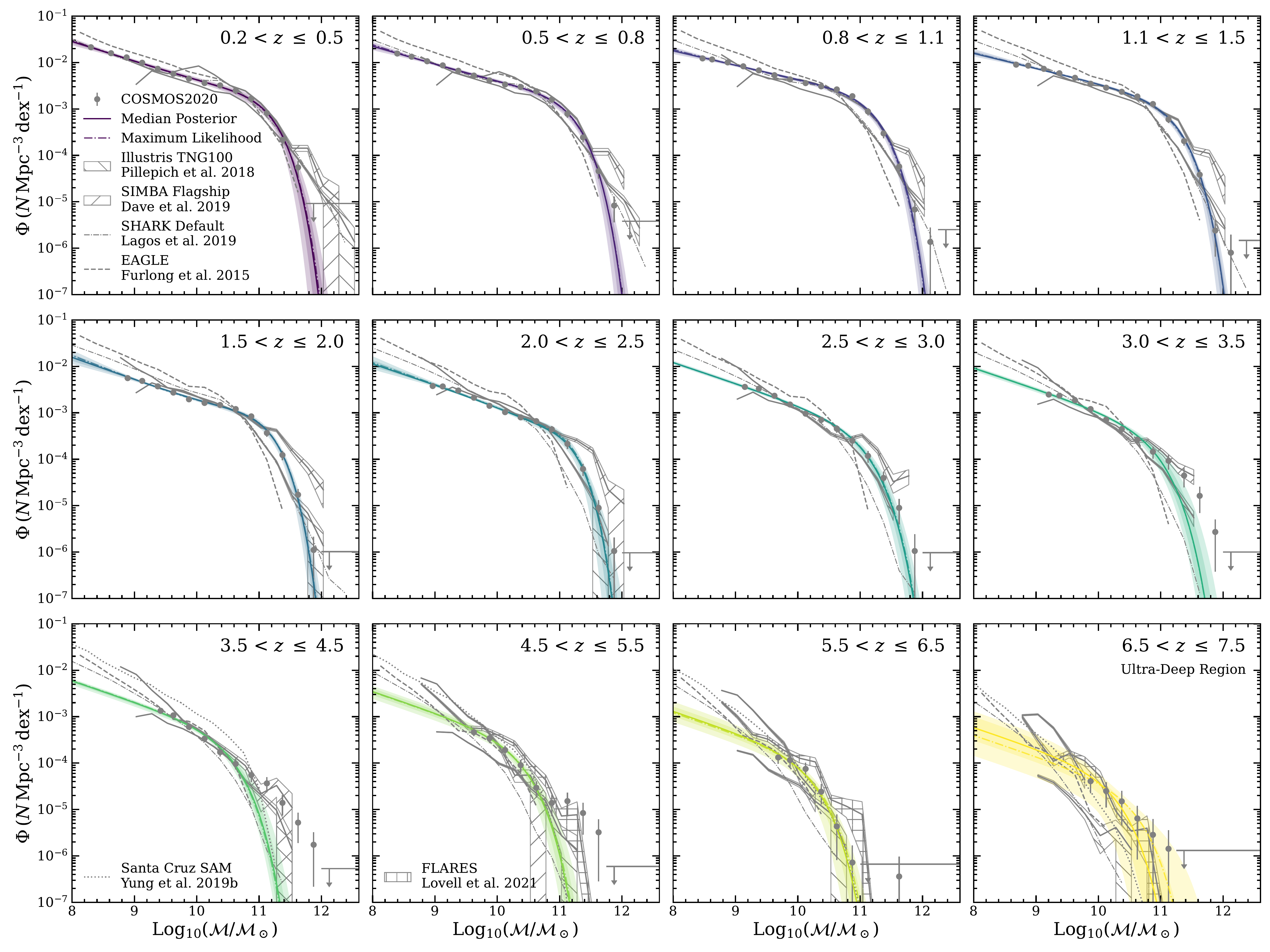}
	\caption{Comparison of observed and inferred galaxy stellar mass function (gray points, and colored curve with 1 and 2$\sigma$ envelopes, respectively) to the reference flavors of four simulations: \textsc{TNG100} of the \textsc{IllustrisTNG} project \citep{Pillepich2018_ILLUSTRIS},  \textsc{Eagle} \citep[][]{Furlong2015_EAGLE},  \textsc{Shark} \citep[Default;][]{Lagos2018_SHARK}, \textsc{Santa Cruz} \citep{Yung2019a, Yung2019b}, and \textsc{Simba} \citep[Flagship;][]{Dave2019_SIMBA}. To note, we do not apply any artificial normalization to any model or observation. Upper limits for empty bins are shown by the horizontal gray line with an arrow. Mass incomplete measurements are not shown. 
	} 
     \label{fig:comp_sims}
\end{figure*}

Observational constraints on the shape and evolution of the SMF may be useful in their own right, but we can also gain meaningful insight by comparing them with SMFs produced by galaxy formation simulations. Fig.~\ref{fig:comp_sims} shows the SMF constraints from this work and the inferred SMF derived by the binned MCMC fits evaluated at maximum likelihood, with the flagship or reference versions of simulated hydrodynamical SMFs overlaid: \textsc{TNG100} of the \textsc{IllustrisTNG} project\footnote{We choose TNG100 as a compromise between resolution and volume, see \citealt{Pillepich2018_ILLUSTRIS, Donnari2021a, Donnari2021b}.}, \textsc{Eagle}, \textsc{Simba} and \textsc{Flares} \citep[][respectively]{Pillepich2018_ILLUSTRIS, Furlong2015_EAGLE, Dave2019_SIMBA, Lovell2021_FLARES}. It is worth noting that \textsc{Flares} uses the same code and model as \textsc{Eagle} (specifically they adopted the strong AGN feedback \textsc{Eagle} model introduced in \citealt{Crain15_eagle}), but sample regions that span a wider dynamical range, and therefore has an much larger effective volume than \textsc{Eagle}. We also include the semi-analytical results from \textsc{Shark} \citep{Lagos2018_SHARK} as well as those of \textsc{Santa Cruz} (\citealt{Yung2019a, Yung2019b}, see also \citealt{Somerville2015}). To note, we do not apply any artificial normalization (aside from \textit{h} considerations) to any model or observation such that direct comparisons are possible from the figures alone.

Since our observed SMF measurements are affected by Eddington bias, it is preferable to compare the SMF directly predicted by the simulations assuming no errors to our inferred SMF with considerations as to the assumptions made in our Eddington bias correction. We acknowledge that each simulation has multiple flavors with variations to their physical recipes. Although these variations provide additional insight, care must be taken when making these more complicated comparisons. As the goal of this work is to provide constraints on the observed and inferred SMF, we reserve comparisons to the variations of simulated SMFs for future work.

Overall, \textsc{Flares},  \textsc{Shark}, and  \textsc{Eagle} perform best below $\mathcal{M}^*$, with \textsc{Simba}, \textsc{Flares}, and the \textsc{Santa Cruz} SAM performing best above $\mathcal{M}^*$. At high-$z$ ($z\gtrsim4$), we find the best agreement with \textsc{Simba}, \textsc{Illustris} TNG100, and the \textsc{Santa Cruz} SAM. At low-$z$  ($z\lesssim1.5$), however, there is a slight preference toward \textsc{Simba} as \textsc{Illustris} TNG100 underestimates the number densities at $\mathcal{M}^*$ for $z\leq1.5$. The situation is different at $1.5<z\leq3.0$ where \textsc{Simba} significantly overestimates the high-$\mathcal{M}$ end. While this could be explained by a systematic bias in observed masses or 
a missing high-$\mathcal{M}$ population (see Section~\ref{subsec:massivegal}), it could also be due to \textsc{Simba} potentially over-grouping several separate dark matter haloes into one massive halo and/or insufficient AGN feedback at early times. Over this same range, it is apparent that \textsc{Eagle} suffers from volume limitations and thereby does not contain the most massive galaxies. This is most clear from the comparison between {\sc Eagle} and {\sc Flares}, which are based on the same physical recipes, but the latter sample much rarer overdensities, and hence captures the high-$\mathcal{M}$ population.  Both the \textsc{Shark} and Santar Cruz SAMs fare better as their semi-analytical prescriptions are able to produce high-$\mathcal{M}$ systems. In some $z$-bins, both  \textsc{Shark} and  \textsc{Eagle} assemble low-$\mathcal{M}$ systems too early relative to the observed SMF where there are direct constraints (most apparent at $1\lesssim\,z\,\lesssim 2$). We find a lesser degree of agreement at $z>3.5$: while both  \textsc{Shark} and  \textsc{Eagle} underproduce the number of galaxies at all $\mathcal{M}$, there is significant scatter between \textsc{Illustris} TNG, \textsc{Simba}, \textsc{Flares}, and \textsc{Santa Cruz}. Meanwhile the volume limitations of \textsc{Illustris} TNG100 become significant at $z>6$ (see \citealt{Pillepich2018_ILLUSTRIS} for comparisons with TNG300).

It is remarkable that all of the simulations\footnote{With the exception \textsc{Flares} and the \textsc{Santa Cruz} SAM which are limited to $z>4.5$ and $z>4.0$, respectively.} reproduce the $0.2<z<0.5$ SMF, and yet at higher redhshifts display severe disagreement with our observations and each other with number densities differing by more than a factor of ten. While this should not come as a surprise given that simulations are typically tuned to reach an end state in agreement with the local Universe (despite known disagreements with the observed sSFR history and merger rates, see \citealt{Popesso2022}, \citealt{Conselice2022}), it suggests that their initial conditions and early evolutionary behavior are strikingly different such that variously tuned physical recipes and initial conditions can produce the same end state. This highlights the need for continued observations to improve constraints on the SMF at high-$z$ where simulations can be critically tested, and their physical thereby recipes refined. In addition to the treatment of AGN and associated outflows \citep[e.g.,][]{Debuhr2012, Richardson2016, Mitchell2020}, modeling of dust attenuation at high-$z$ is a particularly relevant concern at $z\gtrsim4$ where the dust content of galaxies has been largely unknown, and in turn makes stellar mass estimates based on rest-frame UV/optical light increasingly uncertain. This, coupled with uncertainties in dust production mechanisms at high-$z$, can lead to simulations diverging significantly (from each other and observations), and has often been cited as a major contributor to such discrepancies \citep[see discussions in e.g.,][]{Kokorev2021}. Although future observations will benefit from improved more comprehensive mass uncertainties, our current constraints indicate that while it appears that simulations are able to reproduce the observed abundance of high-$\mathcal{M}$ galaxies and so the production of early low-$\mathcal{M}$ systems needs improvement.

\subsection{Abundant massive galaxies at $z\sim3-5$}
\label{subsec:massivegal}

One of the most striking results highlighted in Fig.~\ref{fig:comp_lit_total} is the high number density of massive $\mathcal{M}>10^{11}\,\Msol{}$ galaxies not only at $z>3.5$. Although few in number, their identification in COSMOS2020 is a direct consequence of utilizing the larger 1.27\,deg\textsuperscript{2} now accessible with deep, homogeneous NIR coverage. While no $\mathcal{M}>10^{11}\,\mathcal{M}_\odot$ galaxies are observed at $z>5$, their growth since then has been similar to galaxies at other masses, as shown in Fig.~\ref{fig:phivz}. The majority of $\mathcal{M}>10^{11}\,\Msol{}$ systems are star-forming at $z>1$, as shown by Fig.~\ref{fig:comp_lit_sfqg} and quantified in Fig.~\ref{fig:qg_frac_evol}, with only $z<1$ systems at $\mathcal{M}\approx10^{10.5-11}\,\Msol{}$ being equally divided between star-forming and quiescent states. We find evidence for a sustained population of massive quiescent systems at $z>2$, but their number densities are dwarfed by that of star-forming systems by a factor of $\sim10$. The existence of these massive quiescent systems seems to defy the timescales expected for mass assembly \citep{Steinhardt2016, Faisst2017, Schreiber2018b, Cecchi2019}, and so their tremendously early formation and growth are a topic of great interest \citep{Caputi2011, Toft2017, Hill2017, Carnall2018, Tanaka2019, Valentino20_QGz4, Whitaker2021, Akhshik2022, Marsan2022}, including a more focused investigation into the origins of similarly selected massive galaxies from COSMOS2020 by \citet{Gould2023}.

\begin{figure}
  \centering
  \includegraphics[width=\columnwidth]{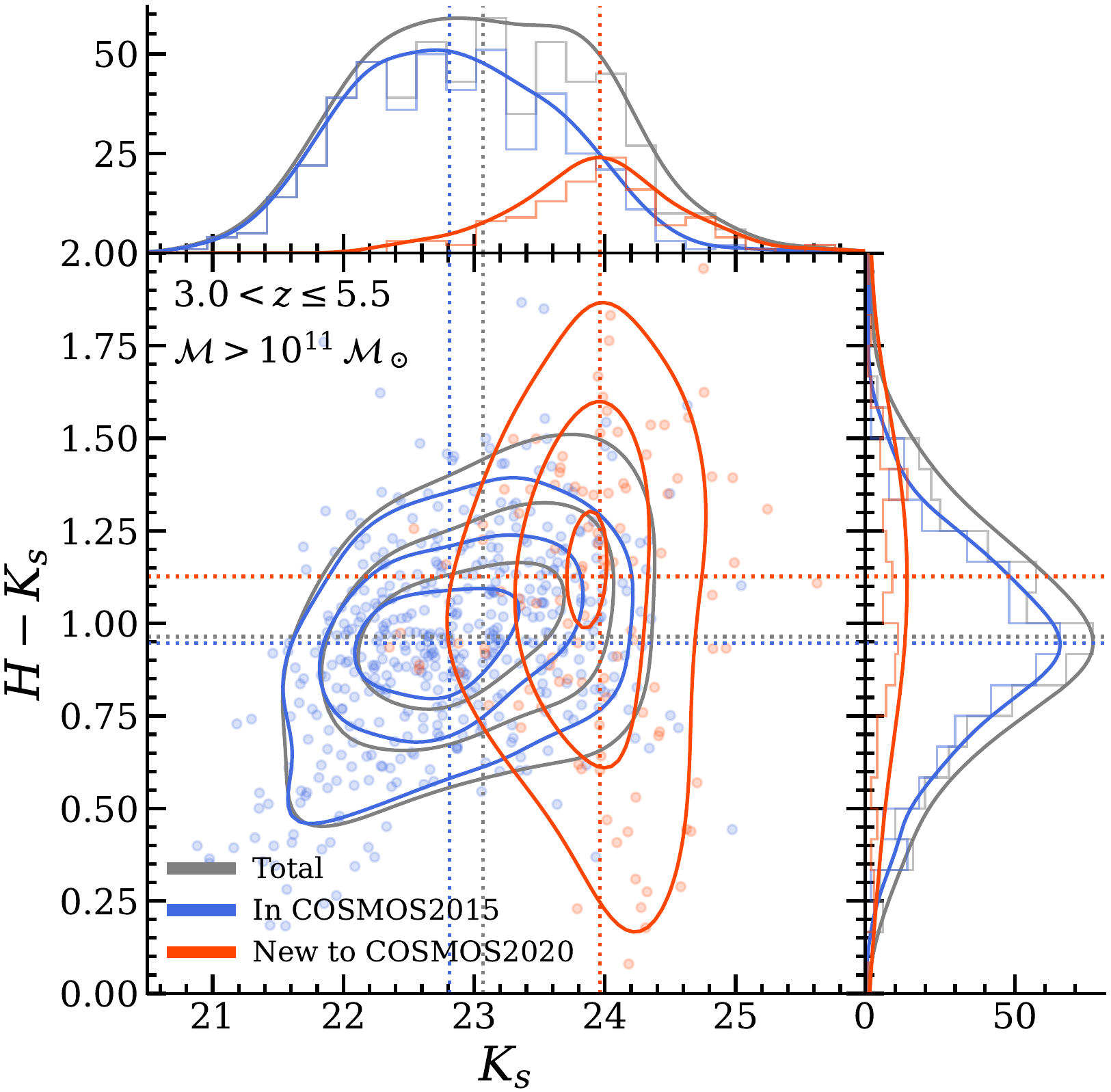}
  \caption{$K_s$ magnitude and $H-K_s$ colors of 125 new massive $\mathcal{M}>10^{11}\,\Msol{}$ galaxy candidates $3.0<z\leq5.5$ found in COSMOS2020 (orange), relative to the total sample used in this work (gray) and the 444 galaxies found in COSMOS2015 (blue). Measurements are taken from COSMOS2020, although they are similar to those from COSMOS2015, where matched. Number densities are summarised using gaussian kernel density estimators to produce smoothed distributions and contours for each sample. Median $K_s$ and $H-K_s$ values for each sample are shown by colored dotted lines.
  }
  \label{fig:newsources}
\end{figure}

Despite nearly identical sample selections, we still find greater number densities of massive galaxies compared to \citet{Davidzon17_mass} at $3<z\leq5.5$. By comparing to COSMOS2015 (matched within 0.6\arcsec), we find that 78\% of $\mathcal{M}>10^{11}\,\Msol{}$ galaxies in our sample are also found in COSMOS2015 where they are exclusively high-$z$ (91\% agree within $\Delta z\pm0.5$) and high-mass (78\% agree within $\Delta \mathcal{M}\pm0.5\,\Msol{}$). The remaining 23\% are new sources found only in COSMOS2020. As shown by Fig.~\ref{fig:newsources}, they are predominantly faint, having a median $K_s$ magnitude $\approx24.2\,{\rm AB}$ compared to the median sample brightness $K_s\approx23.4\,{\rm AB}$ with a photometric uncertainty $0.05-0.1\,{\rm AB}$ (see Fig.~10 of \citealt{Weaver2022_catalog}). They are generally as massive as the already known sources from COSMOS2015 and follow the same $\mathcal{M}$ distribution. They are also remarkably red, having a median $H-K_s$ color of $\approx1.2\,{\rm AB}$ compared to that of the sample overall ($\approx0.8\,\,{\rm AB}$). It seems unlikely that they are now found because of the better de-blending by \cfarmer{} as visual inspection shows that they are generally isolated sources and are detected by \texttt{Source Extractor} in the \textsc{Classic} version of the catalog. Therefore the most probable explanation for the new faint, red sources is that the deeper UltraVISTA $YJHK_s$ (as well as HSC $iz$) have enabled the detection of these faint red sources which in COSMOS2015 could not be detected. Interestingly, we find that their distribution in stellar mass is consistent with that of the total sample such that number densities at all masses $\mathcal{M}>10^{11}\,\Msol{}$ are proportionately represented.

Visual inspection of photometry and SED fits indicate that nearly all of the $3<z\leq5.5$ $\mathcal{M}>10^{11}\,\Msol{}$ galaxies have red colors. About 75\% of them are selected as star-forming by $NUVrJ$, $\gtrsim80\%$ of which are likely attenuated by a thick screen of dust ($A_V>1$; compared to only $\sim10\%$ across entire sample). The remaining $\sim25\%$ are classified as quiescent. The red colors appear to be genuine and not driven by blends, as confirmed by visually inspecting the imaging for each galaxy. Their broad-band photometry lacks the strong spectral features that contribute to a secure \photoz{}: there is no detectable flux contamination by nebular emission lines and both the Lyman and Balmer breaks are weak. However, they are surprisingly well fit by \lephare{} (typical $\chi^{2}_N\approx1.5$). Their likelihood redshift distributions $\mathcal{L}(z)$ are narrow as $>68\%$ of the probability is typically contained within $\Delta\,z\approx1$, with similarly well constrained $\mathcal{L}(\mathcal{M}\,|\,z)$. Recently, \citet{Lower2020} has shown how the relatively simple parametric star formation histories assumed by most template-based SED fitting codes are susceptible to biases on the order of 0.5\,dex in mass, which suggests that these uncertainties are likely underestimated \citep[see also][]{Michalowski2012, Michalowski2014}. 

One possible explanation for smooth SEDs with a power-law slope is contamination by AGN. The COSMOS2020 results computed by \lephare{} include classifications for sources with strong X-ray detections\footnote{Sources with x-ray counterparts (i.e., AGN) are classified in \citet{Weaver2022_catalog} by \texttt{lp\_type=2}.} determined from crossmatching to \citet{civano_chandra_2016} sources within 0.6\arcsec radius. In our sample we only use those sources identified as galaxies, which excludes these X-ray detections. Their inclusion would only serve to increase these already surprisingly large number densities. We do not attempt to quantify this, as the expected accretion disc light from Type 1 Seyferts and quasars make estimates of \photoz{} and $\mathcal{M}$ unreliable and susceptible to catastrophic failures. See \citet{Weaver2022_catalog} and \citet{Salvato2011} for details.

Identification of AGN (including X-ray faint AGN) from broadband colors has been explored in the literature \citep{Stern2005, Donley2012, Hviding2022}, and their discussion is a standard component of SMF studies at these redshifts \citep[see][]{Grazian2015, Davidzon17_mass}, although a consensus has yet to be reached. In general, AGN selection criteria rely on colors derived from (near-)infrared wavelengths from most notably \textit{Spitzer}/IRAC. While the \citeauthor{Donley2012} criteria have been used successfully at $z<3$, they require constraints in all four IRAC bands, which is not the case for COSMOS2020 where channel~3 and channel~4 lack sufficient depth to detect these sources. Even if deeper IRAC images could be taken, the selection criteria rely on the four bands sampling the continuum shape at restframe $2-10\,\mu$m but at $z>3$ instead sample the rest-frame stellar bulk at $\lesssim2\,\mu$m. While the MIPS $24\,\mu$m data from S-COSMOS \citep{Sanders_2007} is an attractive solution, it also is too shallow \citep[20.2\,AB at 3$\sigma$,][]{jin_super-deblended_2018} to fully constrain the restframe $2-10\,\mu$m continuum at $3<z\leq5$: only galaxies $H\lesssim20$ with large AGN fractions can be positively identified, and both low-fraction AGN at $H\approx20$ and normal galaxies at $H<20$ cannot be classified with MIPS. Full spectral fitting including far-infrared measurements is challenging without the constraints from channel~3, channel~4, and $24\,\mu$m, and attempts to gain further insight using \textsc{Stardust} \citep{Kokorev2021} were broadly unsuccessful with tentative contamination found to be on the order of $10-30$\%. We note, importantly, that removing potentially contaminated sources provides no significant change to the SMF at these epochs. 

Despite these challenges, we are able to leverage the elementary AGN template fitting from \lephare{} to statistically assess the spectral similarity between the best-fit AGN and galaxy templates for each source, limited to rest-frame UV/optical light. While only 10\% of galaxies in the total sample are best-fit with AGN templates, this fraction grows to 30\% for all mass complete galaxies $3.5<z<5.5$ and then to 50\% for those with $\mathcal{M}>10^{10.8}\,\Msol{}$ with broad wings, having $>80\%$ of these sources statistically indistinguishable as either galaxy or AGN ($|\Delta\chi^2|<0.5$). Turning to further infrared data of individual sources, we find 15\% of the sample are detected by VLA-COSMOS \citep{smolcic17_radio}. While not conclusive, we find 5\% of these massive $\mathcal{M}>10^{11}\,\Msol{}$ galaxies at $3<z\leq5$ are detected in the ALMA maps of A$^3$COSMOS \citep{Liu2019_A3COSMOS}, which currently covers $\sim$5\% of the COSMOS field, highlighting the need for further observations. While similar sample statistics for individual sources have been extrapolated in the literature \citep[e.g.,][]{Santini2021}, anticipated surveys such as the TolTEC Ultra-Deep Galaxy Survey \citep{Pope2019} will expand and deepen the far-infrared/(sub)mm coverage in COSMOS so that these samples can be more conclusively investigated.

A recent X-ray and radio stacking analysis of similarly selected $z>1.5$, $\mathcal{M}>10^{10}\,\Msol{}$ COSMOS2020 galaxies by \citet{Ito2022} (selected from independent SED fitting with \texttt{MIZUKI}, \citealt{Tanaka2015}) revealed that low-luminosity AGN are likely ubiquitous in massive quiescent galaxies from $1.5<z\le5$, even if individually they are not X-ray or radio detected. They also estimate the AGN contribution to optical/NIR continuum and find that at these redshifts, e.g., the rest-frame $B$-band luminosities of their quiescent galaxies are a factor of 30$\times$ larger than the expected rest-frame AGN luminosity. \citeauthor{Ito2022} also find that the AGN luminosities for quiescent galaxies are significantly larger than those of star-forming galaxies. Together, their findings provide further confidence that our redshifts and stellar mass estimates for our X-ray undetected sample are not strongly contaminated by AGN light. 

The lack of obvious systematics or likely only weak AGN contamination increases our confidence that these sources, or at least a part of them, are truly massive at $z\gtrsim3$. In agreement with \citet{Ito2022}, we also find that $>60\%$ of $\mathcal{M}>10^{11}\,\Msol{}$ galaxies appear to be star-forming, and it is likely that at least some of them are also dust obscured \citep[DSFGs;][and references therein]{Casey2014, Zavala2021}. Since there are a number of sources found in the deeper NIR images of COSMOS2020, it makes sense that we are now more sensitive to fainter red sources than ever before (see Fig.~\ref{fig:newsources}). It is precisely this class of galaxy which are efficiently captured by infrared facilities such as \textit{Spitzer}, ALMA, and \textit{JWST} until now being optically ``dark'' \citep[e.g.,][]{Schreiber2018, Wang2019,Gruppioni2020,Sun2021,Fudamoto2021,Shu2022}. If genuine, their existence not only points at an incompleteness particularly in the massive end of SMFs reported in previous studies lacking the necessarily deep infrared data over degree scales, but also highlights the sensitive interplay between the shape of their SEDs and the selection function consisting of the bands, their depths, and the detection methodology (see Fig.~3 of \citealt{Weaver2022_catalog}). Although detailed simulations will be explored in future work, qualitatively this could explain why the excess of sources relative to a Schechter at $z\approx3-5$ diminishes at $z\approx6$ as similarly red sources become too faint to still be detected. Optically dark galaxies selected from 2\,mm ALMA observations in COSMOS from \citet{Casey2021} and \citet{Manning2022} are constrained to similar redshifts, stellar masses, and number densities. Importantly, \citeauthor{Manning2022} studied two systems with star formation rates above $200\,\Msol{}\,{\rm yr}^{-1}$ but a gas depletion timescale $<1\,{\rm Gyr}$, suggesting rapid star formation cessation and a potential transition to massive quiescent galaxies by $z\sim3-4$. \citet{Casey2021} find that DSFGs are responsible for $\sim30$\% of the integrated star formation rate density at $3<z<6$ and that 2\,mm selection is an efficient way to identify larger samples in future surveys (see also \citealt{Cooper2022}). In an empirically based numerical model, \citet{Long2022} uses known DSFG population properties to predict the DSFG stellar mass function out to $z\sim5$. They find a shallow power law extension beyond the knee of traditional star-forming Schechter SMFs, and posit that 60-80\% of massive ($\mathcal{M}\,>10^{11}$\,$\mathcal{M}_\odot$) star-forming galaxies at z\,$\approx3-6$ are significantly dust-obscured and not captured in previous SMF studies \citep[see also][]{Martis2016}. This is in line with our aforementioned results on significantly attenuated \textit{NUVrJ}-selected massive star-forming galaxies at similar redshifts.

\begin{figure}
  \centering
  \includegraphics[width=\columnwidth]{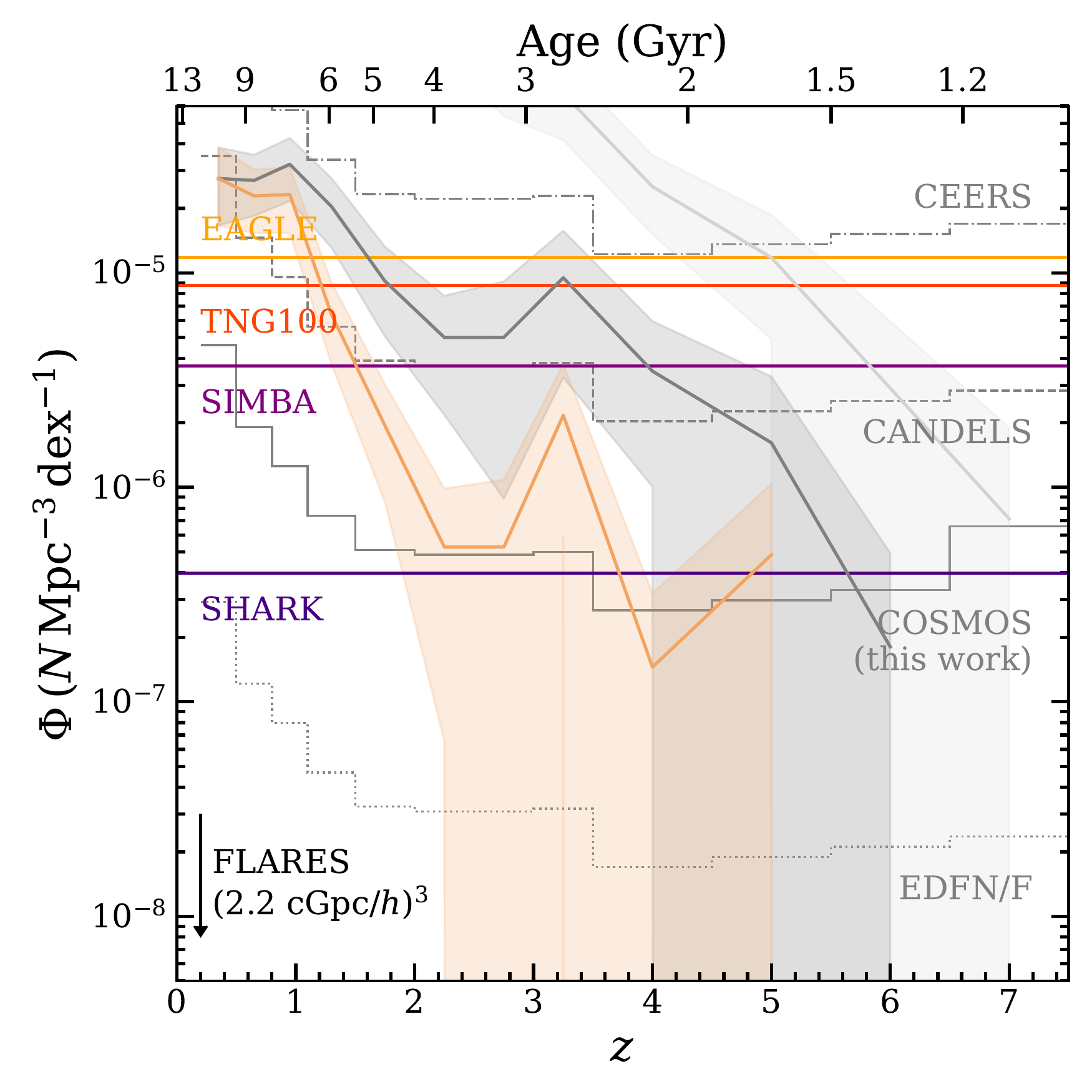}
  \caption{Comparison of observed number densities and upper limits on the rarity probed by various observed (gray steps) and simulated (colored lines) volumes. Number densities correspond to $10^{11.0}<\mathcal{M}<10^{11.5}\,\Msol{}$ and $10^{11.5}\mathcal{M}<10^{12.0}\,\Msol{}$ from the total sample (light and dark gray, respectively), and $10^{11.5}\mathcal{M}<10^{12.0}\,\Msol{}$ from the quiescent sample (orange) from Fig.~\ref{fig:phivz}. $1\,\sigma$ upper limits are computed following \citet{Gehrels1986}, which for the observed volumes are dependent on widths of redshift bins.
  }
  \label{fig:volumelimit}
\end{figure}

As introduced in Section~\ref{subsec:compsims} and shown in Fig.~\ref{fig:comp_sims}, there is generally good agreement between number densities of massive galaxies found in several simulations and in this work. This may be surprising given the range of physical recipes utilized in these simulations, and may suggest that several different tuning of physical recipes (mainly those of radiative and mechanical AGN feedback) can reproduce observations. However, from $2.5<z\leq5.5$, we observe galaxies with mass $1.8\times$ larger than the most massive galaxies in any of these simulations. While bias and underestimated mass uncertainties may contribute, the simulations are usually considered to be limited by their volume. Fig.~\ref{fig:volumelimit} compares the observed number densities of massive galaxies (including quiescent) to the upper limits set by volumes of different surveys as well as simulations. The fully hydrodynamical codes (\textsc{Eagle}, TNG100, and \textsc{Simba}) have the smallest volumes comparable to CANDELS and are similar to the observed number densities of $\mathcal{M}>10^{11}\,\Msol{}$ galaxies. However, the fact that they contain a large enough volume to catch the most massive galaxy seen in our observations may suggest that their volumes are adequate, but that their DM halo growth physics are not providing the large halos from which they can grow. Alternatively, recent observations of a highly star-forming galaxy at $z=6.9$ \citep[SFR$\approx2900\,\Msol{}\,{\rm yr}^{-1}$,][]{Marrone2018} place close to a maximally massive DM halo not seen in current simulations -- populations expected to evolve into the most massive systems at $z\approx2-4$ \citep{Lower2022}. This is especially true of \textsc{Shark}, whose volume is similar to that of COSMOS itself. More massive galaxies, if they exist, are not likely to be found in COSMOS, and so larger volumes (such as that probed by the two Euclid Deep Fields North and Fornax, as well as \textit{Roman}) will be required, see McPartland et al. in preparation).

Naturally, massive galaxies in the $z>3$ Universe are of great interest as targets for \textit{JWST}. The widest deep-field ERS program is the 100\,arcmin\textsuperscript{2} Cosmic Evolution Early Release Science Survey \citep[CEERS;][]{Finkelstein_CEERS}, and based on our estimates it stands to find approximately 22, 16, 5 for $\mathcal{M}>10^{10.5}\,\Msol{}$, and 6, 4, 2 for $>10^{11.0}\,\Msol{}$) galaxies at $3<z\leq3.5$, $3.5<z\leq4.5$, and $4.5<z\leq5.5$, respectively, although the small area will equate to a stronger density bias from cosmic variance. The widest galaxy survey field of any GO program is the 0.6\,deg\textsuperscript{2} COSMOS-Web \citep{Kartaltepe_COSMOSWeb}, which is expected to find 493 (137), 340, (94), 103 (40) galaxies for the same redshift ranges (and masses). Although \textit{JWST} will be instrumental in studying these sources in exquisite spatial resolution and with efficient spectroscopy \citep{2021jwst.prop.2198B, 2021jwst.prop.2285C, 2021jwst.prop.2565G}, ground-based NIR observations that can efficiently identify them in wide-area surveys will retain their importance. 

\subsection{Rise of quiescent galaxies}
\label{subsec:rise_qg}

As shown by Fig.~\ref{fig:comp_lit_sfqg}, low-$z$ quiescent galaxies are well described by a two component Schechter function whose low-$\mathcal{M}$ component diminishes rapidly from $z\approx0.2\rightarrow1$ (see Fig.~\ref{fig:param_evol}). Simultaneously our sample becomes less mass complete with redshift, doing so more rapidly for quiescent systems due to their red color characteristic of their high mass-to-light ratios. Despite the considerable uncertainties on the completeness of our low-$\mathcal{M}$ quiescent sample outlined in Section~\ref{subsec:completeness}, it seems likely that the shape of the quiescent stellar mass function in this work and in previous literature in fact does turn down at low-masses, with selection effects playing a comparably minor role (see also \citealt{Ilbert_2010}). However, there are hints of $\mathcal{M}\lesssim10^{9}\,\Msol{}$ quiescent systems at $z\gtrsim1.5$ (beyond our mass limit) reported by \citet{Santini2022}, albeit tentative from a $\sim130\times$ smaller effective area from the 33.4~arcmin\textsuperscript{2} Hubble Frontier Field parallels. 

Given their low apparent brightness and rarity, the present work is the first to consistently quantify the number densities of $\mathcal{M}\approx10^{9.5}\,\Msol{}$ low-mass quiescent systems at $1.5<z\leq4.5$ at high signal-to-noise, based on the deepest NIR data taken over such a homogeneously measured degree-scale area required to find them in sufficient numbers. As seen in Fig.~\ref{fig:phivz}, the rate of growth in the number density of low-mass ($\mathcal{M}\lesssim10^{10}\,\Msol{}$) quiescent galaxies has been seemingly rapid over the past $\sim10$ billion years. Still, they constitute only a minor fraction of low-mass sources by $z\sim0.2$, and by extrapolating the number densities from $z\approx1\,-\,2$ one may expect to find none within COSMOS by $z\sim3$ (see Fig.~\ref{fig:phivz}). This is typically interpreted by the phenomenological model of \citet{peng10_quenching} to mean that the processes which act to halt star formation cessation in low-$\mathcal{M}$ systems are inefficient at early times. For example the lack of virialized at $z>2\!-\!3$ structure makes influence from environmental effects unlikely. 

As shown by Fig.~\ref{fig:phivz}, the apparent lack of growth in the abundance of massive systems at $z<1$ is the result of a decline in star-forming galaxies simultaneous with an increase in quiescent number densities. As noted by \citet{ilbert_mass_2013}, this decrease in the star-forming population is consistent with star formation cessation becoming extremely efficient, to the extent that massive star-forming galaxies are becoming quiescent faster than they can be replaced. Therefore, the mass assembly of massive star-forming systems at $z<1$ is slower than the cessation of their star formation. However, there is also a slowing down in the rate of growth in the number density of massive quiescent systems themselves. While this may suggest high incidences of dry mergers or even rejuvination, it must also relate to evolving demographics: from $z\approx1.5\rightarrow0.3$ there are fewer and fewer high-$\mathcal{M}$ star-forming galaxies \textit{available} to become quiescent. While the number density growth of massive systems seems to have stalled out, that of lower-mass systems continue to grow; this is the so-called phenomenon of ``downsizing with time'' \citep{Cowie1996, Neistein2006, Fontanot2009}.

The quiescent mass function at $\mathcal{M}>10^{11}\Msol{}$ does not change much from $z\approx4.5\rightarrow2.0$, with surprisingly elevated quiescent fractions (Fig.~\ref{fig:qg_frac_evol}) being above 20\% at $z<5$. Fig.~\ref{fig:phivz} shows why: while the number density of star-forming galaxies at $z\approx5$ is lower than at $z\approx2$, quiescent galaxies are roughly constant in density over this same range. However, we caution that their number densities are only marginally above what should be possible to find within the nominal volume of COSMOS at these redshifts, and so it is plausible that they are dominated by noise and/or \photoz{} bias. The question of whether or not this behavior is genuine can only be explored in future large volume surveys, as demonstrated by Fig.~\ref{fig:volumelimit}. Even though the most massive galaxies at $z\gtrsim3-4$ are typically too faint to obtain continuum features, spectroscopic follow-up and supporting SED fitting will continue to provide valuable insights \citep{Gobat2012, Schreiber18_QGz4, Valentino20_QGz4, Glazebrook2017}.

\begin{figure}
  \centering
  \resizebox{\hsize}{!}{\includegraphics{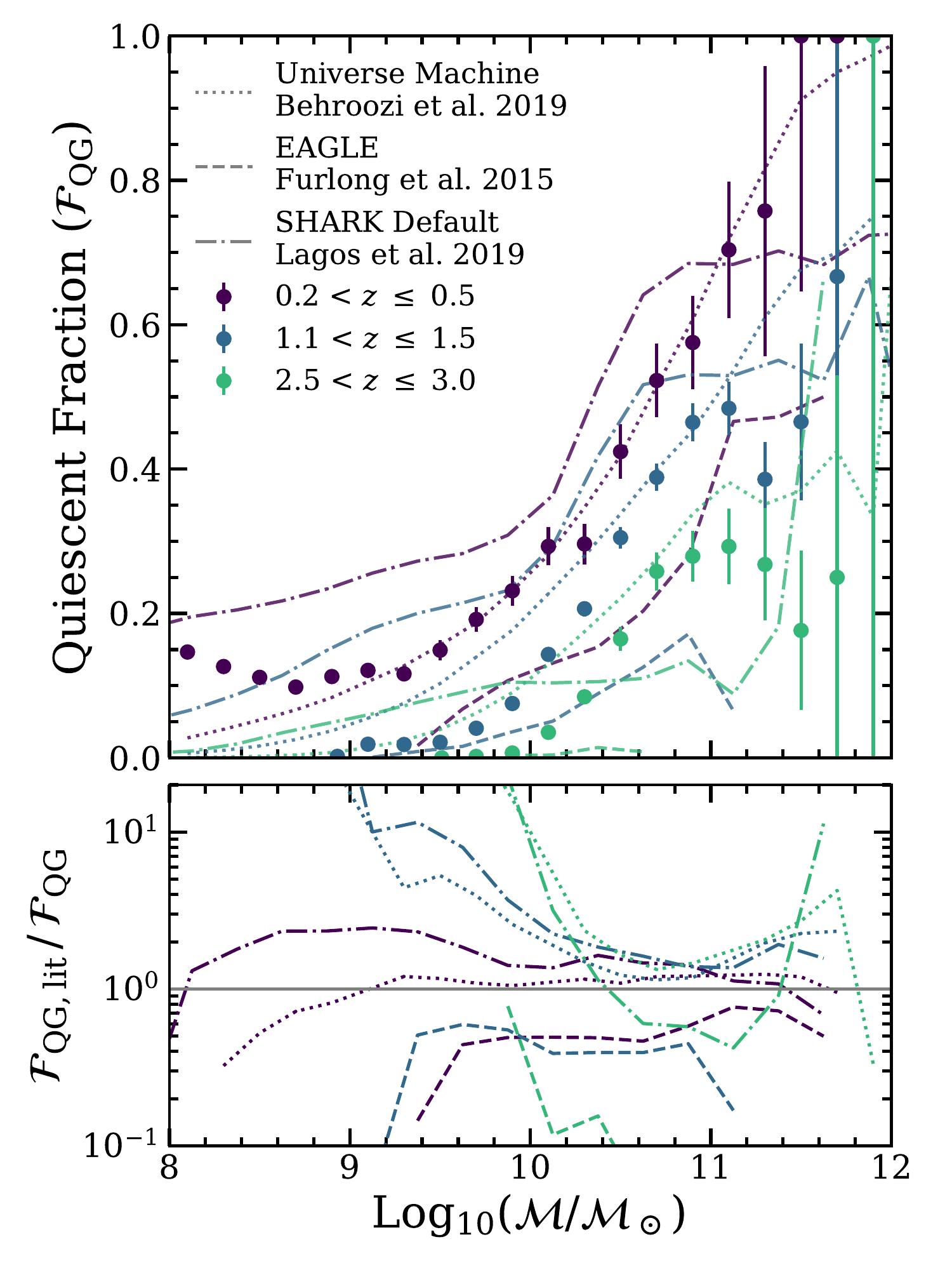}}
  \caption{Quiescent mass fractions. \textit{Upper:} Fraction of quiescent galaxies as a function of mass for three redshift ranges compared with predictions from \textsc{Eagle} \citep{Furlong2015_EAGLE} and  \textsc{Shark} \citep{Lagos2018_SHARK} as well as measurements from the empirically calibrated \textsc{Universe Machine} \citep{Behroozi2019}. \textit{Lower}: Fractional difference between this work and literature predictions/measurements. See Appendix~\ref{app:qg_fraction} for more details.}
  \label{fig:qg_frac}
\end{figure}

Fig.~\ref{fig:qg_frac} shows the fraction of quiescent galaxies in bins of mass for three epochs: $z\approx0.3$, 1.3, and 2.7. A key advantage of examining fractions is that they are less sensitive to the overall normalization of the simulation and biases from observations (e.g., in simulations: overproduction of all galaxy masses; in observations: systematics in effective survey volume), and provide additional insight which is obscured by simply comparing mass functions. Although comparisons to \textsc{Universe Machine} have been already discussed in Section~\ref{subsec:compsims}, we introduce two samples of quiescent galaxies which we selected from  \textsc{Eagle} and  \textsc{Shark} with an $NUVrJ$ selection consistent with our methodology. While  \textsc{Eagle} underproduces quiescent systems by $10-20$\% at all masses, \textsc{Shark} overproduces them in all but the most massive bins at $z\approx0.3-1.3$ but underproduces them at $z\approx2.7$. Roughly, the fraction of quiescent galaxies in \textsc{Shark} at $z\approx1.3$ matches the observed fractions at $z\approx0.3$. This may suggest that \textsc{Shark}'s physical recipes that halt star formation in lower mass galaxies are too aggressive. These same systems are also seemingly overproduced (see Fig.~\ref{fig:comp_sims}), and so may be assembling \textit{and} maturing too early. For additional figures and details, see Appendix~\ref{app:qg_fraction}.

\subsection{Dark matter halo connection}

The mass assembly of a galaxy is inherently connected to the dark matter halo in which it formed and grew (see \citealt{Wechsler2018_ARAA} for a review). Yet, stellar masses $\mathcal{M}$ have been observed to be $<2\%$ of their halo mass $\mathcal{M}_{\rm h}$, which point to galaxy formation as a strikingly inefficient process \citep{Mandelbaum2006, conroy_connecting_2009, behroozi_comprehensive_2010}. This has led to investigations into the role of dark matter halo mass in influencing star formation cessation, including promoting thermal heating and/or gas expulsion by AGN \citep{Shankar2006, Main2017}, as well as virial shock heating of in-falling cold gas whose Jeans mass inhibits the formation of star-forming molecular clouds \citep[so-called hot-halo mode,][]{Birnboim2003}. Hence, the gravitational influence of the halo mass on the cold gas reservoir regulates the ability of a galaxy to form stars, and hence the stellar-to-halo mass relationship (SHMR). Therefore it is no surprise that there is a similar evolution between the specific mass increase rate of the halo by accretion (sMIR$\equiv\mathcal{M}_{\rm h}^{-1}\partial\mathcal{M}_{\rm h}/\partial t$, \citealt{Neistein2008, Saintonge2013}) and the instantaneous mass growth by star formation (sSFR) \citep[see discussions in][]{Lilly2013}. While in this work we restrict ourselves to comparisons with theoretical HMFs, another paper in this series computes a self consistent SHMR based on measuring the halo occupation distribution directly from angular correlation functions and SMFs of COSMOS2020 galaxies \citep{Shuntov2022}. For an investigation into the SHMR split by star-forming and quiescent samples, see \citet{Cowley2019}, which is also based on galaxies from the COSMOS field.

\begin{figure*}
  \centering
  \resizebox{\hsize}{!}{\includegraphics{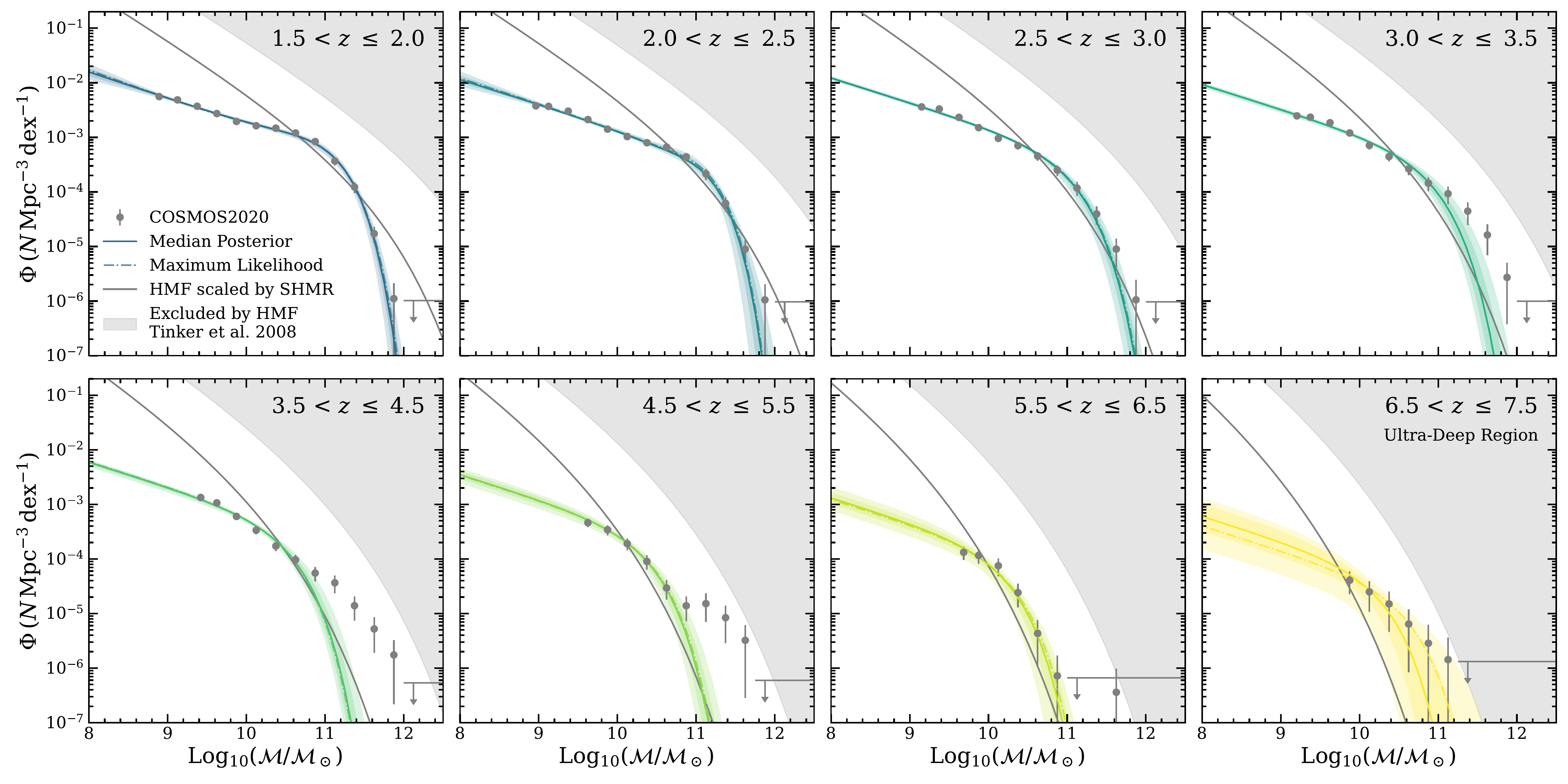}}
  \caption{Comparison of the inferred $z>1.5$ galaxy stellar mass function (colored curve with 1 and 2\,$\sigma$ envelopes) fitted to observed measurements (gray points). We scale the \citet{Tinker2008} halo mass function (HMF; lower bound of the gray shaded region) by 0.018 corresponding to the stellar-to-halo mass relation at $z=0$ and $\mathcal{M}_{\rm h} = \mathcal{M}_{\rm h}^*$ \citep{Behroozi2013} to produce an idealized SMF (gray curve). Upper limits on the SMF at each redshift are derived from the HMF assuming a fixed baryon fraction ($0.166$). Upper limits for empty bins are shown by the horizontal gray line with an arrow. Mass incomplete measurements are not shown.}
  \label{fig:hmf}
\end{figure*}

As shown in Fig.~\ref{fig:hmf}, we compare our observed and inferred SMFs to the halo mass function (HMF) of \citet{Tinker2008}\footnote{The \citet{Tinker2008} HMF is computed according to our cosmology ($\sigma_8=0.82$) at the mid-point in each $z$-bin using \textsc{Colussus} \citep{Diemer2018_COLOSSUS}, which explicitly takes into account that these mass functions derived were originally derived from spherical overdensities, which is not universal with redshift (see Equations 3-8 of \citeauthor{Tinker2008}). We adopt the definition for a halo as the DM mass contained in a region $200\times$ the mean matter density, and found no significant differences from using other definitions (e.g., friends-of-friends, spherical overdensity, virial radius).} from $z=1.5\rightarrow7.5$. We choose not to show $z<1.5$ as these comparisons have been thoroughly explored by previous investigations \citep[e.g.,][]{Davidzon17_mass, legrand19_hod} and we do not observe any significant differences. The effects of feedback can be seen in the first panel of Fig.~\ref{fig:hmf} at $1.5<z\leq2.0$ that explains the relatively lower number densities of both low- and high-$\mathcal{M}$ systems, with those around $\mathcal{M}^*$ being most similar to HMF. This apparent tension is a well known feature and lies at the foundation of the contemporary galaxy evolution paradigm, involving halting star formation by secular (internal) and/or environmental (external) action on the gas reservoir such as thermal heating, dynamical turbulence, and/or removal. The growth of low-mass $\mathcal{M}<\mathcal{M}^*$ galaxies can be impeded by secular (e.g., supernovae and stellar winds) and environmental processes (e.g., ram-pressure stripping, thermal evaporation). Similarly, secular and environmental processes can also impede the growth of massive $\mathcal{M}>\mathcal{M}^*$ galaxies, albeit from different driving forces such as AGN and major mergers, respectively \citep[see][]{peng10_quenching, Peng2012, Peng2015, Wechsler2018_ARAA, Schreiber2020_ARAA}. In this context, the characteristic knee at $\mathcal{M}^*$ is the result of a build-up of massive galaxies which can no longer sustain mass growth. At $z\approx3$, the low-mass end is still considerably lower than the scaled HMF but the number density of massive systems comes into agreement. Although the stellar mass function slightly lies above the SHMR-scaled HMF at some points, we caution that this should not be taken as a challenge to theory as it assumes that the SHMR at $z=0$ is appropriate at higher-$z$ \citep[which is unlikely; see][]{legrand19_hod} and small modifications can reconcile this difference. We note that Fig.~18 of \citeauthor{Davidzon17_mass} finds no such offset using the same scaling, but we are unable to reproduce their precise result.

We derive an upper limit for the baryonic matter distribution by rescaling the HMF by $\Omega_{\rm b}/\Omega_{\rm m}=0.166$, which for our adopted cosmology is redshift independent, and assume a 100\% efficient baryon-to-stellar mass conversion. This is the maximum SMF physically allowed under our simple assumptions. This upper limit becomes relevant especially at $z>3.5$ where our observed number densities exceed those inferred by the Schechter model. While a large Eddington bias or selection systematic could explain this excess (see Section~\ref{subsec:massivegal}), we stress that our inferred SMF assumes the applicability of Schechter formalism, which cannot accurately describe the observed number densities at $3.5<z\leq5.5$. Nevertheless, the inferred stellar mass function agrees well with the SHMR-scaled HMF up to $z\approx7$. This suggests that the most efficient haloes during these early epochs are not around $\mathcal{M}^*$, but rather the most massive ones, and with little observed evolution consistent with the findings from \citet{Stefanon2021}. This phenomenon has also been observed in some local massive spiral galaxies \citep[e.g.,][]{DiTeodoro2022}, which have seemingly matured without significant star formation cessation. One interpretation is that this high star formation efficiency in massive haloes is the result of diminished feedback from AGN, with stellar mass growing similarly to the host halo at early times. Indeed, this is consistent with findings of inefficient radiative AGN feedback from simulations \citep{Kaviraj2017, laigle19_horizonAGN, Roos2015, Bieri2017, Habouzit2022}, as well as FIR/radio observations of AGN activity at $z>3$ \citep{Maiolino2012, Cicone2014, Cicone2015, Padovani2015, Vito2018}. 

At no point do our mass functions, observed or inferred, exceed this upper limit. Therefore we do not report evidence of ``impossibly early galaxies'' introduced by \citet{Steinhardt2016} who point out that there appears to be too many massive galaxies at $z>4$ compared to the dark matter haloes that should host them. However at $z>6.5$, where \citeauthor{Steinhardt2016} predicts that the effect will be most obvious, we report observed number densities approaching this upper limit and in clear excess of the SHMR-scaled HMF. We caution that these sources are the most vulnerable to misclassification and bias, being susceptible to blending in addition to being constrained by only a handful of NIR bands yielding proportionately uncertain Schechter fits. Extrapolation to $\mathcal{M}\gtrsim10^{11.5}\,\mathcal{M}_\odot$ would place their number density below that which can be probed in a volume contained by the $0.716\deg^2$ area of the Ultra-Deep region, and so COSMOS2020 is unlikely to find them if they exist. While they are not ``impossibly early galaxies'', their surprising abundance hint that explorations of $z>7$ with future deep, large-volume surveys may provide the evidence necessary to firmly challenge theoretical frameworks.

\section{Summary \& conclusions}
\label{sec:conclusions}

Following on from COSMOS2020 \photoz{} catalog \citep{Weaver2022_catalog}, we study the shape and evolution of the galaxy stellar mass function. Our measurements span three decades in mass (at $z\approx0.2$) across 10 billion years of cosmic history ($z=7.5\rightarrow0.2$), including the most mass complete sample of quiescent galaxies at $z>2$ enabled by our unprecedented, homogeneous NIR depths across an effective 1.27\,deg\textsuperscript{2}. We probe a volume nearly $2\times$ that of \citet{Davidzon17_mass} which not only improves sample statistics but also finds new galaxies of still greater mass at all redshifts. Complementary deep IRAC coverage has allowed us to directly measure the stellar bulk, and hence galaxy stellar mass, in a less biased way and to higher redshifts compared to $K_s$-based measurements. We developed a robust, mass-dependent error budget with contributions from poisson, stellar mass, and cosmic variance, and account for the effects of Eddington bias by fitting a kernel-convolved Schechter function to our observed SMF. We use three fitting techniques, including the continuity model of \citet{Leja2019}, finding good agreement with literature measurements with smaller bin-to-bin variance with $z$. We stress that our derived parameters are dependent on the assumed Eddington correction, and while the inferred SMF evaluated at maximum likelihood and associated parameters are robust, parameters (and their uncertainties) derived from the median of posterior distributions can be unreliable if constraints are weak. To make these results more transparent and to facilitate comparisons, we have made the object IDs and key measurements presented in this work available to download\footnote{ \url{https://doi.org/10.5281/zenodo.7808832}}.

Although literature comparisons on the shape of the SMF at fixed redshift show good agreement, the novel advantage of COSMOS2020 is the extended historical baseline over which the mass functions (as well as many other properties) can be consistently measured. Not only do we examine the evolution of the integrated mass density $\rho_*$ over this time, we also examine the remarkably consistent rate of growth in the number densities of systems of different masses from $z\approx7\rightarrow1$, whereupon the most massive star-forming galaxies become quiescent faster than they can be replaced. Similarly, we find evidence for the sharp rise in low-mass quiescent systems consistent with the phenomenological model of \citet{peng10_quenching} probed to $z\approx2.5$ where our sample becomes incomplete. Although tentative, we also find evidence for a sustained $>20\%$ fraction of high-mass quiescent systems from $z\approx5\rightarrow2$.

Furthermore, we highlight three main results:

\begin{itemize}
\item[--] Comparisons with several hydrodynamical simulations and semi-analytical models indicate an exceptional degree of agreement for the most massive galaxies out to high-$z$. This comes despite the surprisingly high number densities of massive galaxies at $z\approx3-5$ in excess of a Schechter function, suggesting that existing physical recipes are assembling massive $\mathcal{M}\approx10^{10-11.5}\,\Msol{}$ systems in sufficient quantity at early times. In order to explore star formation cessation and feedback modes, we identify quiescent galaxies out to $z\approx5.5$ by means of a $NUVrJ$ selection and compare them to consistently selected quiescent samples produced by two simulation codes, finding evidence for delayed assembly of low-mass quiescent systems in  \textsc{Eagle}, and too rapid assembly in the \textsc{Shark}.  
    
\item[--]  A closer examination of these massive systems reveals that a quarter are not found in COSMOS2015. Not only are they $K_s$-faint, but their extremely red colors challenge SED fitting templates. We find no strong evidence for AGN contamination, although we stress the need for future infrared facilities with deep surveys capable of measuring the rest-frame MIR light at $z\approx3-5$. Recent findings of optically dark galaxies from IRAC and ALMA suggest that previous studies have missed contributions from dust-obscured star-forming galaxies. Their brightness, redshifts, mass, and number densities are consistent with our findings, suggesting that the $K_s\sim26$ depth of UltraVISTA DR4 may indeed be sufficient to reach out into the tail end of this population missed by previous optical-NIR selections. Further work is required to conclusively establish the nature of these massive galaxy candidates.
    
\item[--]  Lastly, we investigate the connection to dark matter halos by comparing both our observed and inferred SMFs to constraints provided by the HMF. While we confirm the divegence of the SMF from the HMF at both low- and high-mass regimes which has been historically intepreted as evidence for feedback processes, the massive end of the inferred, Schechter-fit SMF comes into agreement with the HMF at $z\gtrsim2$. While we find no evidence of tension which would challenge theoretical models, our observed number densities at $z\approx3-5$ approach the upper limit for fully efficient star formation in the most massive halos. Larger volume surveys containing even more massive systems, if they exist, may be able to challenge these models, especially at $z\gtrsim6-7$.
\end{itemize}

The launch of \textit{JWST} has opened the door on a new era, and it will soon be flanked by efficient survey facilities from space ($Euclid$, $Roman$) and the ground ($Rubin$). While massive quiescent systems may exist at $z\sim5$ and perhaps even at earlier times \citep{Mawatari2016, Mawatari2020}, their identification is beyond the reach of COSMOS2020. They may be identified soon by deep degree-scale \textit{JWST} surveys i.e., COSMOS-Web \citep{Kartaltepe_COSMOSWeb, Casey2022} and possibly by narrower ones including
the \textit{JWST} Advanced Deep Extragalactic Survey \citep[JADES,][]{Eisenstein_JADES, Ferruit_JADES}, 
the Cosmic Evolution Early Release Science Survey \citep[CEERS,][]{Finkelstein_CEERS},
the Next Generation Deep Extragalactic Exploratory Public Survey \citep[NGDEEP,][]{Finkelstein_WDEEP},
and Ultra-deep NIRCam and NIRSpec Observations Before the Epoch of Reionization \citep[UNCOVER,][]{Labbe_UNCOVER, Bezanson2022} to name a few.

While low mass quiescent systems at $z\gtrsim2.5$ may be found by \textit{JWST} (with hints emerging from \citealt{Marchesini2022}), a truly precise quantification of their number density and demographics is likely to remain a future objective. By extrapolating these observations, $<1$\% of $\mathcal{M}\approx10^{9.5-10.0}\,\Msol{}$ are expected to be quiescent by $z\sim2.5$, become even rarer at earlier times. While identifying even one quiescent, low-mass system at $z>2$ in the absence of virialized structure would present a significant challenge to the paradigm of \citet{peng10_quenching}, performing a statistically meaningful survey of them will require incredibly deep, degree-scale NIR surveys, placing them out of reach by current facilities. While the deepest degree-scale surveys from \textit{JWST} (COSMOS-Web), \textit{Roman}, and \textit{Euclid} stand to establish the rarity of these systems at $z\approx2-3$, it seems that no currently planned survey will be able to definitively quantify their contribution at $z\gtrsim3$.

While explorations of mass-selected samples at $z>7$ are being be made possible by \textit{JWST}, the most UV-luminous sources at these epochs will likely be missed by small area programs and yet are crucial to a comprehensive study of the ionizing UV budget \citep{Kauffmann2022, Donnan2022}. Following up known samples with \textit{JWST} will not only allow us to measure the star formation rate and dust content of the first ultra-luminous galaxies from deep within the epoch of reionization, but also to directly identify the progenitors of $z\sim3-4$ massive galaxies while still in their formation stage \citep[e.g.,][]{Weaver2021_jwst}. Yet despite the incredible promises of highly resolved IR imaging and spectroscopy from \textit{JWST}, identifying the rarest and potentially most informative populations that can challenge and thereby improve galaxy formation models will remain the domain of wide-field surveys. 

\begin{acknowledgements}
The authors thank 
Vadim Ruskov,
Sidney Lower,
Desika Narayanan,
Stephen Wilkins,
William Roper,
Lukas Furtak,
and
Pratika Dayal
for helpful discussions. 
We are also grateful for the many helpful and constructive comments from the anonymous referee.

The Cosmic Dawn Center (DAWN) is funded by the Danish National Research Foundation under grant No. 140. ST and JW acknowledge support from the European Research Council (ERC) Consolidator Grant funding scheme (project ConTExt, grant No. 648179). OI acknowledges the funding of the French Agence Nationale de la Recherche for the project iMAGE (grant ANR-22-CE31-0007). HJMcC acknowledges support from the PNCG. ID has received funding from the European Union’s Horizon 2020 research and innovation program under the Marie Sk\l{}odowska-Curie grant agreement No. 896225.
This work used the CANDIDE computer system at the IAP supported by grants from the PNCG, CNES, and the DIM-ACAV and maintained by S. Rouberol. BMJ is supported in part by Independent Research Fund Denmark grant DFF - 7014-00017. GEM acknowledges the Villum Fonden research grant 13160 “Gas to stars, stars to dust: tracing star formation across cosmic time”. DR acknowledges support from the National Science Foundation under grant numbers AST-1614213 and AST-1910107. YP acknowledges National Science Foundation of China (NSFC) Grant No. 12125301, 12192220, 12192222, and the science research grants from the China Manned Space Project with NO. CMS-CSST-2021-A07.

The authors wish to recognize and acknowledge the very significant cultural role and reverence that the summit of Mauna Kea has always had within the indigenous Hawaiian community.  We are most fortunate to have the opportunity to conduct observations from this mountain.
This work is based on data products from observations made with ESO Telescopes at the La Silla Paranal Observatory under ESO program ID 179.A-2005 and on data products produced by CALET and the Cambridge Astronomy Survey Unit on behalf of the UltraVISTA consortium. This work is based in part on observations made with the NASA/ESA \textit{Hubble} Space Telescope, obtained from the Data Archive at the Space Telescope Science Institute, which is operated by the Association of Universities for Research in Astronomy, Inc., under NASA contract NAS 5-26555. Some of the data presented herein were obtained at the W.M. Keck Observatory, which is operated as a scientific partnership among the California Institute of Technology, the University of California and the National Aeronautics and Space Administration. The Observatory was made possible by the generous financial support of the W.M. Keck Foundation.

This research is also partly supported by the Centre National d'Etudes Spatiales (CNES). These data were obtained and processed as part of the CFHT Large Area U-band Deep Survey (CLAUDS), which is a collaboration between astronomers from Canada, France, and China described in Sawicki et al. (2019, [MNRAS 489, 5202]).  CLAUDS is based on observations obtained with MegaPrime/ MegaCam, a joint project of CFHT and CEA/DAPNIA, at the CFHT which is operated by the National Research Council (NRC) of Canada, the Institut National des Science de l’Univers of the Centre National de la Recherche Scientifique (CNRS) of France, and the University of Hawaii. CLAUDS uses data obtained in part through the Telescope Access Program (TAP), which has been funded by the National Astronomical Observatories, Chinese Academy of Sciences, and the Special Fund for Astronomy from the Ministry of Finance of China. CLAUDS uses data products from TERAPIX and the Canadian Astronomy Data Centre (CADC) and was carried out using resources from Compute Canada and Canadian Advanced Network For Astrophysical Research (CANFAR).

\end{acknowledgements}


\bibliographystyle{aa}
\bibliography{GSMF_2022.bib}

\begin{appendix}

\section{Sample selection}
\label{app:cuts}

\begin{figure}
  \centering
  \resizebox{\hsize}{!}{\includegraphics{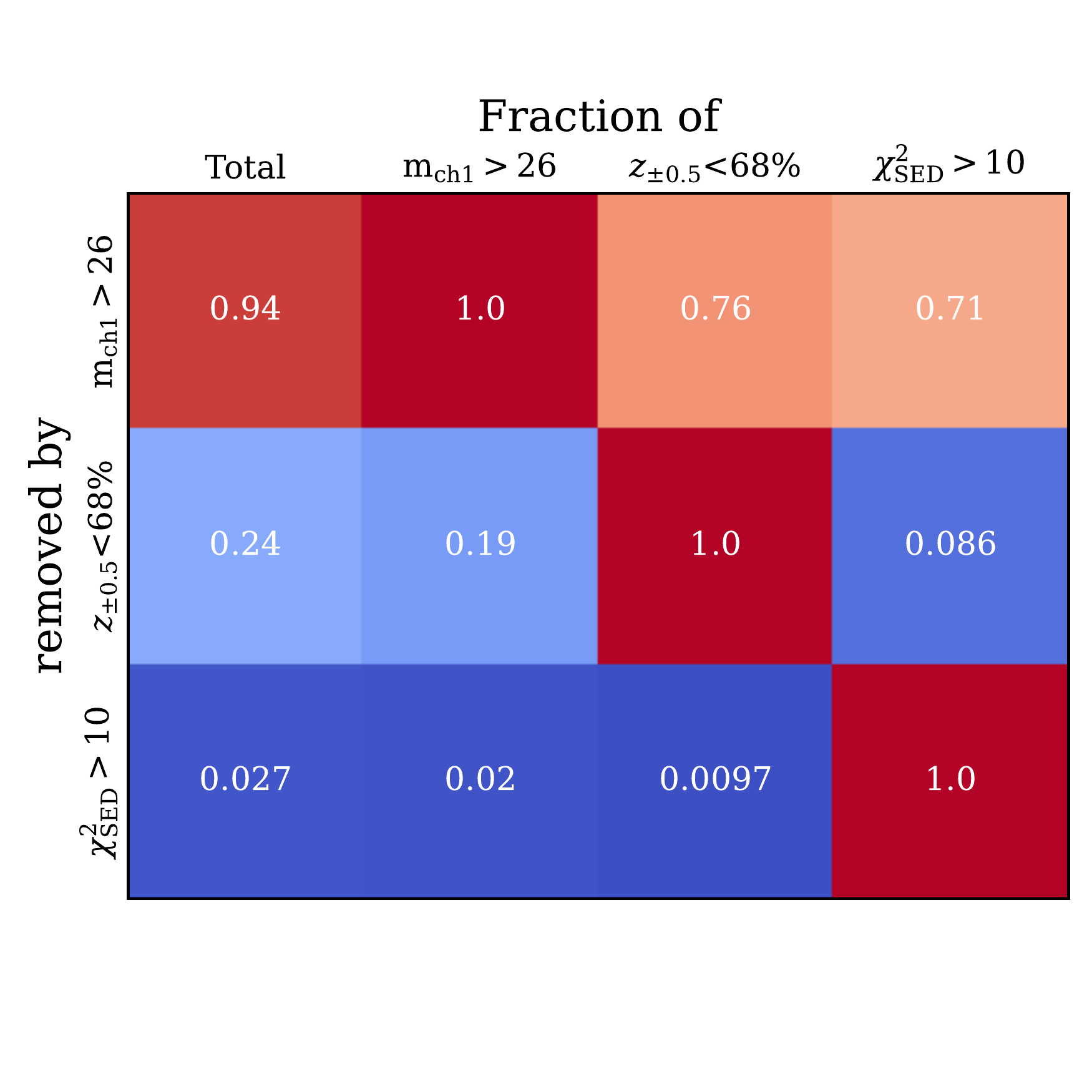}}
  \caption{Fraction of sources removed by a single criterion that are also removed by another, after applying the primary criteria on area and \photoz{}. E.g., 71\% of $\chi_{\rm SED}^2\,>\,10$ sources have $m_{\rm ch1}\,>\,26$.}
  \label{fig:cuts}
\end{figure}

As discussed in Section~\ref{sec:selection}, we select sources from the 1.27\,deg\textsuperscript{2} \texttt{COMBINED} region for sources $0.2<z\leq6.5$, and impose an additional restriction of the ultra-deep stripe for sources $6.5<z\leq7.5$. The three remaining criteria are aimed to ensure secure \photoz{} and $\mathcal{M}$ based on the IRAC ch1 magnitude, SED $\chi^2$ fit quality, and $\mathcal{L}(z)$. As shown in Fig.~\ref{fig:cuts}, there is considerable overlap between these three criteria, with m$_{\rm ch1}>26$ accounting for 94\% of the sample removed by combining these three. 

\section{Further comparisons with simulations}
\label{app:qg_fraction}

\begin{figure*}[t]
	\centering
	\includegraphics[width=17cm]{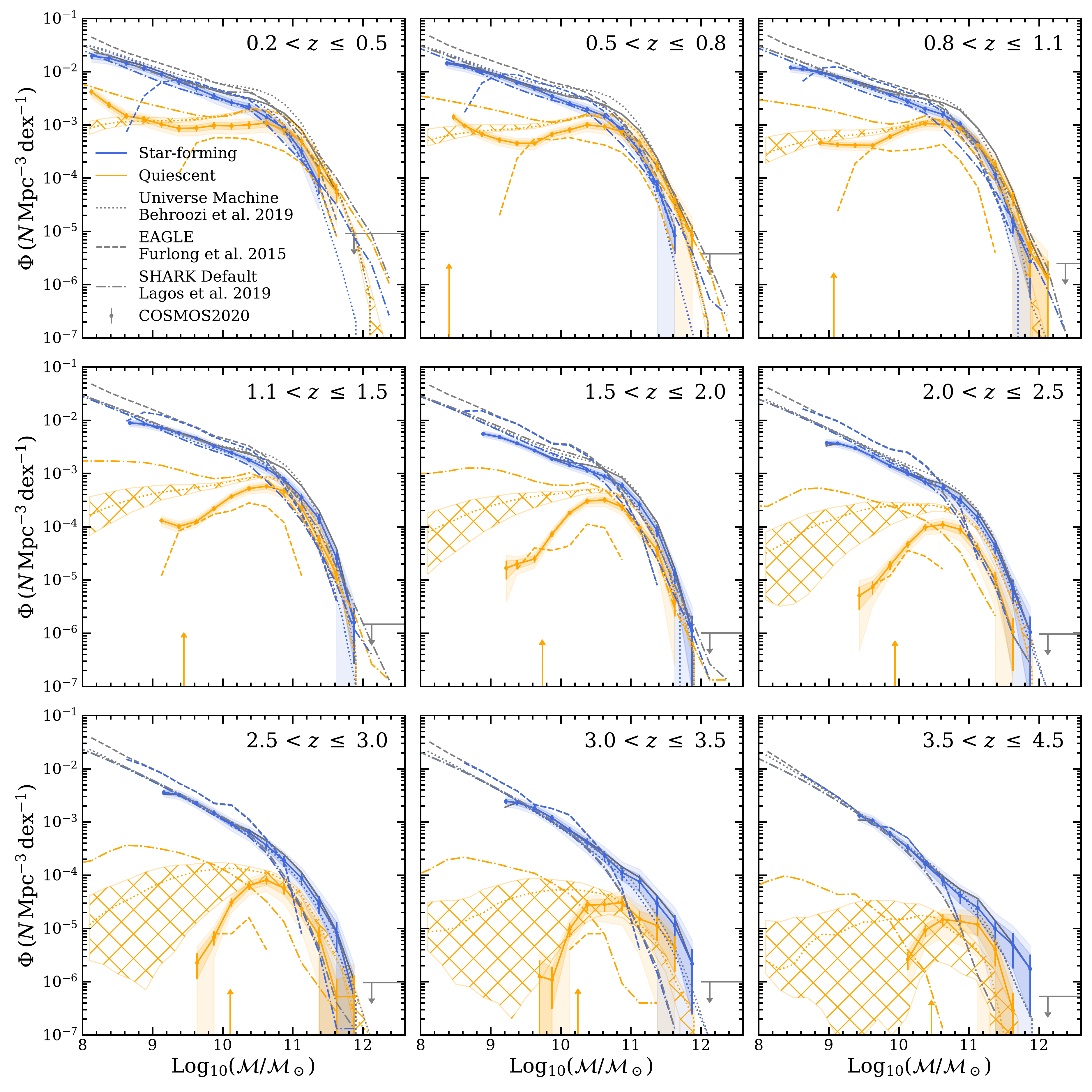}
	\caption{Galaxy stellar mass functions for total (gray), star-forming (blue), and quiescent (orange) samples from this work compared to simulations. We include those from \textsc{Universe Machine} \citet[][sSFR$<10^{-11}\,\Msol{}\,$yr$^{-1}$]{Behroozi2019} shown by dotted lines with hatched uncertainty envelopes, as well as consistently $NUVrJ$-selected samples from  \textsc{Eagle} \citep{Furlong2015_EAGLE} and  \textsc{Shark} \citep{Lagos2018_SHARK} shown by the dashed and dot-dashed lines, respectively. Upper limits for empty bins are shown by the horizontal gray line with an arrow. Mass incomplete measurements are not shown. 
	} 
	\label{fig:sfqg_sims}
\end{figure*}

\begin{figure*}[t]
	\centering
	\includegraphics[width=17cm]{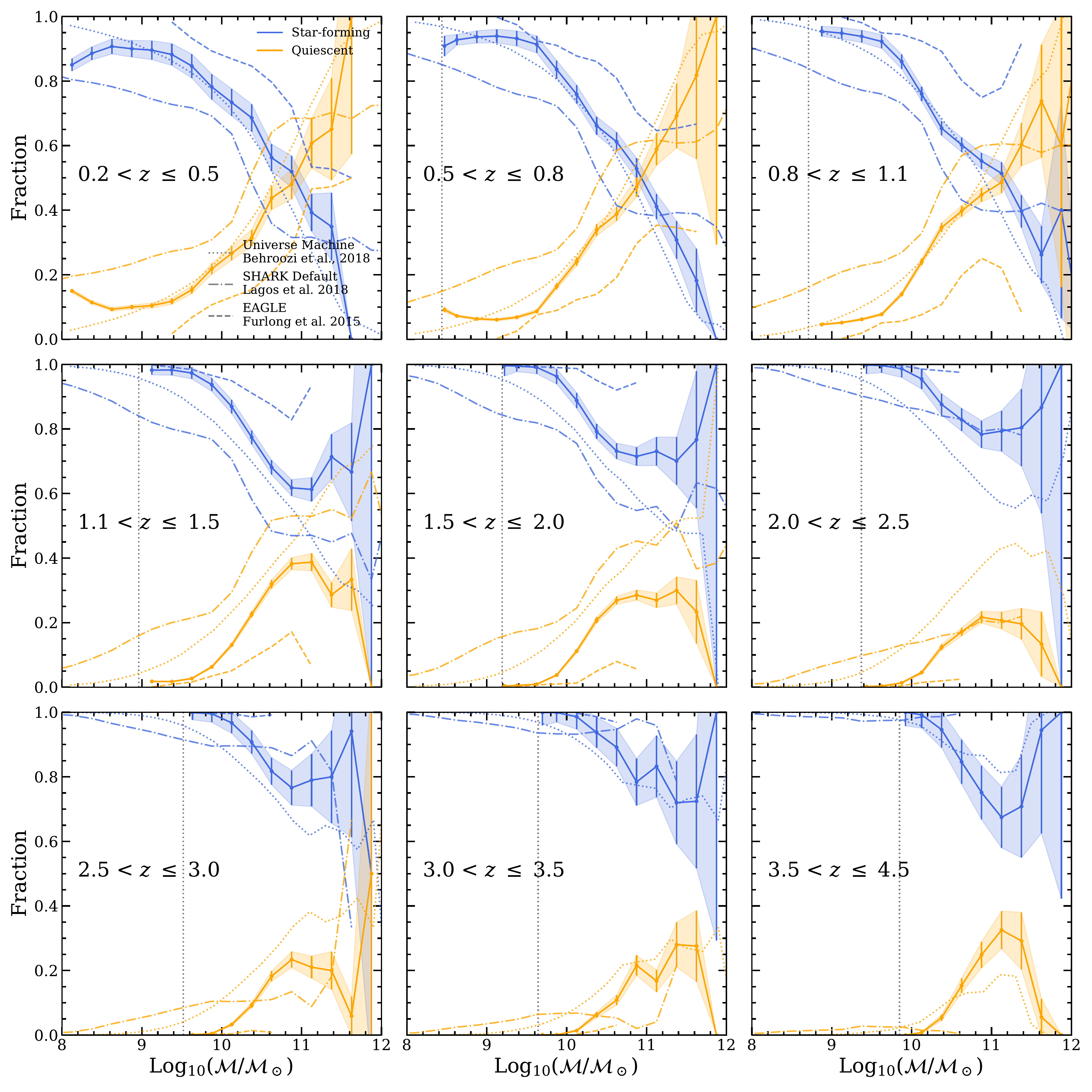}
	\caption{Fraction of star-forming (blue) and quiescent (orange) galaxy samples as a function of mass at for three redshift ranges compared to simulations. We include measurements from \textsc{Universe Machine} \citet[][sSFR$<10^{-11}\,\Msol{}\,$yr$^{-1}$]{Behroozi2019} shown by dotted lines, as well as those of consistently $NUVrJ$-selected samples from \textsc{Eagle} \citep{Furlong2015_EAGLE} and  \textsc{Shark} \citep{Lagos2018_SHARK}.}
	\label{fig:sfqg_frac}
\end{figure*}

Here we include additional figures comparing the total, star-forming, and quiescent mass functions (Fig.~\ref{fig:sfqg_sims}) and quiescent fractions (Fig.~\ref{fig:sfqg_frac}) from this work to those of \textsc{Universe Machine} \citet[][sSFR$<10^{-11}\,\Msol{}\,$yr$^{-1}$]{Behroozi2019} as well as those of consistently $NUVrJ$-selected quiescent galaxies from  \textsc{Eagle} \citep{Furlong2015_EAGLE} and  \textsc{Shark} \citep{Lagos2018_SHARK}. 

\section{Fitting Results}
\label{app:fitting}

\begin{figure*}[t]
	\centering
	\includegraphics[width=17cm]{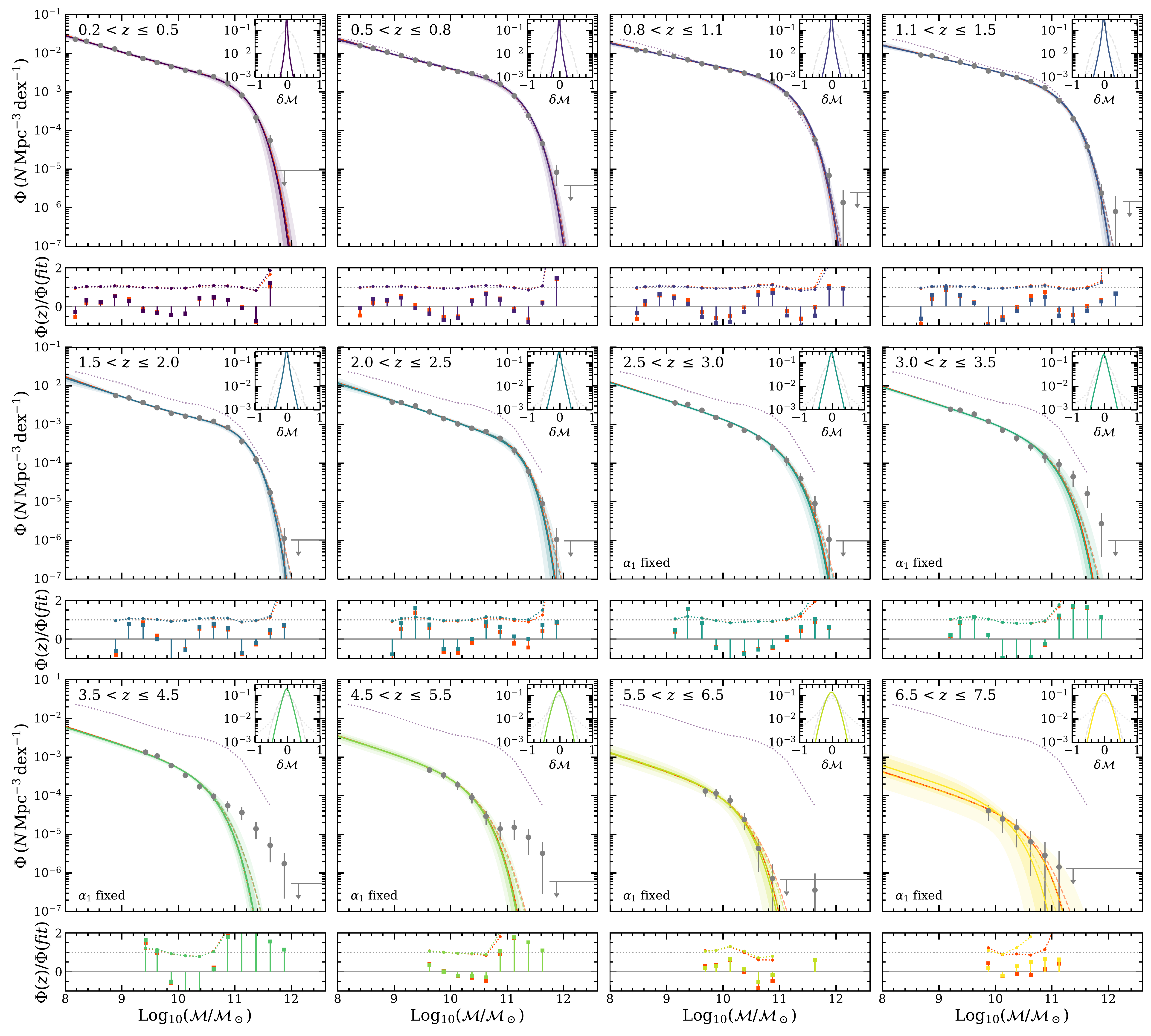}
	\caption{Results of fitting a double ($z<3$) and single Schechter ($z\geq3$) functional forms to the observed galaxy stellar mass function. Measurements are binned in redshift and inferred from both $\chi^2$ minimization (red) and likelihood methods (evolving colors) taken at the median of each parameter posterior distribution. In both cases, Schechter functions are convolved with parameterized, redshift-dependent kernels in $\delta\,M$ (upper right subpanels) to account for Eddington bias (dashed lines) which is then removed in the corrected fit (solid lines). For completeness, the corrected maximum likelihood solutions are also indicated (dash-dot colored curves). In the lower panels, fitting residuals between the uncorrected kernel convolved fits and the data are shown by both the relative fractional difference (dashed curve) and the difference weighted by the uncertainty (square points). Corresponding parameters are shown in Table~\ref{table:fit_total}.
	}
	\label{fig:fit_total}
\end{figure*}

\begin{figure*}[t]
	\centering
	\includegraphics[width=17cm]{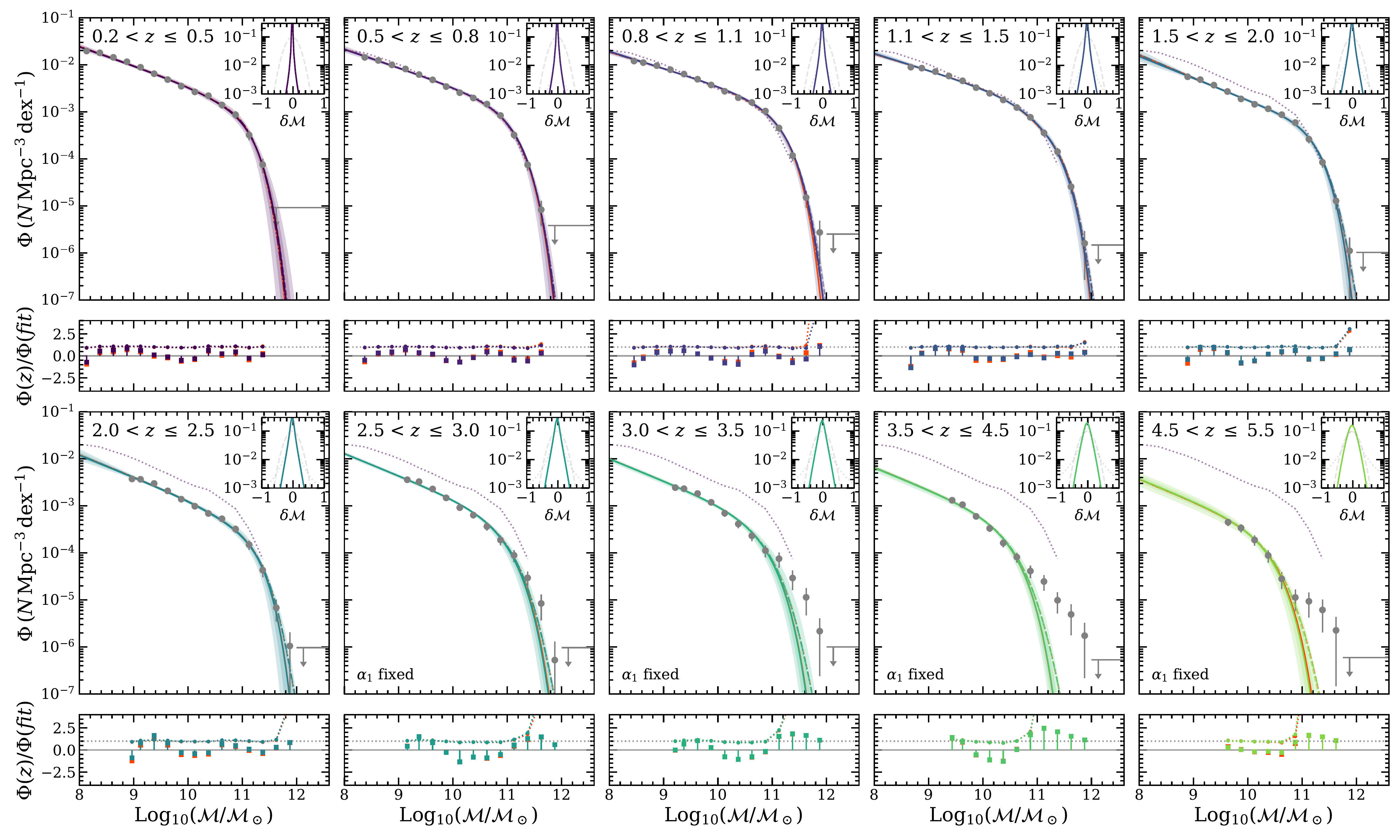}
	\caption{Results of fitting a double ($z\leq3.0$) and single Schechter ($z>3.0$) functional forms to the star-forming sample. The figure follows the same layout as Fig.~\ref{fig:fit_total}. Corresponding parameters are shown in Table~\ref{table:fit_sfg}.
	}
	\label{fig:fit_sfg}
\end{figure*}

\begin{figure*}[t]
	\centering
	\includegraphics[width=17cm]{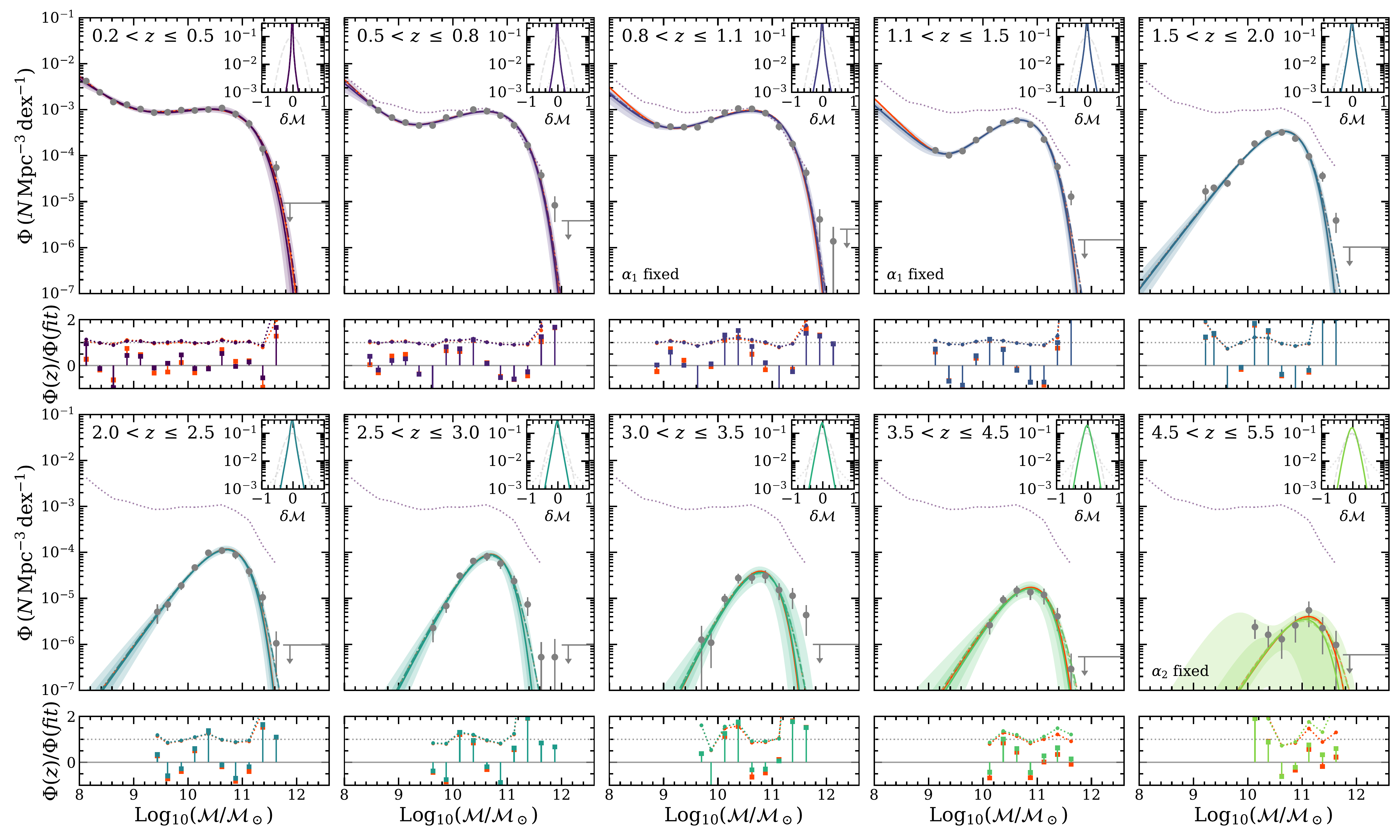}
	\caption{Results of fitting a double ($z\leq1.5$) and single Schechter ($z>1.5$) functional forms to the quiescent sample. The figure follows the same layout as Fig.~\ref{fig:fit_total}. Corresponding parameters are shown in Table~\ref{table:fit_qg}.}
	\label{fig:fit_qg}
\end{figure*}

\begin{figure*}[t]
	\centering
	\includegraphics[width=17cm]{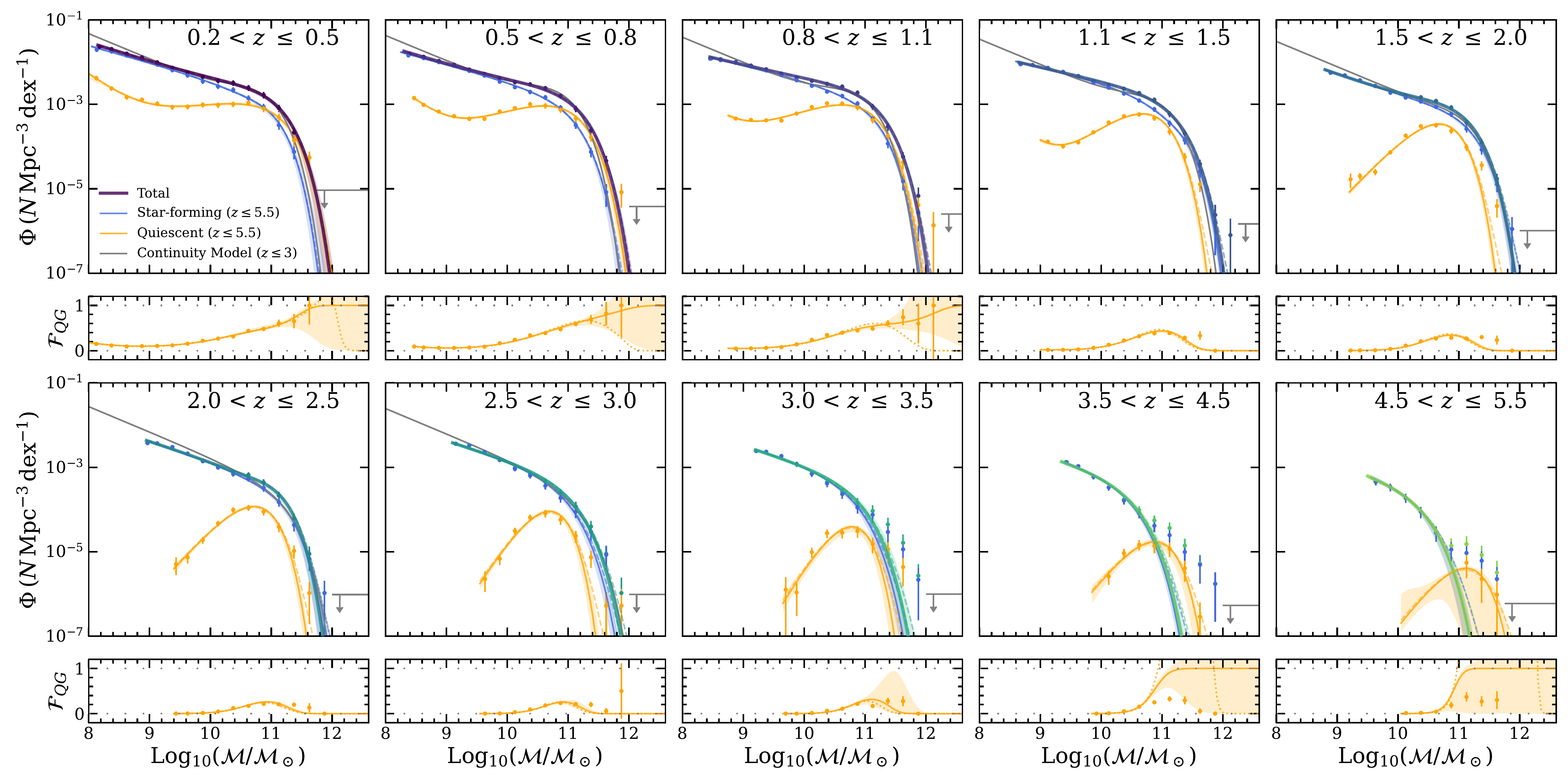}
	\caption{Best fit Schechter models corresponding to maximum likelihood for the total, star-forming, and quiescent samples (solid colored curves) based on the observed data (colored points). Fits for the median posterior will be similar for symmetric parameter posterior distributions, which are found in all but the last bin shown here. Kernel convolved fits are shown by the dashed curves. Fits measured using the continuity model are shown in gray solid curves. Lower panels indicate the fraction of quiescent galaxies $\mathcal{F}_{\rm QG}\equiv\Phi_{\rm QG}\,/\,(\Phi_{\rm SF} + \Phi_{\rm QG}$) as a function of ${\rm log}_{10}(\mathcal{M}/\mathcal{M}_\odot)$ for the data (colored points) and maximum likelihood fits (solid curve). Since the star-forming and quiescent fits are not required to sum to the total fit, changing the definition to $\mathcal{F}_{\rm QG}\equiv\Phi_{\rm QG}\,/\,\Phi_{\rm Total}$ as reported by \citet{ilbert_mass_2013} and \citet{Davidzon17_mass} produces the dotted curves with noticeably lower $\mathcal{F}_{\rm QG}$ at high $\mathcal{M}$.}
	\label{fig:fit_summary_alt}
\end{figure*}

Here we include the inferred total, star-forming, and quiescent mass functions (Figs.~\ref{fig:fit_total}, \ref{fig:fit_sfg}, and \ref{fig:fit_qg}) optimized by both $\chi^2$ and Likelihood maximization; their derived parameters are contained in Tables~\ref{table:fit_total}, \ref{table:fit_sfg}, and \ref{table:fit_qg}, respectively. These results are summarized together in Fig.~\ref{fig:fit_summary_alt}, along with quiescent fractions as a function of mass (defined three ways). Shown in the same figure are the SMFs realized by evaluation of the continuity model \citep[based on][]{Leja2019}, with related parameters contained in Table~\ref{table:fit_cont}. For those interested in parameter covariances, we have made the full Markov chains available to download\footnote{ \url{https://doi.org/10.5281/zenodo.7808832}}.


\renewcommand{\arraystretch}{1.4}

\begin{table*}
\caption{Double ($z\leq3$) and single ($z>3$) Schechter parameters derived for the total mass complete sample.}
\footnotesize
\begin{threeparttable}
\resizebox{\textwidth}{!}{%
\begin{tabular}[]{ccccccc}
\hline
$z$-bin  & ${\rm Log}_{\rm 10}\,\mathcal{M}^*$ & $\alpha_1$ & $\Phi_1\,\times\,10^{-3}$ & $\alpha_2$ & $\Phi_2\,\times\,10^{-3}$ & $\rho_{*}(\mathcal{M}\leq10^8)\,\times\,10^{7}$ \\
  & ($\mathcal{M}_\odot$) &   & ($\mathrm{Mpc}^{-3}\,\mathrm{dex}^{-1}$) &   & ($\mathrm{Mpc}^{-3}\,\mathrm{dex}^{-1}$) & ($\mathcal{M}_{\odot}\,\mathrm{Mpc}^{-3}$) \\
\hline
Likelihood Fit\tnote{a} &&&&& \\
\hline
$0.2<z\leq0.5$ & $10.89_{-0.14}^{+0.14} [10.92]$ & $-1.42_{-0.06}^{+0.05} [-1.46]$ & $0.73_{-0.27}^{+0.25} [0.59]$ & $-0.46_{-0.46}^{+0.50} [-0.66]$ & $1.09_{-0.54}^{+0.50} [1.12]$ & $15.91_{-1.43}^{+1.50} [16.03]$ \\
$0.5<z\leq0.8$ & $10.96_{-0.10}^{+0.10} [10.98]$ & $-1.39_{-0.07}^{+0.05} [-1.45]$ & $0.66_{-0.27}^{+0.22} [0.48]$ & $-0.61_{-0.39}^{+0.46} [-0.79]$ & $0.83_{-0.43}^{+0.37} [0.95]$ & $15.49_{-1.17}^{+1.25} [15.44]$ \\
$0.8<z\leq1.1$ & $11.02_{-0.09}^{+0.08} [11.01]$ & $-1.32_{-0.06}^{+0.04} [-1.35]$ & $0.84_{-0.31}^{+0.20} [0.71]$ & $-0.63_{-0.44}^{+0.48} [-0.72]$ & $0.66_{-0.42}^{+0.34} [0.74]$ & $17.51_{-1.27}^{+1.33} [16.80]$ \\
$1.1<z\leq1.5$ & $11.00_{-0.11}^{+0.07} [11.00]$ & $-1.33_{-0.05}^{+0.04} [-1.34]$ & $0.72_{-0.23}^{+0.15} [0.68]$ & $-0.51_{-0.62}^{+0.62} [-0.57]$ & $0.34_{-0.25}^{+0.30} [0.34]$ & $12.64_{-0.82}^{+0.86} [12.27]$ \\
$1.5<z\leq2.0$ & $10.86_{-0.08}^{+0.07} [10.89]$ & $-1.48_{-0.09}^{+0.07} [-1.52]$ & $0.29_{-0.11}^{+0.11} [0.23]$ & $-0.43_{-0.31}^{+0.37} [-0.56]$ & $0.64_{-0.16}^{+0.13} [0.65]$ & $7.62_{-0.63}^{+0.62} [7.63]$ \\
$2.0<z\leq2.5$ & $10.78_{-0.14}^{+0.16} [10.79]$ & $-1.46_{-0.06}^{+0.05} [-1.47]$ & $0.27_{-0.08}^{+0.09} [0.26]$ & $0.07_{-0.60}^{+0.58} [0.07]$ & $0.27_{-0.12}^{+0.12} [0.31]$ & $4.25_{-0.54}^{+0.58} [4.52]$ \\
$2.5<z\leq3.0$ & $10.97_{-0.07}^{+0.06} [10.98]$ & $-1.46$ & $0.24_{-0.02}^{+0.03} [0.23]$ & -- & -- & $3.51_{-0.33}^{+0.33} [3.54]$ \\
$3.0<z\leq3.5$ & $10.83_{-0.09}^{+0.11} [10.83]$ & $-1.46$ & $0.21_{-0.03}^{+0.03} [0.21]$ & -- & -- & $2.22_{-0.28}^{+0.24} [2.22]$ \\
$3.5<z\leq4.5$ & $10.46_{-0.06}^{+0.09} [10.44]$ & $-1.46$ & $0.20_{-0.03}^{+0.03} [0.20]$ & -- & -- & $0.87_{-0.08}^{+0.08} [0.87]$ \\
$4.5<z\leq5.5$ & $10.30_{-0.10}^{+0.10} [10.32]$ & $-1.46$ & $0.14_{-0.03}^{+0.04} [0.13]$ & -- & -- & $0.42_{-0.06}^{+0.07} [0.42]$ \\
$5.5<z\leq6.5$ & $10.14_{-0.12}^{+0.10} [10.18]$ & $-1.46$ & $0.06_{-0.02}^{+0.03} [0.05]$ & -- & -- & $0.13_{-0.03}^{+0.03} [0.13]$ \\
$6.5<z\leq7.5$ & $10.18_{-0.27}^{+0.37} [10.49]$ & $-1.46$ & $0.03_{-0.02}^{+0.03} [0.01]$ & -- & -- & $0.06_{-0.02}^{+0.03} [0.06]$ \\
\hline
$\chi^2$ Fit &&&&&& \\
\hline
$0.2<z\leq0.5$ & $10.93\pm0.09(0.14)$ & $-1.45\pm0.05(0.08)$ & $0.60\pm0.21(0.35)$ & $-0.66\pm0.33(0.54)$ & $1.07\pm0.32(0.51)$ & $16.15_{-5.44}^{+7.52}$ \\
$0.5<z\leq0.8$ & $10.98\pm0.08(0.11)$ & $-1.44\pm0.08(0.12)$ & $0.48\pm0.29(0.41)$ & $-0.79\pm0.36(0.50)$ & $0.92\pm0.27(0.37)$ & $16.73_{-5.98}^{+8.09}$ \\
$0.8<z\leq1.1$ & $11.01\pm0.09(0.11)$ & $-1.35\pm0.07(0.09)$ & $0.70\pm0.35(0.47)$ & $-0.72\pm0.45(0.61)$ & $0.77\pm0.34(0.46)$ & $18.89_{-7.72}^{+9.13}$ \\
$1.1<z\leq1.5$ & $11.01\pm0.10(0.13)$ & $-1.34\pm0.04(0.06)$ & $0.69\pm0.18(0.25)$ & $-0.56\pm0.62(0.84)$ & $0.32\pm0.24(0.33)$ & $11.97_{-4.12}^{+6.51}$ \\
$1.5<z\leq2.0$ & $10.88\pm0.08(0.08)$ & $-1.52\pm0.10(0.11)$ & $0.23\pm0.12(0.13)$ & $-0.55\pm0.34(0.37)$ & $0.67\pm0.15(0.16)$ & $7.72_{-2.82}^{+3.32}$ \\
$2.0<z\leq2.5$ & $10.80\pm0.14(0.15)$ & $-1.46\pm0.06(0.06)$ & $0.26\pm0.09(0.10)$ & $0.04\pm0.63(0.66)$ & $0.29\pm0.11(0.11)$ & $4.48_{-2.04}^{+4.07}$ \\
$2.5<z\leq3.0$ & $11.00\pm0.06(0.06)$ & $-1.46$ & $0.22\pm0.02(0.02)$ & -- & -- & $3.69_{-0.47}^{+0.50}$ \\
$3.0<z\leq3.5$ & $10.85\pm0.14(0.12)$ & $-1.46$ & $0.20\pm0.04(0.04)$ & -- & -- & $2.36_{-0.95}^{+1.20}$ \\
$3.5<z\leq4.5$ & $10.46\pm0.18(0.10)$ & $-1.46$ & $0.20\pm0.07(0.04)$ & -- & -- & $0.81_{-0.36}^{+0.50}$ \\
$4.5<z\leq5.5$ & $10.33\pm0.12(0.10)$ & $-1.46$ & $0.13\pm0.04(0.03)$ & -- & -- & $0.46_{-0.16}^{+0.18}$ \\
$5.5<z\leq6.5$ & $10.19\pm0.05(0.09)$ & $-1.46$ & $0.05\pm0.01(0.02)$ & -- & -- & $0.13_{-0.03}^{+0.03}$ \\
$6.5<z\leq7.5$ & $10.50\pm0.14(0.36)$ & $-1.46$ & $0.01\pm0.00(0.01)$ & -- & -- & $0.05_{-0.02}^{+0.05}$ \\
\hline
\end{tabular}
}
\begin{tablenotes}
\item[a] For the Likelihood fit, values are shown for the median posterior distributions with even-tailed $68\%$ range and the values corresponding to the maximum likelihood solution in brackets. 
\item[b] For the $\chi^2$ regression fit, uncertainties on parameter values are shown multiplied by $\sqrt{\chi^2_N}$ with formal uncertainties following in brackets.
\end{tablenotes}
\end{threeparttable}
\label{table:fit_total}
\end{table*}

\begin{table*}
\caption{Double ($z\leq3$) and single ($z>3$) Schechter parameters derived for the star-forming mass complete subsample.}
\footnotesize
\begin{threeparttable}
\resizebox{\textwidth}{!}{
\begin{tabular}{ccccccc}
\hline
$z$-bin  & ${\rm Log}_{\rm 10}\,\mathcal{M}^*$ & $\alpha_1$ & $\Phi_1\,\times\,10^{-3}$ & $\alpha_2$ & $\Phi_2\,\times\,10^{-3}$ & $\rho_{*}(\mathcal{M}\leq10^8)\,\times\,10^{7}$ \\
  & ($\mathcal{M}_\odot$) &   & ($\mathrm{Mpc}^{-3}\,\mathrm{dex}^{-1}$) &   & ($\mathrm{Mpc}^{-3}\,\mathrm{dex}^{-1}$) & ($\mathcal{M}_{\odot}\,\mathrm{Mpc}^{-3}$) \\
\hline
Likelihood Fit\tnote{a} &&&&& \\
\hline
$0.2<z\leq0.5$ & $10.73_{-0.15}^{+0.17} [10.67]$ & $-1.41_{-0.04}^{+0.03} [-1.41]$ & $0.80_{-0.22}^{+0.24} [0.87]$ & $-0.02_{-0.79}^{+0.61} [0.14]$ & $0.49_{-0.33}^{+0.32} [0.63]$ & $8.98_{-0.82}^{+1.00} [9.19]$ \\
$0.5<z\leq0.8$ & $10.83_{-0.13}^{+0.10} [10.83]$ & $-1.39_{-0.05}^{+0.03} [-1.40]$ & $0.71_{-0.19}^{+0.17} [0.70]$ & $-0.32_{-0.75}^{+0.60} [-0.36]$ & $0.36_{-0.25}^{+0.28} [0.32]$ & $9.05_{-0.67}^{+0.67} [8.89]$ \\
$0.8<z\leq1.1$ & $10.92_{-0.10}^{+0.08} [10.82]$ & $-1.36_{-0.03}^{+0.02} [-1.35]$ & $0.73_{-0.16}^{+0.10} [0.84]$ & $-0.36_{-0.73}^{+0.49} [0.06]$ & $0.29_{-0.21}^{+0.27} [0.39]$ & $10.52_{-0.70}^{+0.74} [10.25]$ \\
$1.1<z\leq1.5$ & $10.94_{-0.09}^{+0.11} [10.87]$ & $-1.36_{-0.03}^{+0.02} [-1.35]$ & $0.65_{-0.13}^{+0.11} [0.74]$ & $0.37_{-1.11}^{+0.59} [0.79]$ & $0.11_{-0.08}^{+0.11} [0.13]$ & $8.95_{-0.56}^{+0.64} [9.00]$ \\
$1.5<z\leq2.0$ & $10.90_{-0.11}^{+0.12} [10.93]$ & $-1.47_{-0.07}^{+0.05} [-1.49]$ & $0.29_{-0.10}^{+0.09} [0.25]$ & $-0.42_{-0.52}^{+0.52} [-0.56]$ & $0.31_{-0.15}^{+0.14} [0.31]$ & $5.92_{-0.46}^{+0.46} [5.98]$ \\
$2.0<z\leq2.5$ & $10.85_{-0.16}^{+0.17} [10.86]$ & $-1.47_{-0.05}^{+0.05} [-1.47]$ & $0.24_{-0.07}^{+0.09} [0.24]$ & $0.01_{-0.85}^{+0.66} [-0.03]$ & $0.13_{-0.09}^{+0.10} [0.13]$ & $3.60_{-0.36}^{+0.37} [3.69]$ \\
$2.5<z\leq3.0$ & $10.87_{-0.07}^{+0.07} [10.88]$ & $-1.47$ & $0.25_{-0.03}^{+0.03} [0.25]$ & -- & -- & $3.00_{-0.28}^{+0.23} [3.03]$ \\
$3.0<z\leq3.5$ & $10.74_{-0.08}^{+0.09} [10.75]$ & $-1.47$ & $0.22_{-0.03}^{+0.03} [0.22]$ & -- & -- & $1.98_{-0.21}^{+0.28} [1.98]$ \\
$3.5<z\leq4.5$ & $10.42_{-0.06}^{+0.08} [10.41]$ & $-1.47$ & $0.20_{-0.03}^{+0.03} [0.21]$ & -- & -- & $0.85_{-0.08}^{+0.09} [0.85]$ \\
$4.5<z\leq5.5$ & $10.30_{-0.10}^{+0.10} [10.32]$ & $-1.47$ & $0.14_{-0.03}^{+0.04} [0.13]$ & -- & -- & $0.42_{-0.06}^{+0.07} [0.42]$ \\
\hline
$\chi^2$ Fit &&&&&& \\
\hline
$0.2<z\leq0.5$ & $10.69\pm0.10(0.17)$ & $-1.41\pm0.02(0.04)$ & $0.84\pm0.16(0.25)$ & $0.05\pm0.44(0.70)$ & $0.60\pm0.21(0.33)$ & $9.05_{-2.76}^{+4.02}$ \\
$0.5<z\leq0.8$ & $10.81\pm0.11(0.18)$ & $-1.40\pm0.02(0.04)$ & $0.72\pm0.13(0.23)$ & $-0.28\pm0.49(0.83)$ & $0.35\pm0.21(0.36)$ & $8.62_{-2.57}^{+3.58}$ \\
$0.8<z\leq1.1$ & $10.82\pm0.12(0.16)$ & $-1.35\pm0.03(0.03)$ & $0.85\pm0.16(0.21)$ & $0.07\pm0.52(0.70)$ & $0.39\pm0.21(0.29)$ & $10.77_{-3.41}^{+4.73}$ \\
$1.1<z\leq1.5$ & $10.88\pm0.09(0.12)$ & $-1.35\pm0.02(0.03)$ & $0.73\pm0.11(0.15)$ & $0.83\pm0.45(0.61)$ & $0.11\pm0.07(0.10)$ & $9.66_{-2.76}^{+3.06}$ \\
$1.5<z\leq2.0$ & $10.92\pm0.10(0.12)$ & $-1.49\pm0.06(0.08)$ & $0.26\pm0.10(0.13)$ & $-0.51\pm0.47(0.60)$ & $0.32\pm0.11(0.14)$ & $5.80_{-2.35}^{+2.61}$ \\
$2.0<z\leq2.5$ & $10.85\pm0.22(0.23)$ & $-1.47\pm0.06(0.06)$ & $0.24\pm0.10(0.11)$ & $0.01\pm0.90(0.95)$ & $0.13\pm0.11(0.12)$ & $3.77_{-1.97}^{+4.20}$ \\
$2.5<z\leq3.0$ & $10.89\pm0.08(0.07)$ & $-1.47$ & $0.24\pm0.03(0.03)$ & -- & -- & $2.95_{-0.60}^{+0.72}$ \\
$3.0<z\leq3.5$ & $10.75\pm0.12(0.09)$ & $-1.47$ & $0.22\pm0.04(0.03)$ & -- & -- & $2.00_{-0.62}^{+1.24}$ \\
$3.5<z\leq4.5$ & $10.41\pm0.14(0.08)$ & $-1.47$ & $0.21\pm0.07(0.04)$ & -- & -- & $0.84_{-0.32}^{+0.35}$ \\
$4.5<z\leq5.5$ & $10.33\pm0.10(0.10)$ & $-1.47$ & $0.13\pm0.04(0.03)$ & -- & -- & $0.42_{-0.13}^{+0.21}$ \\
\hline
\end{tabular}
}
\begin{tablenotes}
\item[a] Values are provided as described in Table~\ref{table:fit_total}.
\end{tablenotes}
\end{threeparttable}
  \label{table:fit_sfg}
\end{table*}

\begin{table*}
\footnotesize
\caption{Double ($z\leq1.5$) and single ($z>1.5$) Schechter parameters derived for the quiescent mass complete subsample.}
\footnotesize
\begin{threeparttable}
\resizebox{\textwidth}{!}{%
\begin{tabular}{ccccccc}
\hline
$z$-bin  & ${\rm Log}_{\rm 10}\,\mathcal{M}^*$ & $\alpha_1$ & $\Phi_1\,\times\,10^{-5}$ & $\alpha_2$ & $\Phi_2\,\times\,10^{-5}$ & $\rho_{*}(\mathcal{M}\leq10^8)\,\times\,10^{7}$ \\
  & ($\mathcal{M}_\odot$) &   & ($\mathrm{Mpc}^{-3}\,\mathrm{dex}^{-1}$) &   & ($\mathrm{Mpc}^{-3}\,\mathrm{dex}^{-1}$) & ($\mathcal{M}_{\odot}\,\mathrm{Mpc}^{-3}$) \\
\hline
Likelihood Fit\tnote{a} &&&&& \\
\hline
$0.2<z\leq0.5$ & $10.90_{-0.08}^{+0.07} [10.97]$ & $-1.82_{-0.18}^{+0.15} [-2.05]$ & $0.80_{-0.56}^{+1.37} [0.17]$ & $-0.63_{-0.12}^{+0.15} [-0.76]$ & $90.81_{-15.04}^{+16.80} [79.76]$ & $6.68_{-0.78}^{+0.74} [6.88]$ \\
$0.5<z\leq0.8$ & $10.88_{-0.05}^{+0.05} [10.90]$ & $-2.01_{-0.17}^{+0.16} [-2.12]$ & $0.19_{-0.12}^{+0.26} [0.11]$ & $-0.46_{-0.07}^{+0.07} [-0.49]$ & $94.44_{-9.74}^{+8.07} [93.41]$ & $6.49_{-0.69}^{+0.59} [6.62]$ \\
$0.8<z\leq1.1$ & $10.89_{-0.04}^{+0.04} [10.87]$ & $-2.01$ & $0.11_{-0.03}^{+0.03} [0.12]$ & $-0.47_{-0.05}^{+0.04} [-0.43]$ & $96.00_{-7.51}^{+4.64} [102.25]$ & $6.67_{-0.65}^{+0.54} [6.81]$ \\
$1.1<z\leq1.5$ & $10.63_{-0.04}^{+0.04} [10.64]$ & $-2.01$ & $0.12_{-0.03}^{+0.03} [0.11]$ & $0.11_{-0.09}^{+0.09} [0.09]$ & $70.06_{-4.47}^{+4.25} [69.96]$ & $3.19_{-0.27}^{+0.29} [3.24]$ \\
$1.5<z\leq2.0$ & $10.48_{-0.05}^{+0.05} [10.49]$ & -- & -- & $0.54_{-0.10}^{+0.11} [0.53]$ & $35.38_{-1.84}^{+1.83} [35.77]$ & $1.47_{-0.16}^{+0.17} [1.49]$ \\
$2.0<z\leq2.5$ & $10.49_{-0.06}^{+0.06} [10.50]$ & -- & -- & $0.66_{-0.15}^{+0.16} [0.64]$ & $11.49_{-0.87}^{+0.86} [11.76]$ & $0.53_{-0.07}^{+0.07} [0.55]$ \\
$2.5<z\leq3.0$ & $10.33_{-0.08}^{+0.08} [10.33]$ & -- & -- & $1.30_{-0.26}^{+0.30} [1.31]$ & $5.66_{-1.10}^{+0.91} [5.78]$ & $0.33_{-0.07}^{+0.06} [0.34]$ \\
$3.0<z\leq3.5$ & $10.41_{-0.13}^{+0.15} [10.39]$ & -- & -- & $1.41_{-0.44}^{+0.47} [1.49]$ & $2.12_{-0.70}^{+0.62} [2.13]$ & $0.16_{-0.05}^{+0.06} [0.17]$ \\
$3.5<z\leq4.5$ & $10.58_{-0.12}^{+0.10} [10.61]$ & -- & -- & $0.95_{-0.37}^{+0.46} [0.87]$ & $1.36_{-0.33}^{+0.26} [1.52]$ & $0.10_{-0.03}^{+0.03} [0.11]$ \\
$4.5<z\leq5.5$ & $10.76_{-0.58}^{+0.15} [10.82]$ & -- & -- & $0.95$ & $0.29_{-0.10}^{+0.11} [0.34]$ & $0.03_{-0.03}^{+0.03} [0.04]$ \\
\hline
$\chi^2$ Fit &&&&&& \\
\hline
$0.2<z\leq0.5$ & $10.97\pm0.06(0.08)$ & $-2.04\pm0.17(0.24)$ & $0.18\pm0.21(0.29)$ & $-0.75\pm0.08(0.12)$ & $79.50\pm11.69(16.20)$ & $6.84_{-1.23}^{+1.70}$ \\
$0.5<z\leq0.8$ & $10.90\pm0.05(0.05)$ & $-2.14\pm0.21(0.22)$ & $0.09\pm0.11(0.12)$ & $-0.50\pm0.08(0.09)$ & $92.03\pm11.27(11.33)$ & $6.87_{-1.15}^{+1.64}$ \\
$0.8<z\leq1.1$ & $10.87\pm0.07(0.05)$ & $-2.14$ & $0.07\pm0.03(0.02)$ & $-0.44\pm0.09(0.07)$ & $102.65\pm14.27(11.57)$ & $6.81_{-1.59}^{+1.84}$ \\
$1.1<z\leq1.5$ & $10.65\pm0.06(0.04)$ & $-2.14$ & $0.07\pm0.03(0.02)$ & $0.08\pm0.12(0.09)$ & $69.89\pm6.29(4.54)$ & $3.15_{-0.57}^{+0.60}$ \\
$1.5<z\leq2.0$ & $10.49\pm0.10(0.05)$ & -- & -- & $0.53\pm0.21(0.10)$ & $35.74\pm3.72(1.84)$ & $1.46_{-0.49}^{+0.65}$ \\
$2.0<z\leq2.5$ & $10.50\pm0.07(0.06)$ & -- & -- & $0.64\pm0.17(0.15)$ & $11.76\pm0.90(0.83)$ & $0.52_{-0.13}^{+0.15}$ \\
$2.5<z\leq3.0$ & $10.33\pm0.10(0.08)$ & -- & -- & $1.30\pm0.34(0.28)$ & $5.80\pm1.28(1.03)$ & $0.33_{-0.12}^{+0.20}$ \\
$3.0<z\leq3.5$ & $10.39\pm0.19(0.12)$ & -- & -- & $1.49\pm0.66(0.43)$ & $2.12\pm1.06(0.70)$ & $0.12_{-0.09}^{+0.26}$ \\
$3.5<z\leq4.5$ & $10.61\pm0.08(0.11)$ & -- & -- & $0.88\pm0.30(0.39)$ & $1.51\pm0.21(0.27)$ & $0.11_{-0.04}^{+0.07}$ \\
$4.5<z\leq5.5$ & $10.84\pm0.13(0.11)$ & -- & -- & $0.88$ & $0.35\pm0.11(0.10)$ & $0.04_{-0.01}^{+0.03}$ \\
\hline
\end{tabular}
}
\begin{tablenotes}
\item[a] Values are provided as described in Table~\ref{table:fit_total}. The scaling factor on $\Phi_1$ and $\Phi_2$ is now $10^{-5}$.
\end{tablenotes}
\end{threeparttable}
  \label{table:fit_qg}
\end{table*}

\begin{table*}
\footnotesize
\caption{Double Schechter function parameters ($z\leq3$) derived for the total sample from the continuity model fitting.}
\footnotesize
\begin{threeparttable}
\resizebox{\textwidth}{!}{%
\begin{tabular}{ccccccc}
\hline
$z$-fix  & ${\rm Log}_{\rm 10}\,\mathcal{M}^*$ & $\alpha_1$ & $\Phi_1\,\times\,10^{-3}$ & $\alpha_2$ & $\Phi_2\,\times\,10^{-3}$ & $\rho_{*}(\mathcal{M}\geq10^8)\,\times\,10^{7}$ \\
  & ($\Msol{}$) &   & ($\mathrm{Mpc}^{-3}\,\mathrm{dex}^{-1}$) &  & ($\mathrm{Mpc}^{-3}\,\mathrm{dex}^{-1}$) & ($\mathcal{M}_{\odot}\,\mathrm{Mpc}^{-3}$) \\
\hline
continuity model\tnote{a} &&&&& \\
\hline
$0.2$ & $10.69_{-0.01}^{+0.03}[10.71]$ & $-0.38_{-0.02}^{+0.04}[-0.38]$ & $1.843_{-0.078}^{+0.162}[1.922]$ & $-1.60_{-5E-5}^{+2E-4}[-1.60]$ & $0.504_{-0.011}^{+0.021}[0.512]$ & $12.25[13.258]$ \\
$1.5$ & $10.82_{-0.01}^{+0.02}[10.84]$ & $-0.38_{-0.02}^{+0.04}[-0.38]$ & $0.874_{-0.024}^{+0.049}[0.878]$ & $-1.60_{-5E-5}^{+2E-4}[-1.60]$ & $0.283_{-0.004}^{+0.009}[0.284]$ & $8.49[8.887]$ \\
$3.0$ & $11.02_{-0.02}^{+0.04}[11.05]$ & $-0.38_{-0.02}^{+0.04}[-0.38]$ & $0.005_{-0.001}^{+0.002}[0.005]$ & $-1.60_{-5E-5}^{+2E-4}[-1.60]$ & $0.148_{-0.005}^{+0.009}[0.151]$ & $2.90[3.172]$ \\
\hline
\end{tabular}
}
\begin{tablenotes}
\item[a] Values are provided as described in Table~\ref{table:fit_total} for the three fixed $z$ points. The parameters $\alpha_1$ and $\alpha_2$ are constant with $z$.
\end{tablenotes}
\end{threeparttable}
  \label{table:fit_cont}
\end{table*}

\end{appendix}
\end{document}